\documentclass[12pt]{article}
\usepackage[table]{xcolor}
\usepackage{amsmath,amssymb,exscale}
\usepackage{graphicx}
\usepackage{epsfig}
\usepackage{multicol}
\usepackage{color}
\usepackage{mathrsfs}
\usepackage{blindtext}
 \usepackage{fancyhdr}
\usepackage{hyperref}
\usepackage{cite}
\usepackage{mathtools}
\usepackage{amsmath}
\usepackage{rotating,slashed,amsmath,charter,xcolor,catchfilebetweentags,ifluatex}

\usepackage{graphicx}
\usepackage{sidecap}

\usepackage[latin1]{inputenc} 
\usepackage{multicol}  
 
 \usepackage{titlesec}
 
 \usepackage{rotating,slashed,xcolor,amsfonts,expdlist,charter}

\numberwithin{equation}{section}

\usepackage{xcolor}
\usepackage{sectsty}

\usepackage{mdframed}
\usepackage{titletoc}

\definecolor{secnum}{RGB}{13,151,225}
\definecolor{ptcbackground}{RGB}{212,237,252}
\definecolor{ptctitle}{RGB}{0,177,235}

\titlecontents{lsection}
  [5.8em]{\sffamily}
  {\color{secnum}\contentslabel{2.3em}\normalcolor}{}
  {\titlerule*[1000pc]{.}\contentspage\\\hspace*{-5.8em}\vspace*{5pt}%
    \color{white}\rule{\dimexpr\textwidth-15.5pt\relax}{1pt}}

\usepackage{hyperref}
\hypersetup{colorlinks,bookmarksopen,bookmarksnumbered,citecolor=blus,
linkcolor=redy,pdfstartview=FitH,urlcolor=blus}
\usepackage{slashed}

\definecolor{blus}{cmyk}{1,0.9,0,0.1}
\definecolor{verdes}{cmyk}{0.99,0,0.59,0.65}
\definecolor{rossos}{cmyk}{0,1,1,0.55}
\definecolor{redy}{cmyk}{0,1,1,0.7}
\definecolor{greeny}{cmyk}{0.99,0,0.59,0.98}
\definecolor{green-go}{cmyk}{0.79,0,0.59,0.5}

\usepackage{titlesec}

\def\Lag{\mathscr{L}}

\newcommand{\gs}{f_0}
\newcommand{\gt}{f_2}

\newcommand{\beq}{\begin{equation}}
\newcommand{\eeq}{\end{equation}}

\def\hhref#1{\href{http://arxiv.org/abs/#1}{arXiv:#1}} 

 \def\Lag{\mathscr{L}}
 
\newcommand{\tmtextbf}[1]{{\bfseries{#1}}}
\newcommand{\tmtextrm}[1]{{\rmfamily{#1}}}
\newcommand{\bp}{\bar M_P}

\def\be{\begin{equation}}
\def\ee{\end{equation}}
\def\ba{\begin{array} }

\def\bac{\begin{array} {c}}
\def\bacc{\begin{array} {cc}}
\def\baccc{\begin{array} {ccc}}
\def\bacccc{\begin{array} {cccc}}
\def\ea{\end{array}}
\def\bea{\begin{eqnarray}}
\def\eea{\end{eqnarray}}

\definecolor{red}{rgb}{1,0,0}

\def\psl{\hbox{\hbox{${p}$}}\kern-1.9mm{\hbox{${/}$}}}
\def\dsl{\hbox{\hbox{${\partial}$}}\kern-2.2mm{\hbox{${/}$}}}
\def\Dsl{\hbox{\hbox{${D}$}}\kern-2.6mm{\hbox{${/}$}}}

\def\Lag{\mathscr{L}}

\newcommand{\gappeq}{{\rlap{{\raise}.5ex\text{\ensuremath{>}}}{{\lower}.5ex\text{\ensuremath{\sim}}}}}
\newcommand{\lappeq}{{\rlap{{\raise}.5ex\text{\ensuremath{<}}}{{\lower}.5ex\text{\ensuremath{\sim}}}}}
\newcommand{\I}{\tmtextrm{1{\kern}-.24em l}}

\begin{document}
\topmargin -1.0cm
\oddsidemargin 0.9cm
\evensidemargin -0.5cm

{\vspace{-1cm}}
\begin{center}

\vspace{-1cm}


 {\Huge \tmtextbf{ 
\color{rossos} Dimensional  Transmutation \\  in 
Gravity and Cosmology}} {\vspace{.5cm}}\\
 
\vspace{1.9cm}

{\large  {\bf Alberto Salvio }
{\em  

\vspace{.4cm}

 Physics Department, University of Rome Tor Vergata, \\ 
via della Ricerca Scientifica, I-00133 Rome, Italy\\

\vspace{0.6cm}

I. N. F. N. -  Rome Tor Vergata,\\
via della Ricerca Scientifica, I-00133 Rome, Italy\\ 

\vspace{0.4cm}


\vspace{0.2cm}

 \vspace{0.5cm}
}


\vspace{0.2cm}

}
\vspace{0.cm}

%
%
%
%
%

\end{center}

%
%
\noindent --------------------------------------------------------------------------------------------------------------------------------

\begin{center}
{\bf \large Abstract}
\end{center}

\noindent We review (and extend) the analysis of general theories of all interactions (gravity included) where the mass scales are due to dimensional transmutation. Quantum consistency  requires the presence of terms in the action with four derivatives of the metric.
It is shown, nevertheless, how unitary is achieved and the classical Ostrogradsky instabilities can be avoided. The  four-derivative terms allow us to have a UV complete framework and a naturally small ratio between the Higgs mass and the Planck scale. Moreover, black holes of Einstein gravity with horizons smaller than a certain (microscopic) scale are replaced by horizonless ultracompact objects that are free from any singularity and  have interesting phenomenological applications. We also discuss the predictions that can be compared with observations of the microwave background radiation anisotropies and find that this scenario is viable and can be tested with future data.  Finally, how strong phase transitions can emerge in models of this type with approximate scale symmetry and how to test them  with GW detectors is reviewed and explained.

  \vspace{0.4cm}

\noindent --------------------------------------------------------------------------------------------------------------------------------

\vspace{1.1cm}


\vspace{2cm}

\noindent Email: alberto.salvio@roma2.infn.it


\newpage

\tableofcontents


\section{Introduction}\label{Introduction}

Most of the mass of the matter we have observed so far is due to the quantum phenomenon of dimensional transmutation (DT) in QCD. Only few percents of the mass of the proton are due to the quark masses, 
 which come from the Higgs mechanism. It is then natural to ask whether in fact all mass scales (including, among others, the Planck mass and the cosmological constant) can be generated through DT. It turns out, as we discuss in this review, that this is feasible and we refer to such possibility as complete dimensional transmutation (CDT). Theories with CDT are also sometimes called theories with ``classical scale invariance" because, modulo quantum effects,  scale symmetry is realized due to the absence of dimensionful quantities. 
 
 The consistency of these theories at the quantum level demands the presence of all local terms in the fundamental action with {\it dimensionless} parameters in 4D.  These allow us to reabsorb all divergences at all orders in perturbation theory. Should we not include a term of this sort, it would be unavoidably generated by quantum corrections. This implies, among other things, that the gravitational part of the action must include all possible local terms quadratic in the curvature tensors.
 
 There are several motivations for CDT. They will be reviewed in the other sections of this work, but we discuss the main ones already here at the qualitative level. 
 
  Regarding the observational motivations, we can mention the nearly scale invariance of primordial fluctuations. Indeed, this suggests that the underlying theory might be nearly scale invariant, as is the case in CDT when the quantum effects leading to the breaking of scale symmetry are small corrections; quantum physics in this context is then responsible for the small departure from scale invariance observed in the cosmic microwave background (CMB) anisotropies (see Ref.~\cite{Ade:2015lrj} for the most recent data).  
  
  Furthermore, phase transitions in models with approximate scale symmetry (as will be reviewed and explained here) are typically strong and may lead to observable gravitational waves (GWs) at present and future GW detectors. 
  
  There are also several theoretical motivations for considering theories with CDT. It has been noted that nature at the most fundamental level may be described by an action with only dimensionless parameters~\cite{Salvio:2014soa} and, furthermore, may be invariant under Weyl transformations (a local version of scale transformations) in the infinite energy limit provided that all couplings approach UV fixed points~\cite{Salvio:2017qkx}. These fixed points can be free (asymptotic freedom)~\cite{Gross} or interacting (asymptotic safety)~\cite{WeinbergAS}. The UV completion of gravity in this framework occurs thanks to the above-mentioned quadratic-in-curvature terms, which feature four-derivatives of the metric. The physical viability of this type of terms will constitute an important part of  this review.
  
  Moreover, classical scale invariance obviously gives a strong restriction on the possible actions one can write down, which can lead to testable predictions. To have an idea of how strong this requirement is, one can note, for example, that  a realistic theory  of this type can only be formulated in 4D: a gauge theory in higher dimensions necessarily includes dimensionful parameters in the action. So CDT  also provides a nice explanation of the observed number of dimensions. The predictive power becomes even stronger when the above-mentioned requirement of having UV fixed points for all couplings is taken into account. Indeed, some UV fixed points can only be reached if certain couplings acquire specific isolated values in the IR (see e.g.~\cite{Giudice:2014tma}). These fixed points are called ``IR attractive" and the isolated values are genuine predictions.  

Furthermore, another motivation for considering this type of theories is the fact that the terms with four derivatives of the metric that, as already mentioned are necessary in CDT, can also provide  a mechanism for a naturally small ratio between the Higgs and the Planck mass~\cite{Salvio:2014soa,Kannike:2015apa,Salvio:2017qkx}. So they offer a solution to what is known as the ``hierarchy problem". Here we adopt the definition of ``naturally small" given by Hooft~\cite{tHooft:1979rat}: a quantity is naturally small when setting it to zero leads to an enhanced symmetry. This is possible because the metric, having four-derivative terms, leads to two (rather than one) spin-2 particles. One of these particles is identified with the ordinary graviton and the other one mediates an opposite gravitational interaction that erases  the one of the ordinary graviton the more efficiently the higher the energy is. This leads to a scenario of ``softened gravity"~\cite{Giudice:2014tma}, where gravity reproduces Einstein's general relativity (GR) at small energies, but becomes the weaker  the more we move towards the UV. In this context the theory acquires a shift symmetry acting on the Higgs field, which can be softly broken at the scale of the Higgs mass $M_h$.

This work is mainly a review of CDT and its implications in theories of all interactions (gravity included). This scenario will be also referred to as ``agravity"~\cite{Salvio:2014soa}. The focus is on the phenomenological implications in  astrophysics and, in particular, cosmology. Indeed, the properties and implications of DT in flat space and in the absence of gravity are well known and can be found in many text books on quantum field theory (QFT).

Let us provide now an outline of this work (which includes a qualitative summary). 
\begin{itemize}
\item In the next section we give the general field content and the fundamental action for theories with CDT. To construct this gravitational theory, we impose the (strong) equivalence principle, as defined in e.g.~\cite{WeinbergGravity}, which implies that one can set the connection equal to the Levi-Civita one. We also discuss the relation with what is known as the Palatini formulation of gravity, where the connection is independent of the metric.  In the matter sector the general  field content and  action compatible with renormalizability and the no-scale principle is presented. How agravity can hold up to infinite energy~\cite{Salvio:2017qkx} is reviewed there as well.

\item In Sec.~\ref{Dimensional transmutation and gravity}, after a mention of the case without gravity,
 the phenomenon of DT in theories including gravity is discussed. Both perturbative and non-perturbative mechanisms are covered. After the mass scales  are generated in this way, the gravitational part of the effective action includes (besides the terms quadratic in the curvature) the Einstein-Hilbert and the cosmological constant terms. This low-energy effective theory~\cite{Stelle:1976gc,Weinberg:1974tw,Deser:1975nv} will be referred to as ``quadratic gravity", although sometimes in the literature the term ``higher derivative gravity" is also used because terms quadratic in the curvature have more than two derivatives of the gravitational field.
 
 \item Sec.~\ref{The Weyl-squared term}  discusses the physical viability of the terms with four-derivatives of the metric, in particular the one that is obtained through the square of the Weyl tensor. By using a perturbative approach, it will be shown how the theory can avoid the classical instabilities~\cite{Salvio:2019ewf,Gross:2020tph} suggested by the Ostrogradsky theorem~\cite{ostro} and is unitary even beyond the scattering theory: all probabilities are non-negative and they sum up to one at any time. This is possible because the quantum theory must feature here some  unusual characteristics, which emerge once a correct definition of probability is used. These characteristics are derived and described in Secs.~\ref{Calculation of probabilities} and \ref{The Dirac-Pauli canonical variables}. As reviewed in Sec.~\ref{Decays and scattering}, there is a potential violation of causality at microscopic scales, which, however, can certainly respect the observational bounds.
 \item The softening of gravity is reviewed in Sec.~\ref{Softening of gravity}. The application to the hierarchy problem is discussed in Sec.~\ref{Applications to the hierarchy problem}, while Sec.~\ref{Applications to ultracompact astrophysical objects} is devoted to the implications for compact objects of interest for astrophysics. Indeed, as shown there, the softening of gravity below a certain length scale $L_G$ implies that (event) horizons with radii smaller than $L_G$ cannot form and thus the theory predicts horizonless ultracompact objects that are free from any singularity~\cite{Salvio:2019llz}. These can be interesting dark matter (DM) candidates~\cite{Salvio:2019llz,Aydemir:2020xfd,Aydemir:2020pao}. Furthermore, in Sec.~\ref{The cosmological constant problem} we discuss the cosmological constant problem~\cite{Weinberg:1988cp}, which queries why the cosmological constant is so tiny compared to the Planck mass. No solutions to this long-standing problem will be offered here. 
 \item In Sec.~\ref{Inflation} the implications for the early universe are presented. Sec.~\ref{Inflation: the FRW background} regards the classical part of inflation, where the fields are homogeneous and isotropic.  Quantum fluctuations, which generically break homogeneity and isotropy, are then discussed in Sec.~\ref{Inflation: quantum perturbations} where the main predictions of the theory are reviewed. In that section the above-mentioned non-standard aspects of the quantum theory play an essential role. It is also argued that the Ostrogradsky theorem, together with Linde's theory of chaotic inflation~\cite{Linde:1983gd}, provides us with an explanation of why we live in a nearly homogeneous and isotropic universe~\cite{Salvio:2019ewf}.
How reheating after inflation can successfully occur is then briefly reviewed in Sec.~\ref{Reheating}.
\item Sec.~\ref{Phase transitions and gravitational waves} discusses how strong phase transitions can emerge in models with approximate scale symmetry and how to test them  with GW detectors. The main theoretical tools to study phase transitions and the associated production of GWs are also provided. 
\item Finally, Sec.~\ref{Conclusions} offers the concluding remarks, which include a discussion of the open issues.
\end{itemize}

It is appropriate at this point to discuss the relations between the present work and other reviews on related topics. First, there is a famous review by Adler~\cite{Adler:1982ri} on how DT can induce the Einstein-Hilbert term\footnote{See also Refs.~\cite{Elizalde:1994ry} for some subsequent works.} (what is known as ``induced gravity"). This partially overlaps with Sec.~\ref{Dimensional transmutation and gravity}
 (and to a less extent with Sec.~\ref{Decays and scattering}), but Adler's review was written in 1982 and much has been done after then, so a new discussion of those topics seems appropriate. 

Also, there is a recent review by Wetterich on quantum scale symmetry~\cite{Wetterich:2019qzx}. This is different from what we do here: as already mentioned, in CDT quantum physics is responsible for generating all observed mass scales. Furthermore, as  reviewed here in Secs.~\ref{Applications to the hierarchy problem} and~\ref{The cosmological constant problem},  promoting scale invariance to a quantum symmetry is not sufficient {\it per se} to solve neither the hierarchy nor the cosmological constant problems, contrary to what one could naively think.

The most related review in the literature is Ref.~\cite{Salvio:2018crh} on quadratic gravity. That paper does not regard DT and the hierarchy problem and does not provide a detailed discussion of the astrophysical and, in particular, cosmological implications.  So one of the purposes of this work is to fill that gap. Ref.~\cite{Salvio:2018crh}, however, contains some related material regarding the degrees of freedom, the renormalization\footnote{For previous introductions to the renormalization of quadratic gravity see e.g.~\cite{Avramidi:1986mj,Buchbinder:1992rb}.}, the classical stability, the unitarity and the UV completion, which can be applied unaltered or with very small modifications to agravity. Therefore, we refer to Review~\cite{Salvio:2018crh} in several parts of this work in order not to duplicate existing discussions. However, there has been some significant activity on the classical stability and the unitarity since the date of publication of  Ref.~\cite{Salvio:2018crh}. So an update on those topics, which are provided here in Sec.~\ref{The Weyl-squared term}, appears to be useful.

Although this paper is mainly a review, it does contain some original material. For example, the perturbative mechanism to generate the mass scales in the presence of gravity is generalized in Sec.~\ref{Perturbative mechanisms} to the case of multiple scalar fields. A complete and pedagogical proof of the way the non-standard  quantum theory emerges is provided in Sec.~\ref{Calculation of probabilities}: the previous works on this subject~\cite{Salvio:2019wcp,Strumia:2017dvt} do not contain some steps, which, however, can be useful for non-expert readers. Moreover, in Sec.~\ref{The Dirac-Pauli canonical variables} we explain for the first time how to compute quantum averages of observables, which are then used in Sec.~\ref{Predictions} to extract the predictions for the CMB anisotropies. The critical length scale $L_G$, below which gravity is softened, is determined in Sec.~\ref{Softening of gravity} both close and far from the Weyl-invariant regime of Ref.~\cite{Salvio:2017qkx}. In Sec.~\ref{Predictions} we show explicitly how taking into account all the non-standard aspects of the quantum theory allows us to obtain positive power spectra for the quantum fluctuations generated during inflation. The explanation of the nearly homogeneity and isotropy of the universe is given here (see Sec.~\ref{Classical metastability and chaotic inflation}) in a more general class of inflationary scenarios compared to~\cite{Salvio:2019ewf}. Finally, we show (for the first time in the general case) how classical scale invariance supports the validity of perturbation theory to compute the thermal effective potential (Sec.~\ref{Thermal effective potential}), which is needed to study the phase transitions and in turn the GW production.

 \section{General no-scale action}\label{General no-scale action in field theory}

In this section we introduce the general scenario of theories of all interactions (gravity included) featuring a Lagrangian without dimensionful parameters. As already mentioned, we refer to this scenario as agravity. The observed mass scales are then generated by DT as described in Sec.~\ref{Dimensional transmutation and gravity}. 

\subsection{The field content}

The field content we consider includes the metric $g_{\mu\nu}$ and the connection $\Gamma_{\mu \, \nu}^{\,\sigma}$, which allows us to construct covariant derivatives. 

Although $g_{\mu\nu}$ and $\Gamma_{\mu \, \nu}^{\,\sigma}$ could be a priori independent, the (strong) equivalence principle, as discussed in e.g. Ref.~\cite{WeinbergGravity}, relates these two quantities as follows\footnote{In this work we use the signature $\eta_{\mu\nu} = {\rm diag}(+1,-1,-1,-1)$ and repeated indices understand a sum.}
\be \Gamma_{\mu \, \nu}^{\,\sigma} = \frac12 g^{\sigma\rho}\left(\partial_\mu g_{\nu\rho}+\partial_\nu g_{\mu\rho}-\partial_\rho g_{\mu\nu}\right). \ee
In other words, $\Gamma_{\mu \, \nu}^{\,\sigma}$ is the Levi-Civita connection and the theory is called metric. Deviations from the Levi-Civita connection can occur either when such equivalence principle is violated or when the theory is equivalent to a metric one \cite{Koivisto:2005yc}  (this is at the basis of what is known as the Palatini formulation of gravity). A general connection $\mathcal{A}_{\mu \, \nu}^{\,\sigma}$ can always be written in terms of the Levi-Civita one plus a tensor, $\mathcal{T}_{\mu \, \nu}^{\,\sigma}$, that is $\mathcal{A}_{\mu \, \nu}^{\,\sigma} =\Gamma_{\mu \, \nu}^{\,\sigma} + \mathcal{T}_{\mu \, \nu}^{\,\sigma}$.
The extra quantity $\mathcal{T}_{\mu \, \nu}^{\,\sigma}$ generically includes particles with spin higher than 1. In this review we stick to the above-mentioned equivalence principle and we therefore  do not include this extra tensor field (see, however,~\cite{Anselmi:2020opi,Gialamas:2020snr} for a related discussion). 

Our conventions for the Riemann and Ricci tensors and the Ricci scalar are, respectively, $$R_{\mu\nu\,\,\, \sigma}^{\quad \rho} \equiv \partial_{\mu} \Gamma_{\nu \, \sigma}^{\,\rho}- \partial_{\nu} \Gamma_{\mu \, \sigma}^{\,\rho} +  \Gamma_{\mu \, \tau}^{\,\rho}\Gamma_{\nu \,\sigma }^{\,\tau}- \Gamma_{\nu \, \tau}^{\,\rho}\Gamma_{\mu \,\sigma }^{\,\tau}, \quad R_{\mu\nu} \equiv R_{\rho\mu\,\,\, \nu}^{\quad \rho},\quad R\equiv g^{\mu\nu}R_{\mu\nu}.$$
The Weyl tensor $W_{\mu\nu\alpha\beta}$ is then defined by 
\beq W_{\mu\nu\alpha\beta} \equiv R_{\mu\nu\alpha\beta} + \frac{1}{2} (
g_{\mu\beta} R_{\nu\alpha} -g_{\mu\alpha} R_{\nu\beta}+ g_{\nu\alpha} R_{\mu\beta} - g_{\nu\beta}R_{\mu\alpha})
+\frac16 (g_{\mu\alpha}g_{\nu\beta}-g_{\nu\alpha}g_{\mu\beta} )R,
\eeq
where as usual the metric $g_{\mu\nu}$ and its inverse $g^{\mu\nu}$ is used to lower and raise the spacetime indices.

Moreover, we consider a general renormalizable matter sector, which includes real scalars $\phi_a$, Weyl fermions $\psi_j$ and vectors $V^A_\mu$ (with field strength $F_{\mu\nu}^A$). The $V^A_\mu$ are gauge fields (connections of an internal gauge group)
and allows us to construct the covariant derivatives
$$D_\mu \phi_a = \partial_\mu \phi_a+ i \theta^A_{ab} V^A_\mu \phi_b, \qquad D_\mu\psi_j = \partial_\mu \psi_j + i t^A_{jk}V^A_\mu\psi_k + \frac12 \omega^{ab}_\mu \gamma_{ab}  \psi_j, $$ 
where the spin-connection $ \omega^{ab}_\mu$ is defined as usual by 
$ \omega^{ab}_\mu = e^a_{\,\, \nu} \partial_\mu e^{b\nu} + e^a_{\,\, \rho}  \Gamma_{\mu \, \sigma}^{\,\rho}  e^{b\sigma}$
and $\gamma_{ab} \equiv \frac14 [\gamma_a, \gamma_b]$ (with the $\gamma_a$ being the Dirac gamma matrices).  
The gauge couplings are contained in the matrices $\theta^A$ and $t^A$, which are the generators of the (internal) gauge group in the scalar and fermion representation, respectively.

\subsection{The fundamental action and the UV behavior}\label{The action}

The full fundamental action of a general no-scale (ns) theory is 
\be S=\int d^4x  \sqrt{-g} \, \Lag, \qquad \Lag^{\rm ns}=  \mathscr{L}^{\rm ns}_{\rm gravity} + \mathscr{L}^{\rm ns}_{\rm matter}+  \mathscr{L}_{\rm non-minimal} \label{TotAction}, \ee
where $g$ is the determinant of the metric.

Let us describe these three terms in turn.
 The quantity $\mathscr{L}^{\rm ns}_{\rm gravity}$ represents the no-scale gravitational Lagrangian, whose general form is~\cite{Salvio:2018crh}
\be    \boxed{\mathscr{L}^{\rm ns}_{\rm gravity} =  \frac{R^2}{6f_0^2}   - \frac{W^2}{2 f_2^2}-\epsilon_1 G-\epsilon_2\Box R,} \label{Lgravity3}
 \ee 
 where $f_0^2$, $f_2^2$, $\epsilon_1$ and $\epsilon_2$ are four real parameters,
 \be W^2\equiv W_{\mu\nu\rho\sigma}W^{\mu\nu\rho\sigma} = R_{\mu\nu\rho\sigma}R^{\mu\nu\rho\sigma} - 2R_{\mu\nu}R^{\mu\nu} + \frac13 R^2 \ee
 and $G$ is  the Gauss-Bonnet term
  \beq   G\equiv  R_{\mu\nu\rho\sigma}R^{\mu\nu\rho\sigma} - 4 R_{\mu\nu} R^{\mu\nu} + R^2  =  \frac{1}{4}\epsilon^{\mu\nu\rho\sigma}
\epsilon_{\alpha\beta\gamma\delta}R_{\,\,\,\,\,\, \mu\nu}^{\alpha\beta} R_{\,\,\,\,\,\,\rho\sigma}^{\gamma\delta}= \mbox{divs.}, \label{Gdef}
 \eeq
with $\epsilon_{\mu\nu\rho\sigma}$ the antisymmetric Levi-Civita tensor and ``divs" represents the covariant divergence of some current.  Therefore, both $\Box R$ and $G$ are total covariant derivatives and, once inserted in the action, give rise to boundary terms and do not contribute to the field equations. For this reason $\Box R$ and $G$ can be often  ignored; in the applications described in this review such terms will not play any role and therefore  will be neglected (see, however,~\cite{deBerredoPeixoto:2004if,Shapiro:2008sf,Einhorn:2014bka,Salles:2017xsr} for studies of these terms).
The no-scale gravitational Lagrangian given in~(\ref{Lgravity3}) does not contain the Einstein-Hilbert term and the cosmological constant because they involve mass scales. We will see in Sec.~\ref{Dimensional transmutation and gravity} how these terms (which are essential for the phenomenological viability of the theory) can be generated through DT. Once these terms appear in the effective action one can show  that the absence of tachyonic instabilities  requires $f_0^2> 0$ and $f_2^2 >0$ (see Secs~2.2 and~5 of~\cite{Salvio:2018crh} for a detailed explanation), which is why we have introduced these parameters as squares of other parameters, $f_0$ and $f_2$. We therefore  take $f_0$ and $f_2$ real.

The second term in~(\ref{TotAction}), $\mathscr{L}^{\rm ns}_{\rm matter}$, represents the no-scale matter Lagrangian
\be \label{eq:Lmatterns}
\Lag^{\rm ns}_{\rm matter} =  
- \frac14 (F_{\mu\nu}^A)^2 + \frac{D_\mu \phi_a \, D^\mu \phi_a}{2}  + \bar\psi_j i\slashed{D} \psi_j  - \frac12 (Y^a_{ij} \psi_i\psi_j \phi_a + \hbox{h.c.}) 
- V_{\rm ns}(\phi), 
\ee 
where the $Y^a_{ij}$  are the Yukawa couplings  and $V_{\rm ns}(\phi)$ is the no-scale potential,
\be V_{\rm ns}(\phi)= \frac{\lambda_{abcd}}{4!} \phi_a\phi_b\phi_c\phi_d, \label{Vns}\ee
with  $\lambda_{abcd}$ being the quartic couplings. We take the $\lambda_{abcd}$ totally symmetric with respect to the exchange of their indices without loss of generality. In~(\ref{eq:Lmatterns}) all terms are contracted in a gauge-invariant way.
In Sec.~\ref{Dimensional transmutation and gravity} we will see how the mass terms for the matter fields can also be generated (together with the Einstein-Hilbert term and the cosmological constant) through DT.

Finally, $\mathscr{L}_{\rm non-minimal}$ represents the non-minimal couplings between the scalar fields $\phi_a$  and $R$, parameterized by the coefficients $\xi_{ab}$:
\be \mathscr{L}_{\rm non-minimal} = -\frac12 \xi_{ab} \phi_a\phi_b R, \label{Lnonminimal}\ee
where, again, the indices are contracted in a gauge-invariant way. We also take the $\xi_{ab}$ totally symmetric with respect to the exchange of their indices without loss of generality. 
Note that the non-minimal couplings~(\ref{Lnonminimal}) do not involve mass scales and we therefore  include them in the no-scale Lagrangian. 

It is important to know that  all the terms\footnote{The fact that the terms quadratic in the curvature in~(\ref{Lgravity3}) are generated by loops of matter fields was originally showed in~\cite{Utiyama:1962sn}.} in $\mathscr{L}^{\rm ns}_{\rm gravity}$ and  $\mathscr{L}_{\rm non-minimal}$ are generically required by renormalizability: if they are omitted at the classical level, quantum corrections (even including only matter loops) generate them. This is a consequence of the form of the renormalization group equations (RGEs) for the dimensionless couplings, which can be found in their general form at one-loop level in~\cite{Salvio:2014soa,Salvio:2018crh} (see also Refs.~\cite{Fradkin:1981iu,Avramidi:1985ki,Avramidi:1986mj,Codello:2006in,Narain:2012te,Ohta:2013uca,Narain:2013eea,Einhorn:2014bka,Ohta:2015zwa,Jack:2020zgo} for other determinations in particular cases and~\cite{Becker:2019fhi} for the renormalization of composite operators).  Therefore, none of the terms in $\mathscr{L}^{\rm ns}_{\rm gravity}$ and  $\mathscr{L}_{\rm non-minimal}$ can be omitted.

The analysis of the RGEs shows that asymptotic freedom (namely $f_2^2\to 0$ and $f_0^2\to 0$ in the infinite energy limit) requires  $f_2^2>0$ but $f_0^2<0$, which leads to a tachyonic instability; nevertheless, for $f_0^2>0$ the theory can reach infinite energy without hitting any Landau-pole (see Sec.~5 of Ref.~\cite{Salvio:2018crh} for a review and references to the original works). A key step to show this result was made in~\cite{Salvio:2017qkx} (see also references therein), by using a perturbative expansion  in $1/f_0$ (which is the appropriate one when $f_0$ runs to large values outside the validity of perturbation theory in $f_0$). We do not reproduce the calculations of Ref.~\cite{Salvio:2017qkx} because they are described in detail there. As shown in~\cite{Salvio:2017qkx},  $f_0$ can grow  up to infinity  in the infinite energy limit without hitting any Landau pole,
 provided that all scalars have asymptotically Weyl-invariant couplings (i.e. $\xi_{ab}\to-\delta_{ab}/6$) and all other couplings approach UV fixed points (which can be free or interacting\footnote{See~\cite{Giudice:2014tma,Holdom:2014hla} for recent discussions of asymptotically free theories and~\cite{Mann:2017wzh,Abel:2017rwl,Pelaggi:2017abg,Wang:2018yer,Sannino:2019sch,Fabbrichesi:2020svm} for phenomenological applications of UV interacting fixed points in the absence of gravity.}). 
  In the infinite-energy limit the theory can therefore  approach conformal gravity (plus a conformal matter sector): a theory that is invariant under Weyl transformations, which is the local version of scale transformations.  The one-loop RGEs of this Weyl invariant theory can be found in Sec.~5.2 of Ref.~\cite{Salvio:2018crh} and were originally derived in~\cite{Salvio:2017qkx}. The idea that one can approach a Weyl-invariant theory at large energy has been investigated in several  articles~\cite{Fradkin:1978yf,Zee:1983mj,Shapiro:1994st,Hamada:2002cm,Hamada:2009hb,Donoghue:2016xnh,Donoghue:2018izj,Alvarez:2017spt}. 
  
  One possible scenario is the  quasi-conformal one: all the dimensionless couplings remain close to the Weyl invariant  values, namely $1/f_0 \approx 0$ and $\xi_{ab}\approx-\delta_{ab}/6$ at all energies, even below the mass scales are generated. This scenario was discussed in Refs.~\cite{Salvio:2017qkx,Salvio:2019wcp} and reviewed in Sec.~5.2 of Ref.~\cite{Salvio:2018crh}. The possibility that $f_0$ runs instead to small values in the IR, but still goes to $\infty$ in the UV,~\cite{Salvio:2017qkx} has not been explored in detail and therefore  has been left (so far) for future research. 

\section{Dimensional transmutation and gravity}\label{Dimensional transmutation and gravity}

In an interacting QFT the renormalization group (RG) approach shows that, generically, a physical mass scale appears even if the classical Lagrangian does not contain any dimensionful parameter. 
This phenomenon is the DT.

The simplest way to understand DT is to look at the RGE of some coupling $\lambda$,
\be \mu\frac{d\lambda}{d\mu} = \beta(\lambda), \ee
and rewrite it as
\be \frac{\mu}{\bar\mu} = \exp\left(\int_{\lambda(\bar\mu)}^{\lambda(\mu)}\frac{d\lambda'}{\beta(\lambda')}\right),\ee
where $\bar\mu$ is some (arbitrarily chosen) reference energy (for example $\bar\mu =1$~GeV). By introducing now some (arbitrarily chosen) reference value for $\lambda$ too, which we denote $\lambda^*$ (for example, $\lambda^*=1$) we have 
\be \mu \exp\left(-\int_{\lambda^*}^{\lambda(\mu)}\frac{d\lambda'}{\beta(\lambda')}\right)=\bar\mu \exp\left(-\int_{\lambda^*}^{\lambda(\bar\mu)}\frac{d\lambda'}{\beta(\lambda')}\right).\label{PhysMass}\ee
The mass scale on the left-hand-side of this expression is therefore  independent of $\mu$ and can be a physical mass scale. 

\subsection{Perturbative mechanisms}\label{Perturbative mechanisms}

We start by reviewing the perturbative approach to DT in the case of non-gravitational models. The first work on this topic was the Coleman-Weinberg (CW) paper~\cite{Coleman:1973jx}. This article considered only a single scalar field, but later Gildener and Weinberg~\cite{Gildener:1976ih} generalized the analysis to an arbitrary number of scalars. 

The basic idea is that, since at the quantum level the couplings depend on the energy 
as dictated by the RG, there can be some specific energy scale 
at which the no-scale potential $V_{\rm ns}$ in Eq.~(\ref{Vns}) develops a flat direction. Such flat direction can be written as $\phi_a = \nu_a \chi$, where $\nu_a$ are the components of a unit vector $\nu$, i.e.~$\nu_a \nu_a =1$, and $\chi$ is a single scalar field 
that parameterizes this direction. 
RG-improving $V_{\rm ns}$ along the flat direction 
we therefore  obtain
\be V_{\rm ns}(\nu \chi) = \frac{\lambda_\chi (\chi)}{4}\chi^4  \label{Vvarphi}\ee 
where 
\be \lambda_\chi(\chi) \equiv\frac1{3!} \lambda_{abcd}(\chi)\nu_a \nu_b \nu_c \nu_d. \label{lambdaphi}\ee
  Given that $V_{\rm ns}(0)=0$, having a flat direction along $\nu$ for $\chi$ equal to some specific value $\chi_{\rm CW}$ means 
 \be \lambda_\chi(\chi_{\rm CW})\equiv\lambda_{abcd}(\chi_{\rm CW})\nu_a \nu_b\nu_c\nu_d=0. \ee
Now, around  $\chi_{\rm CW}$ we can write 
\be V_{\rm ns}(\nu \chi) = \frac{\bar\beta_{\lambda_\chi}}{4}\left(\ln\frac{\chi}{\chi_0} -\frac14\right)\chi^4 \equiv V_{\rm CW}(\chi) , \label{CWpot}\ee 
where we  defined 
\be \beta_{\lambda_\chi} \equiv  \chi\frac{d\lambda_{\chi}}{d\chi}, \qquad \bar\beta_{\lambda_\chi} \equiv \left[\beta_{\lambda_\chi} \right]_{\chi=\chi_{\rm CW}}, \qquad \chi_0\equiv \frac{\chi_{\rm CW}}{e^{1/4}}.\ee
  The expression $V_{\rm CW}$  in~(\ref{CWpot}) is known as the Coleman-Weinberg potential. Note that $V_{\rm CW}$ has a stationary point at $\chi=\chi_0$, which is a point of minimum when $\bar\beta_{\lambda_\chi}>0$. Therefore, when the conditions 
  \beq\left\{
\begin{array}{rcll}
\lambda_\chi(\chi_{\rm CW})  &=& 0 & \hbox{(flat direction).}\\
\beta_{\lambda_\chi}(\chi_{\rm CW})  &>& 0 & \hbox{(minimum condition),}
\end{array}\right.
\label{eq:CWgen}
\eeq
 are satisfied quantum corrections generate a minimum of the potential at a non-vanishing value of $\chi$ given by $\chi_{\rm CW}/e^{1/4}$. 
 
 This non-trivial minimum can generically break global and/or local symmetries and thus generate the  particle masses. Consider for example a term in $\Lag$ of the form 
 \be \Lag_{\chi h}\equiv \frac12 \lambda_{ab} \phi_a\phi_b |h|^2,\label{LvarphiH}\ee 
 where $h$ is the Standard Model (SM) Higgs doublet and the $\lambda_{ab}$ are some of the quartic couplings. By performing again the RG-improvement along $\nu$ we obtain
 \be  \Lag_{\chi H} = \frac12 \lambda_{\chi h}(\chi) \chi^2 |h|^2, \ee
 where 
 \be \lambda_{\chi h}(\chi) \equiv  \lambda_{ab}(\chi) \nu_a\nu_b.\ee
 Thus, by evaluating this term at the minimum $\chi=\chi_0$ we obtain the Higgs squared mass parameter
  \be M_h^2 = \lambda _{\chi h}(\chi_0) \chi_0^2. \ee
 In order to generate the particle masses through the usual Higgs mechanism, we need $M_h^2>0$, namely we have the additional condition
 \be \lambda _{\chi h}(\chi_{\rm CW}/e^{1/4}) >0. \ee
 
 It is possible that several minima are generated radiatively (see Ref.~\cite{Kannike:2020ppf} for a recent study), but for our purposes only one minimum is necessary.  
 
 In the present work we are mainly interested in DT occurring in a theory that does include gravity. A realistic extension of the CW mechanism~\cite{Coleman:1973jx} to gravitational theories was obtained in~\cite{Salvio:2014soa} in the presence of a single scalar field. We include here an arbitrary number of scalar fields. 
 
 The crucial difference with respect to the non-gravitational case is that the Lagrangian effectively includes another $\phi$-dependent term, the non-minimal couplings in~(\ref{Lnonminimal}). Let us consider again a direction in field space $\phi_a = n_a \varphi$, with $n_a$ the components of a unit vector $n$, that is $n_a n_a =1$, not necessarily equal to $\nu$ and $\varphi$ parameterizing this   direction.  Now, define (RG-improving along the  direction $n$)
 \be \lambda_\varphi(\varphi) \equiv \frac1{3!}\lambda_{abcd}(\varphi)n_a n_b n_c n_d, \qquad \xi_\varphi(\varphi) \equiv \xi_{ab}(\varphi)n_an_b. \label{lambdaxiphi}\ee
With these definitions the field equation for a homogeneous scalar $\varphi$ reads
\be \frac{dV_{\rm ns}}{d\varphi} +\frac{R}{2}\frac{d}{d\varphi}\left(\xi_\varphi  \varphi^2\right) = 0, \label{varphiEOM} \ee
where the dependence of $V_{\rm ns}$ on $\varphi$ is given by
\be V_{\rm ns}(n \varphi) = \frac{\lambda_\varphi (\varphi)}{4}\varphi^4.  \label{Vvarphi2}\ee 
On the other hand, the trace of the gravitational field equations gives
\be \xi_\varphi(\varphi) \varphi^2 R+4 V_{\rm ns} = \mathcal{O}(R^2), \label{traceEEOM}\ee
where $\mathcal{O}(R^2)$ represents the contributions due to the first two terms quadratic in the curvature in~(\ref{Lgravity3}). Now, we want to generate the Planck scale and the cosmological constant radiatively. The observed value of the cosmological constant is tiny and we can therefore  neglect the term $\mathcal{O}(R^2)$. Then, by solving Eq.~(\ref{traceEEOM}) for $R$ and plugging the result in Eq.~(\ref{varphiEOM}), we find
\be  \frac{dV_{\rm ns}}{d\varphi}=\frac{2V_{\rm ns}}{\xi_\varphi \varphi^2}\frac{d}{d\varphi}\left(\xi_\varphi  \varphi^2\right). \ee 
Using~(\ref{Vvarphi2})
 this equation can be rewritten as
\be \beta_{\lambda_\varphi} = 2\frac{\lambda_\varphi}{\xi_\varphi} \beta_{\xi_\varphi}, \label{agrGen} \ee 
where 
\be \beta_{\lambda_\varphi}\equiv  \varphi\frac{d\lambda_\varphi}{d\varphi}, \qquad \beta_{\xi_\varphi}\equiv  \varphi\frac{d\xi_\varphi}{d\varphi}.  \ee 

When Eq.~(\ref{agrGen}) is satisfied at some non-vanishing field value $\bar\varphi$  the Planck scale and the cosmological constant $\Lambda$ are generated. Specifically, we have 
\be \bp^2=\xi_\varphi(\bar\varphi) \bar\varphi^2, \qquad  \Lambda= \lambda_\varphi(\bar\varphi) \bar\varphi^4, \label{MLgen}\ee
 where $\bp$ is the reduced Planck mass. Requiring $\bp^2$ to be positive and  $\Lambda$ to match the  observed value (which is negligibly small compared to $\bp^4$)   one obtains the three conditions 
\beq\left\{
\begin{array}{rcll}
\xi_\varphi(\bar\varphi)  &>& 0 & \hbox{(real Planck mass).}\\
\lambda_\varphi(\bar\varphi) &=& 0 & \hbox{(nearly vanishing cosmological constant),}\\
\beta_{\lambda_\varphi}(\bar\varphi) &=& 0 & \hbox{(solution of field  equations),}
\end{array}\right.
\label{eq:agravMPl}
\eeq
where Eq.~(\ref{agrGen}) has been simplified taking into account that $\lambda_\varphi$ nearly vanishes at $\bar\varphi$. These are the necessary and sufficient conditions to generate perturbatively  through DT viable values of the Planck scale and the cosmological constant. 
We will refer to the scalar field $\varphi$ as ``the Planckion" as it is responsible for  the Planck mass. 

It is interesting to compare the conditions for successful DT in the CW mechanism~(\ref{eq:CWgen}) and in gravitational extension~(\ref{eq:agravMPl}). They are significantly different. The only conditions that are superficially similar are the first one in~(\ref{eq:CWgen}) and the second one in~(\ref{eq:agravMPl}). Note, however, that these conditions have a completely different origin. While the former is the requirement of having a flat direction, the latter corresponds to having a negligibly small value of the cosmological constant in the presence of gravity.

Note that in order to avoid the formation of anti de Sitter patches when $\varphi$ undergoes quantum fluctuations around $\bar\varphi$, it is necessary to impose $\lambda_\varphi(\varphi)\geq 0$ for $\varphi$ in a neighbourhood of $\bar\varphi$. Then the running  of $\lambda_\varphi$ is such that $\beta_{\lambda_\varphi}\leq0$ for $\varphi<\bar\varphi$ and $\beta_{\lambda_\varphi}\geq0$ for $\varphi>\bar\varphi$. The scalar $\varphi$ must therefore  interact significantly with some fermions, otherwise one would not have a negative contribution to $\beta_{\lambda_\varphi}$. These fermions can have interesting phenomenological applications.

 If the lightest fermion of this sort has no gauge interactions, then it can also couple to the SM sector
behaving as a right-handed neutrino $\nu_R$. Majorana masses for  $\nu_R$ can be generated via DT~\cite{Humbert:2015epa,Klein:2019iws,Kubo:2020fdd}, for example, through a Yukawa coupling between $\nu_R$ and the CW scalar $\chi$ and/or the Planckion $\varphi$.
The right-handed neutrino can then generate the observed light-neutrino masses via Yukawa couplings with the lepton and the Higgs doublets~\cite{Minkowski:1977sc}
and it can provide baryogenesis via leptogenesis~\cite{Fukugita:1986hr,Giudice:2003jh}. 

The lightest fermion of this sort can instead be a stable DM candidate
if it cannot couple to the SM sector (for example because it has gauge interactions under which the SM fields are neutral)~\cite{Kannike:2015apa}.

Finally, note that the particle masses in the matter sector can be generated through the quartic couplings that couple $\varphi$ and/or $\chi$ to the Higgs fields of the theory. The couplings of $\chi$ to the Higgs fields have already been discussed. Similarly, for $\varphi$, if there is a term of the form~(\ref{LvarphiH}) one obtains the Higgs squared mass parameter 
 \be M_h^2 = \lambda _{\varphi h}(\bar\varphi) \bar\varphi^2 \ee
 where
  \be \lambda_{\varphi h}(\varphi) \equiv  \lambda_{ab}(\varphi) n_an_b.\ee
 and the condition to generate the particle masses through the usual Higgs mechanism is 
 \be \lambda _{\varphi h}(\bar\varphi) >0. \ee

\subsection{Non-perturbative mechanisms}\label{Non-perturbative mechanisms}

If some couplings run outside the regime of validity of perturbation theory DT can still occur. 

Suppose, for example, one deals with an asymptotically free  QFT, for example QCD, and consider $4\pi$ the maximal value (in order of magnitude) of the coupling $\lambda$ compatible with perturbation theory. 
The physical mass in~(\ref{PhysMass}), which we call $\Lambda_{\rm QFT}$, can then be estimated within perturbation theory as
\be \Lambda_{\rm QFT}\sim \bar\mu \exp\left(\int_{\lambda(\bar\mu)}^{4\pi}\frac{d\lambda'}{\beta(\lambda')}\right), \label{PhysMass2}\ee
by setting $\bar\mu$ equal to an energy such that $\lambda(\bar\mu)<4\pi$, which surely exists because the theory is asymptotically free. Taking  for  $\beta(\lambda)$ its one-loop approximation $\beta_1(\lambda)$, one obtains
\be \int_{\lambda(\bar\mu)}^{4\pi}\frac{d\lambda'}{\beta(\lambda')}<0,\ee 
where $\lambda(\bar\mu)<4\pi$  and the asymptotic freedom condition $\beta_1(\lambda)<0$   have been used. This result tells us that $\Lambda_{\rm QFT}$ is exponentially suppressed compared to the scale $\bar\mu$ at which the theory is perturbative. 
For example, in the well-known QCD case, taking $\beta_1(g_s)=-7g_s^3/(4\pi)^2$ (in the SM), where $g_s$ is the QCD gauge coupling, the corresponding physical mass~(\ref{PhysMass2}) reads
\be \Lambda_{\rm QCD} \sim \bar \mu \exp\left(\frac{1}{14}-\frac{(4\pi)^2}{14 g_s(\bar\mu)^2}\right). \ee
For example, by setting  $\bar\mu = 10$~TeV, for which $g_s(\bar\mu)\approx 1$  (see e.g.~\cite{Buttazzo:2013uya}), we obtain the well-known estimate $\Lambda_{\rm QCD}\sim 10^2$~MeV. 

Generically, the physical mass $\Lambda_{\rm QFT}$ generated by DT is transmitted to the gravity sector and can induce the Einstein-Hilbert term. This is at the basis of a mechanism called ``induced gravity" (see~\cite{Adler:1982ri} for a famous review with references to the original works). The original idea behind induced gravity was to generate the kinetic term of the graviton from the dynamics of an ordinary QFT without gravitation. In our framework we already have the kinetic terms of the graviton, they are provided by the quadratic-in-curvature terms in~(\ref{Lgravity3}), but getting the Einstein-Hilbert term is still necessary to have a viable low-energy behavior. For this reason, in recent times there has been some work to generate the Einstein-Hilbert term and the cosmological constant through non-perturbative DT in the no-scale scenario of Sec.~\ref{General no-scale action in field theory} (see Refs.~\cite{Salvio:2017qkx,Donoghue:2017vvl,Donoghue:2018izj}).

\begin{figure}[t]
\begin{center}
\vspace{-2cm}
 $ \includegraphics[scale=0.5]{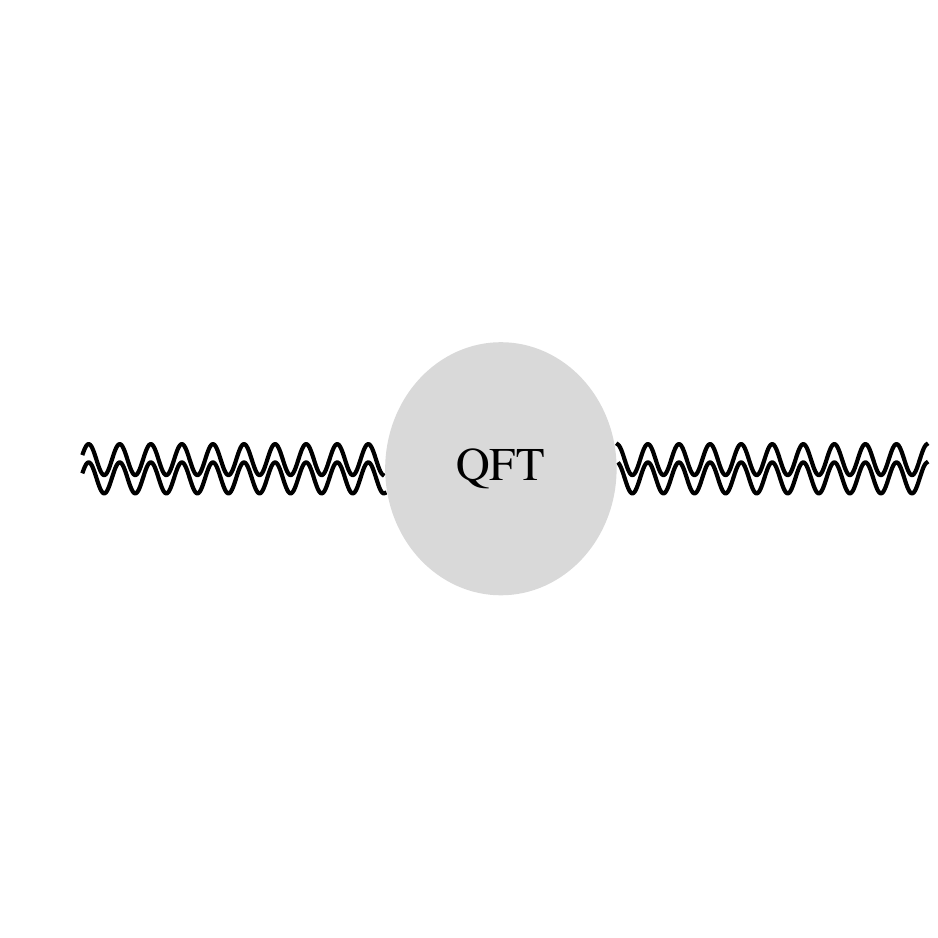}\qquad\qquad\qquad\includegraphics[scale=0.5]{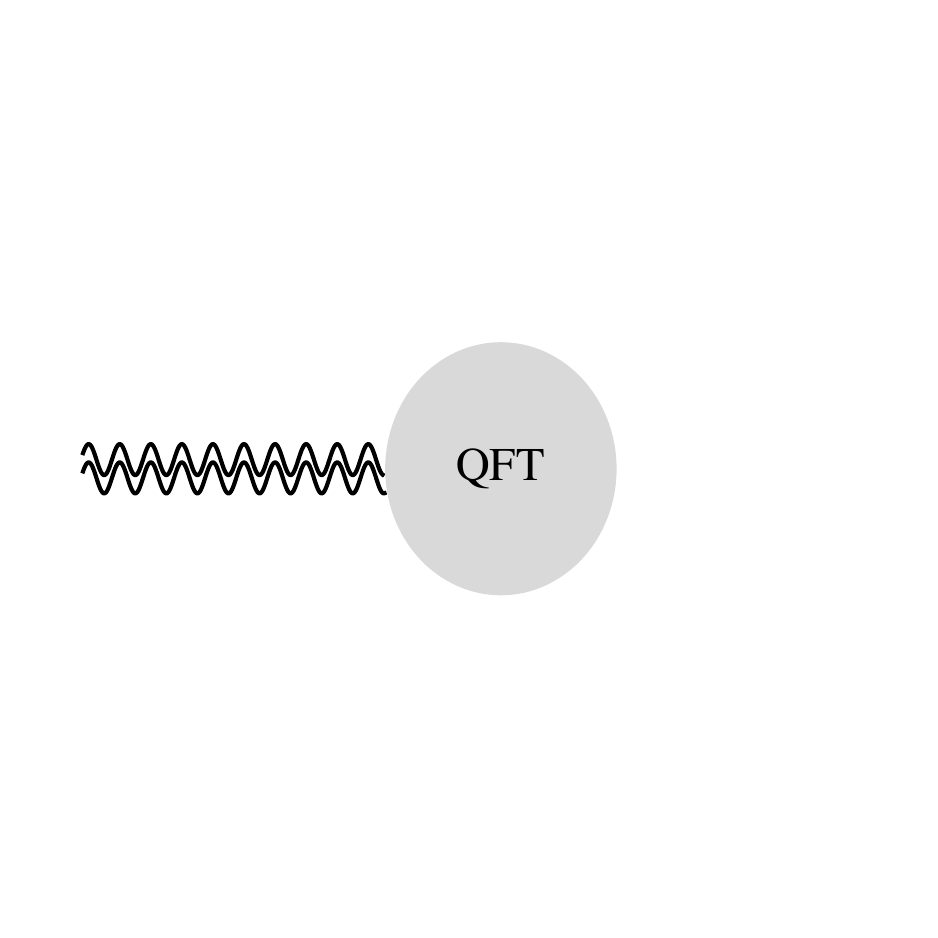} $
  \end{center}
  \vspace{-2cm}
   \caption{\em Diagrams inducing the Planck scale (left diagram) and the cosmological constant (right diagram). The double wavy lines represent the graviton.}
\label{induced}

\end{figure}

The induced value of $\bp$ and the cosmological constant $\Lambda$ in this case are generically expected to be around the scale given by the physical mass $\Lambda_{\rm QFT}$ of some non-perturbative  QFT, that is $\bp\sim \Lambda_{\rm QFT}$ and $\Lambda\sim \Lambda_{\rm QFT}^4$.  In Fig.~\ref{induced} diagrams corresponding to the generation of $\bp$ and $\Lambda$ through the QFT strong dynamics  are shown. By imposing global supersymmetry in such sector, one can, however, obtain a vanishing value of the induced cosmological constant~\cite{Salvio:2017qkx}. Another solution can be to realize the quasi-conformal scenario discussed at the end of Sec.~\ref{The action}: gauge fixing the Weyl symmetry through $g=1$, such that the vacuum energy manifestly does not gravitate, one sees that the problem is avoided (like in unimodular gravity~\cite{Donoghue:2018izj}).  The sign of  $\bp^2$ depends on the details of the QFT in question (such as the gauge group and the matter representations)~\cite{Khuri:1982ts,Adler:1982ri}, but it is known to be positive in QCD-like theories~\cite{Salvio:2017qkx,Donoghue:2017vvl} (see also Ref.~\cite{Kubo:2018kho}). Of course, we need to set $\Lambda_{\rm QFT}$ at a scale several (about 19) orders of magnitude  higher than $\Lambda_{\rm QCD}$ to obtain the measured value of $\bp$. 

Once $\bp$ is induced, the RGEs\footnote{The general form of the RGEs at one-loop level and valid for small couplings ($f_0$, $f_2$, ...) can be found in~\cite{Salvio:2017qkx,Salvio:2018crh}, where references to previous less general results are provided.} of the effective dimensionful parameters can generate the Higgs mass parameter $M_h$ and $\Lambda$. In particular, the one-loop RGE of $M_h^2/\bp^2$ for $f_0,f_2< \mu/\bp\lesssim1$  is~\cite{Salvio:2014soa}
\bea
(4\pi)^2\mu\frac{d}{d\mu} \frac{M_h^2}{\bp^2}&=& -\xi_H [5\gt^4+\gs^4(1+6\xi_H)]
-\frac13 \bigg( \frac{M_h^2}{\bar M_{\rm Pl}^2}\bigg)^2(1+6\xi_H)+\nonumber\\
&& + \frac{M_h^2}{\bar M_{\rm Pl}^2} 
\bigg[C_{M_h}^{\rm matter}+5\gt^2 +\frac{5}{3}\frac{\gt^4}{\gs^2} + \gs^2 \bigg(\frac13 + 6\xi_H + 6 \xi_H^2\bigg)\bigg],\label{eq:RGEm}
\eea
where $\xi_H$ is the non-minimal coupling of the Higgs, which appears in $\mathscr{L}$ as $-\xi_H |h|^2 R$ (a particular term in~(\ref{Lnonminimal})) and $C_{M_h}^{\rm matter}$ parameterizes  the contribution of the matter sector. We see that starting with $M_h=0$ at some energy one ends up with a non-vanishing value of $M_h$ at different energies because of the first term in~(\ref{eq:RGEm}), $-\xi_H [5\gt^4+\gs^4(1+6\xi_H)]$, which is a pure gravitational contribution. Therefore, gravity transmits the DT scale $\Lambda_{\rm QFT}$ even if there are no tree-level couplings between $H$ and the fields of the QFT, whose non-perturbative dynamics generates $\Lambda_{\rm QFT}$. Analogously, by looking at the  one-loop RGE of $\Lambda/\bp^4$ valid for small couplings (see again~\cite{Salvio:2017qkx}), one can see how the cosmological constant can be generated by $\bp$ (and the particle masses) and can be tuned to the observed value.

When we are close to the conformal regime discussed at the end of Sec.~\ref{The action} the scale $\Lambda_{\rm QFT}$ can be transmitted to the SM sector in a similar way. The relevant part of the RGE  reads now~\cite{Salvio:2017qkx}
\be
(4\pi)^2\mu\frac{d  }{d\mu} \frac{M_h^2}{ \bp^2}= \frac{5}{6} \gt^4 + 
\cdots,\qquad \hbox{for}\qquad M_2 < \mu < M_{\rm Pl}.
\ee
 
It is also possible to find other contributions to $M_h$ from the strong dynamics of the QFT by using different portals (other than gravity). For example, in Ref.~\cite{Antipin:2014qva} the scale $\Lambda_{\rm QFT}$ is transmitted to the SM via EW gauge interactions.

\subsection{Quadratic gravity}\label{Quadratic gravity}

The full effective action, which includes both the no-scale terms discussed in Sec.~\ref{The action} and those with dimensionful coefficients, can be found in Sec.~2.1 of Ref.~\cite{Salvio:2018crh}. The resulting effective theory is known as quadratic gravity
(see~\cite{Salvio:2018crh} for a review). The dimensionful coefficients that appear there are considered here as mass scales generated by DT through perturbative and/or non perturbative mechanisms.

As originally discussed  in~\cite{Stelle:1976gc}, the appearance of $\bp$ leads to a non-trivial gravitational mass spectrum that can be obtained by expanding perturbatively\footnote{For a discussion of a possible form of the non-perturbative spectrum see Refs.~\cite{Holdom:2015kbf,Holdom:2016xfn}.} around the flat space: the metric contains the ordinary graviton (a massless spin-two field) a scalar $\zeta$ (due to the $R^2$  term in~(\ref{Lgravity3})) with mass
\be  \boxed{M_0\equiv \frac1{\sqrt{2}} f_0 \bp + ...} \label{M0}\ee
 and a massive spin-two field, a massive graviton, (corresponding to the $W^2$  term in~(\ref{Lgravity3})) with mass
\be  \boxed{M_2\equiv \frac1{\sqrt{2}} f_2 \bp.}\label{M2} \ee
The dots in the expression for $M_0$ represent the possible contribution of other scalars mixing with $\zeta$ (if any), which can be present in specific models. We will explicitly see how the $R^2$ term can be traded with $\zeta$ in Sec.~\ref{The Einstein frame Lagrangian}. A pedagogical calculation of this  gravitational mass spectrum can be found in Sec.~2.3 of~\cite{Salvio:2018crh}.

As mentioned in Sec.~\ref{The action}, the coupling $f_0$ can flow to infinity in the infinite energy limit and the theory approaches conformal gravity plus a conformal matter sector. In this case, the mass $M_0$ becomes large and, as shown in~\cite{Salvio:2017qkx}, the scalar $\zeta$ decouples.

\section{The Weyl-squared term}\label{The Weyl-squared term}

In Sec.~\ref{The action} we have seen that all the terms in $\mathscr{L}^{\rm ns}_{\rm gravity}$ and  $\mathscr{L}_{\rm non-minimal}$ must be included in order for the theory to be consistent at the quantum level. In this section we discuss the important role of one of these terms, the Weyl-squared one  (the term proportional to $W^2$ in~(\ref{Lgravity3})). 


The inclusion of the $W^2$ term allows us to obtain a {\it fully renormalizable theory} including the dynamics of the metric, as rigorously proved by Stelle in~\cite{Stelle:1976gc} (see also~\cite{Weinberg:1974tw,Deser:1975nv} for early conjectures and \cite{Barvinsky:2017zlx}
 for a more recent discussion). This is because the action defined in Sec.~\ref{The action}, with the $W^2$ term included, is the most general one with only dimensionless  parameters; therefore, the only divergences that can be generated in the quantum effective action are proportional to the terms discussed in Sec.~\ref{The action} and so can be reabsorbed in a redefinition of the parameters $f_2$, $f_0$, $\epsilon_1$, $\epsilon_2$,  $Y^a_{ij}$, $\lambda_{abcd}$ and the gauge couplings. The mass scales are then generated through the finite RGEs as discussed in Sec.~\ref{Dimensional transmutation and gravity}. 

However, the $W^2$ term also leads to interesting complications related to classical stability and unitarity, which we discuss in the following two subsections. These complications are basically due to the fact that the kinetic term of the  massive spin-2 field appears with an unusual minus sign (i.e.~it is a ghost), which was originally considered as  an inconsistency of the  $W^2$ term.  But in the last few years there has been some substantial progress in understanding how the theory should be treated. Therefore, we extend here the part of~\cite{Salvio:2018crh}  reviewing  the classical (in)stability and the unitarity of  the theory. As we will see, the theory is viable, but, nevertheless, the abnormal graviton leads to some interesting  non-standard features.

In order to discuss these aspects of the theory, we proceed perturbatively. 
Let us split the metric $g_{\mu\nu}$ as follows:
\be   \boxed{g_{\mu\nu} = g^{\rm cl}_{\mu\nu} + \hat h_{\mu\nu},}\label{metricSplit}\ee
where $g^{\rm cl}_{\mu\nu}$ is a classical background that solves the classical equations of motion (EOMs) and $\hat h_{\mu\nu}$ is a {\it quantum} field describing the fluctuations around the classical background $g^{\rm cl}_{\mu\nu}$.

 Although a non-perturbative approach would be desirable, such an approach is not  currently available. This is partially due to the fact that we do not have a fully non-perturbative understanding of quantum gravitational interactions in agravity. But it is also due to the fact that realistic quantum field theories even describing only non-gravitational interactions do not have currently a non-perturbative formulation. For example, the lattice is currently unable to study fermions with chiral gauge interactions. A fully non-perturbative formulation of the theory is therefore  left as an open challenge for future research.
 
We now consider in turn  classical aspects (related to $g^{\rm cl}_{\mu\nu}$) and  quantum aspects (related to $\hat h_{\mu\nu}$) and we will respectively address the classical stability and unitarity.

\subsection{Classical aspects}\label{Classical aspects}
 
The Weyl-squared term contains four spacetime derivatives.  Therefore,  agravity belongs classically to the set of  theories studied long time ago by Ostrogradsky in~\cite{ostro}: those with Lagrangian $L$ containing more than two time derivatives. Ostrogradsky showed in~\cite{ostro} that these theories have necessarily a classical Hamiltonian that is unbounded from below. This result (known as the Ostrogradsky theorem) can be extended to include an explicit time dependence of the Lagrangian $L$ (see Sec. 4.1.1 of Review~\cite{Salvio:2018crh}), which is relevant for cosmological applications.  We consider a physical system described by a certain number of coordinates $q_i$ and restrict our attention to  Lagrangians that depend on $q_i$, $\dot q_i$, $\ddot q_i$ and, possibly, on time $t$,
\be L(q, \dot q, \ddot q, t), \label{LagOstro}\ee 
where the dot is the derivative w.r.t.~$t$. Note that the case of fields can be obtained by interpreting the index $i$ as a space coordinate $\vec{x}$ and therefore  this setup applies to the agravity Lagrangian too.
The thesis of the Ostrogradsky theorem is the following: the Hamiltonian obtained from a Lagrangian of the form~(\ref{LagOstro}), which depends non-degenerately on $\ddot q$ (namely\footnote{$\partial^2 L/\partial \ddot q^2$ denotes   the Hessian matrix of $L$, whose elements are $\partial^2 L/\partial \ddot q_i\partial \ddot q_j$.} $\det(\partial^2 L/\partial \ddot q^2)\neq 0$), is not bounded from below.

Note that this result applies independently of whether or not the mass scales are present and so it applies both to agravity and to quadratic gravity. The only higher-derivative terms, which can have any relevance from the point of view of the Ostrogradsky theorem, come from the $W^2$ term. Indeed, as reviewed in Sec.~\ref{The Einstein frame Lagrangian}, the $R^2$ term can be traded for an ordinary scalar (which we called $\zeta$ in Sec.~\ref{Quadratic gravity}) and the terms proportional to $G$ and $\Box R$  in~(\ref{Lgravity3}) are total covariant derivatives. 

Now, the fundamental question at the classical level is the following: can we avoid the possible instabilities due to the Ostrogradsky theorem? One obvious way to avoid runaways is to consider only energies much smaller than
$M_2$ such that the $W^2$ term would have a negligible effect on any experimentally observable quantity. However, 
this does not allow us to test the $W^2$ term. Moreover, as we will see in Sec.~\ref{Applications to the hierarchy problem}, a theoretical argument, the naturalness of the Higgs mass,  favors very small values of $f_2$ (that is $\gt\lesssim 10^{-8}$) and so in many inflationary models one would have enough energy to reach $M_2$, as discussed in Sec.~\ref{Inflation}. 

In Ref.~\cite{Salvio:2019ewf} it was shown that the answer to the question above is  positive for quadratic gravity\footnote{This result holds independently of whether or not the mass scales of quadratic gravity are fundamental parameters or are generated through DT as discussed in this review.} even for very small values of $f_2$ (see also Refs.~\cite{Chapiro:2019wua,Gross:2020tph} for related calculations that confirmed this result and~\cite{Salles:2014rua} for a related discussion). The key properties used in~\cite{Salvio:2019ewf} was the fact that the massive spin-2 field decouples as $f_2\to 0$ and is not tachyonic. Here we  highlight the key steps of the argument in~\cite{Salvio:2019ewf} at a more qualitative level. 
  \begin{itemize}
\item First, in the  free-field limit the Hamiltonian of the massive spin-2 field (expanding around the flat spacetime) is
    \be H_{\rm 2} \,\, ~ =  \,\, ~ - \sum_{\alpha=\pm 2, \pm 1, 0} \int d^3 q \left[ P_\alpha^2 + (q^2+M_2^2) Q_\alpha^2 \right] \label{Masspin2H}\ee
    where $Q_\alpha$ and $P_\alpha$ are the associated  canonical variables and conjugate momenta and the spin sum is over $\alpha= \pm 2, \pm 1, 0$ because this massive particle has spin 2. The Ostrogradsky theorem manifests itself here through the overall minus sign. 
    However, despite that sign there are no instabilities in the free-field limit because that sign cancels in the EOMs.
    \item The effective field theory (EFT) approach tells us that at energies much below $M_2$ runaways should not occur even if the massive spin-2 field has an order-one coupling, $f_2 \sim 1$.
       \item The intermediate case $0 < f_2 < 1$ must have intermediate energy thresholds
       (above which the runaways may take place).
        \item   The weak coupling case $f_2 \ll 1$   must therefore  have energy thresholds
        much larger than $M_2$ and so we could see the effect of the massive spin-2 field and avoid the runaways at the same time.
    \end{itemize}

The detailed treatment of~Ref.~\cite{Salvio:2019ewf} leads to the following result. If the typical energies $E$ associated with the derivatives of the spin-2 fields (both the massless graviton and the massive one) satisfy
 \be E\ll  E_2 \equiv \frac{M_2}{\sqrt{f_2}} = \sqrt{\frac{f_2}{2}}\bp  \qquad \mbox{(for the derivatives of the spin-2 fields)}\label{E2} \ee
 and the typical energies $E$ associated with the matter sector (due to derivatives or mass terms of matter fields and/or matter-field values times coupling constants) satisfy
\be E\ll E_m\equiv \sqrt[4]{f_2} \bp \qquad \mbox{(in the matter sector)}\label{Em} \ee
the runaways are avoided\footnote{There are other interacting systems that fulfill the hypothesis of the Ostrogradsky theorem and  nevertheless avoid  the Ostrogradsky runaway solutions (see e.g.~\cite{Pagani:1987ue,Smilga:2004cy,Pavsic:2013noa,Kaparulin:2014vpa,Pavsic:2016ykq,Smilga:2017arl,Kaparulin:2020rqz,Salvio:2019ewf,Gross:2020tph})}. Therefore, $E_2$ and $E_m$ are the energy thresholds we are interested in. Note that when $f_2\sim 1$ the energy thresholds $E_2$ and $E_m$ reduce to $M_2\sim \bp$, as previously discussed.  Generically these energies are not constants of motion, so in order to avoid runaways, one should impose Conditions~(\ref{E2}) and~(\ref{Em}) at the space and time boundaries. Once this has been done, the Ostrogradsky theorem leads to an acceptable metastability in the context of agravity, rather than to an instability.

By taking $f_2$ not too small, we can make $E_2$ and $E_m$ large enough to accomodate a completely realistic cosmology;  in Sec.~\ref{Inflation} we will see that we can have testable predictions of the $W^2$ term at the same time. 

\subsection{Quantum aspects}\label{Quantum aspects}

Although the argument above gives a satisfactory solution to the {\it classical} Ostrogradsky problem it is still needed to address  possible issues at the quantum level. One reason is the fact that quantum effects might lead to tunneling above the energy thresholds even if the classical fields satisfy the bounds in~(\ref{E2}) and~(\ref{Em}). 

\subsubsection{Trading negative energy with an indefinite metric}\label{Trading negative energy with an indefinite metric}

As we have mentioned at the beginning of Sec.~\ref{The Weyl-squared term}, the inclusion of the $W^2$ term allows the quantum theory to be renormalizable. This can be achieved having the energy of all particles (including the massive spin-2 one)  positive; a pedagogical explanation of why this is possible can be found in Sec.~3.1 of~\cite{Salvio:2018crh}. Therefore, the quantum tunneling over the classical thresholds~(\ref{E2}) and~(\ref{Em}) does not occur as the energy is bounded from below.

 However, as originally noted in~\cite{Stelle:1976gc}, one must include an indefinite metric on the  space of states. To illustrate this result one can focus on a simple {\it classically}-negative Hamiltonian,
\be \hat H= -\frac12\left(P^2+\omega^2Q^2\right). \label{simpleH}\ee
Indeed, as we have seen in Eq.~(\ref{Masspin2H}), the various spin components of the massive spin-2 particle have Hamiltonian of this form ($\omega$ in this case is analogous to $\sqrt{q^2+M_2^2}$ in Eq.~(\ref{Masspin2H})). The classical negative energy can be traded with an indefinite metric, with respect to which $H$, $Q$ and $P$ (as well as the momentum and the generators of the Lorentz group) are self-adjoint, $\hat H^\dagger =\hat H$, $Q^\dagger = Q$, $P^\dagger = P$, ...\, . This can be achieved by keeping the standard canonical commutator 
\be [Q,P]=i, \label{CanComm}\ee
 but exchanging the role of the annihilation $a$ and creation  $a^\dagger $  operators of the harmonic oscillator: 
\be \hat a=\sqrt{\frac{\omega}{2}}\left(Q-i\frac{P}{\omega}\right), \qquad \hat a^\dagger=\sqrt{\frac{\omega}{2}}\left(Q+i\frac{P}{\omega}\right) \label{aad}\ee
(in the standard definition of $\hat a$ and $\hat  a^\dagger$ the relative signs between the $Q$-term and the $P$-term are exchanged). 
Indeed, as shown in detail in Sec.~4.2.1 of~\cite{Salvio:2018crh}, the eigenstates $|n\rangle$ of $H$,  obtained as usual by applying $n$ times $\hat a^\dagger$ on the vacuum $|0\rangle$, have eigenvalues $\omega(n+1/2)>0$, but some of the inner products computed with the above-mentioned metric are negative\footnote{For two generic states $|\psi\rangle$ and $|\phi\rangle$, the symbol $\langle\phi|\psi\rangle$ is used to denote the inner product computed with the indefinite metric, which, as usual, is linear with respect to its right argument $|\psi\rangle$ and antilinear with respect to its left argument $|\phi\rangle$.}:
\be \langle n'|n\rangle = (-1)^n \delta_{nn'}. \label{IndfMetricnnp}\ee
Eqs.~(\ref{CanComm}) and~(\ref{aad}) imply an unusual minus sign in the commutator between $a$ and $a^\dagger$:
\be [\hat a,\hat a^\dagger] = -1. \label{UnusComm}\ee

\subsubsection{Calculation of probabilities and unitarity}\label{Calculation of probabilities}

The result in~(\ref{IndfMetricnnp}) leads to an interpretational complication as in quantum mechanics the positivity of the metric is related to the positivity of probabilities. It must be noted, however, that one does not have to use (and, as we will see, must not use) the indefinite metric in the Born rule to compute probabilities~~\cite{Bender:2002vv,Bender:2007nj,Bender:2007wu,Salvio:2015gsi,Mannheim:2015hto,Mannheim:2017apd,Strumia:2017dvt, Raidal:2016wop}.

To find the correct norms to compute  probabilities we start by a definition of observables that generalizes the one usually given in quantum mechanics to avoid any reference to a specific norm. We define ``observable" any linear operator $A$ with a complete set of eigenstates\footnote{The full Hilbert space in agravity (and in quadratic gravity) can be constructed in a way that the canonical  coordinate operators $q_i$ (corresponding to fields), their conjugate momenta $p_i$ (corresponding to derivatives of fields) as well as the Hamiltonian  have complete sets of eigenstates at {\it any} order in perturbation theory.}
$\{|a\rangle \}$~\cite{Salvio:2018crh}:
  for any state $|\psi \rangle$ there is a decomposition\footnote{We include the case of continuous sets $\{|a\rangle\}$ by thinking an integral of the form $\int da~ c(a) |a\rangle$ to be a specific case of the sum in~(\ref{psiDec}). } 
  \be |\psi \rangle = \sum_a c_a|a\rangle\label{psiDec}\ee 
  for some coefficients $c_a$. 
Moreover, we interpret  $|a\rangle$ as the state where $A$ assumes certainly the value corresponding to $a$, which we call $\alpha_a$. This is the deterministic part of the Born rule. The $c_a$ are assumed to carry the information on the  probability distribution $p_a$  in case $|\psi\rangle$ is a generic state (not necessarily an eigenstate of $A$). This is certainly possible, but there is some redundancy because the $c_a$ are complex numbers, while the $p_a$ should be such that  $p_a\geq 0$ and $\sum_a p_a =1$. In particular, as usual one can consider the $c_a$ physically equivalent to $\bar\nu c_a$, where $\bar\nu$ is an arbitrary complex number.

A correct way to approach this situation is to recall how experiments are performed. As we will see this leads to a modification of some postulate of standard quantum mechanics.  Experimentalists prepare a large number  of times $N$ the same state, so one   considers the direct product of $ |\psi\rangle$ with itself $N$ times:
\be |\Psi_N\rangle \equiv  \bar\nu^N|\psi\rangle ... |\psi\rangle = \sum_{a_1 ... a_N} d_{a_1} ...  \, d_{a_N} |a_1\rangle ... |a_N\rangle,\label{PsiNexp}\ee 
  where we have chosen (without loss of generality) the normalization factor $\bar\nu= 1/\sqrt{\sum_b|c_b|^2}$ and correspondingly defined the normalized coefficients $d_{a}\equiv \bar\nu c_a$.

The reason why~(\ref{PsiNexp}) is a good way of representing the full system is because~(\ref{PsiNexp}) has enough information to represent the full probability distribution of the $N$ uncorrelated tries (of the random variables $a_1, a_2, ..., a_N$). Indeed, the probability distribution of $N$ uncorrelated variables is of the form $p_{a_1}...p_{a_N}$ where $0\leq p_i\leq 1$ and so the coefficients $d_{a_1}...d_{a_N}$ are more than enough to represent any probability distribution of this sort.

The next step is to define frequency operators $F_a$, which counts the number  of  times $N_a$ the value $a$ appears in the direct product $|a_1\rangle ... |a_N\rangle$:  
$$ F_a |a_1\rangle ... |a_N\rangle \equiv \frac{N_a}{N}  |a_1\rangle ... |a_N\rangle.  $$
This defines the $F_a$ because it gives their effect on the elements of a basis in the big space of the $N$ repeated experiments.

We now show in the context of indefinite-metric theories that 
$$ \lim_{N\to\infty} F_a   |\Psi_N\rangle  -p_a  |\Psi_N\rangle = 0, $$ 
where the $p_a$ are  given by the Born rule
\be p
 _a \equiv \frac{|c_a|^2}{\sum_b|c_b|^2}, \label{BornRule}\ee
more precisely, we show that all coefficients in the basis $|a_1\rangle ... |a_N\rangle$ of both $F_a   |\Psi_N\rangle$ and $p_a  |\Psi_N\rangle$ converge to the same quantities and so $|\Psi_N\rangle$ tends to an eigenstate of the frequency operators $F_a$ with eigenvalues $p_a$.   We extend a previous proof~\cite{Strumia:2017dvt}, which considered only a two-state system, $|1\rangle, |2\rangle$ (see also~\cite{Salvio:2019wcp} for a subsequent discussion). To perform the proof we introduce a positive metric on the space of states, which is defined (for an arbitrary pair of states $|\psi\rangle$ and $|\phi\rangle$) as follows:
\be\langle \phi|\psi\rangle_A \equiv \langle \phi|P_A|\psi\rangle, \label{Anorm}\ee
where $P_A$, which we call norm operator, is a linear operator that depends on the observable $A$  and is defined by
\be \langle a'|a\rangle_A\equiv\langle a'|P_A|a\rangle \equiv \delta_{a'a}. \label{aapprod}\ee
So $P_A$ is self adjoint with respect to the indefinite metric, $P_A^\dagger = P_A$. 
We refer to the metric in~(\ref{Anorm}) as the $A$-norm.
This corresponds to a positive metric in the big space of the $N$ repeated experiments. To see this explicitly it is convenient to introduce a more compact notation: we replace $a_1, ..., a_N$ with a compound index $\alpha$ and call $N_{\alpha,a}$ (rather than $N_a$) the number of  
 $a_1, ..., a_N$  equal to $a$ (to emphasize that this number actually depends on $\alpha$) and $|\alpha\rangle\equiv |a_1\rangle ... |a_N\rangle$. Note that for any $\alpha$ we have $\sum_a N_{\alpha,a} = N$. Now, for an arbitrary pair of states 
\be  |\Psi_{1N} \rangle= \sum_{\alpha} D_{1\alpha} |\alpha\rangle, \qquad | \Psi_{2N} \rangle = \sum_{\alpha} D_{2\alpha} |\alpha\rangle, \ee
where the $D_{1\alpha}$ and  the $D_{2\alpha}$ are arbitrary complex numbers, the positive metric is given by the following expression:
\be \langle \Psi_{N2}|\Psi_{N1} \rangle_A\equiv \sum_{\alpha'\alpha} (D_{2\alpha'})^* D_{1\alpha}\langle \alpha'|\alpha\rangle_A  = \sum_{\alpha} (D_{2\alpha})^*D_{1\alpha}, \label{BigInner}\ee 
where
\be \langle \alpha'|\alpha\rangle_A \equiv \langle a_1'|a_1\rangle_A ... \langle a_N'|a_N\rangle_A =\delta_{a_1'a_1}...\delta_{a_N'a_N} \ee
has been used.
The inner product~(\ref{BigInner}), for $| \Psi_{2N} \rangle =| \Psi_{1N} \rangle$, is clearly non-negative   and equal to zero only for $|\Psi_{1N} \rangle = 0$. 

The convergence of $F_a   |\Psi_N\rangle$  to  $p_a  |\Psi_N\rangle$ for $N\to \infty$ (in the sense specified above) with $|\Psi_N\rangle$ given in~(\ref{PsiNexp}) can now be established by showing that the $A$-norm of the big state $(F_a-p_a)   |\Psi_N\rangle$  goes to zero as $N\to \infty$.

  To perform this step we rewrite the state $|\Psi_N\rangle$ in~(\ref{PsiNexp})  as
 \be |\Psi_N\rangle =  \sum_\alpha  \left(\prod_a d_a^{N_{\alpha,a}}\right) |\alpha\rangle
  \ee
  and  so, defining $|\Psi_N^a\rangle\equiv (F_a-p_a)  |\Psi_N\rangle$,
  \be |\Psi_N^a\rangle =  \sum_\alpha  \left(\prod_{a'} d_{a'}^{N_{\alpha,a'}}\right) \left(\frac{N_{\alpha,a}}{N} - p_a\right)|\alpha\rangle.
  \ee
 Using~(\ref{BigInner}), the $A$-norm of $(F_a-p_a)  |\Psi_N\rangle$ is therefore  given by
  \be \langle\Psi_N^a|\Psi_N^a\rangle_A =  \sum_\alpha  \left(\prod_{a'} p_{a'}^{N_{\alpha,a'}}\right)\left(\frac{N_{\alpha,a}}{N} -p_a\right)^2. \ee
  Note that the generic term inside the sum over $\alpha$ depends on $\alpha$ only through $N_{\alpha, b}$. We can therefore  substitute the sum over $\alpha$ with the sum over $N_1=1,2, ...$, $N_2=1,2,...$,   with the constraint $\sum_a N_a = N$, and multiply the generic term by the multinomial coefficient $N!/(N_1!N_2!...)$
  (indeed, the multinomial coefficient is precisely the number of $\alpha$ such that $N_{\alpha, a}=N_a$ for given $N_a$):
 \be \langle\Psi_N^a|\Psi_N^a\rangle_A= \sum_{\substack{N_1, N_2, ... \\ N_1+N_2+... = N} } 
\left(\prod_{a'} p_{a'}^{N_{a'}}\right)\left(\frac{N_{a}}{N} -p_a\right)^2\frac{N!}{N_1!N_2!...}.\label{NotCS}\ee
 The $A$-norm of  
 $(F_a- p_a)|\Psi_N\rangle$ can now be rewritten in a clever way by using the multinomial theorem
   \be \sum_{\substack{N_1, N_2, ... \\ N_1+N_2+... = N} } \left(\prod_{a'} p_{a'}^{N_{a'}}\right)\frac{N!}{N_1!N_2!...} = \left(\sum_{a'} p_{a'}\right)^N.\ee
 Since
 \be  p_a \frac{\partial}{\partial p_a}  \left(\sum_{a'} p_{a'}\right)^N=    \sum_{\substack{N_1, N_2, ... \\ N_1+N_2+... = N} } N_a \left(\prod_{a'} p_{a'}^{N_{a'}}\right)\frac{N!}{N_1!N_2!...}.
\ee
we obtain
\bea  &&\langle\Psi_N^a|\Psi_N^a\rangle_A=  \left[\frac1{N^2}\left(p_a \frac{\partial}{\partial p_a} \right)^2 - \frac{2}{N}p_a^2 \frac{\partial}{\partial p_a} +p_a^2 \right] \left(\sum_{a'} p_{a'}\right)^N\nonumber \\ = && \frac{N-1}{N} p_a^2\left(\sum_{a'} p_{a'}\right)^{N-2} +\frac{p_a}{N} \left(\sum_{a'} p_{a'}\right)^{N-1} -2 p_a^2\left(\sum_{a'} p_{a'}\right)^{N-1}  + p_a^2 \left(\sum_{a'} p_{a'}\right)^N. \eea
Using now $\sum_a p_a=1$, which follows immediately from~(\ref{BornRule}),
\be \langle\Psi_N^a|\Psi_N^a\rangle_A = \frac{p_a(1-p_a)}{N} \xrightarrow[N \to \infty]{}  0. \ee

This concludes our proof that $F_a   |\Psi_N\rangle$ goes to  $p_a  |\Psi_N\rangle$ for $N\to \infty$ (in the sense specified above).  Therefore, the $|\Psi_N\rangle$ given in~(\ref{PsiNexp}), which represents the $N$ repeated experiments, goes to an eigenstate of the frequency operators $F_a$ with eigenvalue given by the Born-rule $p_a$ in (\ref{BornRule}), which we therefore   take as the physically realized probability distribution.

The Born rule~(\ref{BornRule}) can be written in terms of the $A$-norm as 
\be \boxed{p_a = \frac{|\langle a|\psi\rangle_A|^2}{\langle\psi|\psi\rangle_A}.} \label{ProbNorm}\ee
We see that in order to compute the probabilities, we must use a norm  that depends on the observable $A$, the $A$-norm defined in terms of the norm operators in~(\ref{Anorm}). This is a point where the theory does deviates from standard quantum mechanics, where the existence of a unique metric to compute the probabilities is postulated. 
The argument for the Born rule we have presented is an extension of those given in~\cite{BornRule} in the case of standard quantum mechanics. From  Eq.~(\ref{BornRule}) it follows immediately that all probabilities are non-negative 
and that they sum up to one at any time (the theory is unitary), unlike what one could have guessed from the presence of an indefinite metric. This is because the probability must be computed with the {\it positively defined} $A$-norms.

An example of $A$-norm can be found in the simple model with single canonical coordinate $Q$ and Hamiltonian in~(\ref{simpleH}). For the eigeinstates $|n\rangle$ of $H$ we simply define the $H$-norm operator 
\be P_H|n\rangle\equiv (-1)^n|n\rangle,\label{PHdef}\ee which gives the positive $H$-norm
\be \langle n'|n\rangle_H = \delta_{nn'}, \label{nnpprod}\ee
where~(\ref{IndfMetricnnp}) has been used. In the case of agravity, this result tells us that $P_H$ leaves the state with an even number  of (including zero) massive spin-2 particles invariant. Therefore, at energies much below $M_2$ the theory nicely reduces to standard quantum mechanics: no massive spin-2 particles can be created and the norm operator disappears. Deviations from standard quantum mechanics can only occur for energies of order (or above) $M_2$. 

\subsubsection{The Dirac-Pauli canonical variables}\label{The Dirac-Pauli canonical variables}

Beyond the presence of observable-dependent norms to compute probabilities, the quantum theory of agravity differs from standard quantum mechanics in another aspect, which we describe now. 

We have seen in Eq.~(\ref{aapprod}) that a generic observable $A$ admits an orthonormal basis of eigenstates and so must be an $A$-normal operator (that is a normal operator with respect to the $A$-norm). So $A$ can be split as $A=A_h+ A_a$, where $A_h$  ($A_a$) is an $A$-(anti)Hermitian operator (namely  an (anti)Hermitian operator with respect to the $A$-norm). Moreover,  $A_h$  and $A_a$  commute, $[A_h,A_a]=0$. Therefore, the study of $A$ can always be traced back to the separate study of $A_h$ and $A_a$. As a result, an observable can always be seen as an operator that has either a purely real or a purely imaginary spectrum, namely
\be A|a\rangle = \lambda_a |a\rangle, \qquad \lambda_a = \alpha_a \quad \mbox{or}\quad \lambda_a = i\alpha_a \quad (\mbox{with}~\alpha_a~\mbox{real}) \ee
The first case is the one that is realized in standard quantum mechanics. In the quantum theory of agravity there should also be some operators $A$ with  imaginary spectra, $\lambda_a = i\alpha_a$ (see below).  It is important to note that this does not imply that some measurable values of the observable in question must be imaginary. We can still have all measurable values real by simply identifying them with the $\alpha_a$, not with the $\lambda_a=i\alpha_a$ (as we do from now on).
In this case the quantum average is also real because it is defined by 
\be \langle A\rangle \equiv \sum_a \alpha_a p_a, \ee
with the $p_a$ given by the $A$-norm Born rule in~(\ref{ProbNorm}). Since the $p_a$ depend on the state $|\psi\rangle$ one is considering, as usual the quantum average depends on $|\psi\rangle$. But in the standard case one has 
\be \langle A\rangle = \frac{\langle\psi|A|\psi\rangle_A}{\langle\psi|\psi\rangle_A}\in \mathbb{R} \qquad (\mbox{for a real spectrum}), \ee
while in the non-standard one
\be \langle A\rangle = -i\frac{\langle\psi|A|\psi\rangle_A}{\langle\psi|\psi\rangle_A}\in \mathbb{R} \qquad (\mbox{for an imaginary spectrum}). \label{QADP} \ee
Consequently, the quantum average of $A^2$, defined by
\be \langle A^2\rangle \equiv \sum_a \alpha_a^2 p_a \geq 0, \ee
reads, respectively,
\be \langle A^2\rangle = \frac{\langle\psi|A^2|\psi\rangle_A}{\langle\psi|\psi\rangle_A}\geq 0 \qquad (\mbox{for a real spectrum}), \ee
and
\be \langle A^2\rangle = -\frac{\langle\psi|A^2|\psi\rangle_A}{\langle\psi|\psi\rangle_A}\geq 0 \qquad (\mbox{for an imaginary spectrum}).\label{QA2DP} \ee

To show that the non-standard case should also be realized, we now focus on the simple model of Sec.~\ref{Trading negative energy with an indefinite metric}, which represents the various components of the massive spin-2 field in the Fourier transform on the space coordinate. 
The case in which $Q$ has an imaginary spectrum   was first discussed by Pauli~\cite{Pauli} for Lagrangians with at most 2 time-derivatives, elaborating on a previous work by Dirac~\cite{Dirac}. In the rest of this work we will therefore  refer to a canonical variable $Q$ with purely imaginary eigenvalues,
\be Q|x\rangle = i x |x\rangle \qquad (x~ \mbox{real})\label{DPdef}\ee
 as a Dirac-Pauli variable. Note that the observable values of $Q$ are identified with  $x$, which is real. 
 
 An introduction to Dirac-Pauli variables can be found in~\cite{Salvio:2018crh} (see also~\cite{Salvio:2015gsi} for a previous discussion and~\cite{Bender:2007wu,Bender:2008gh} for related works). As explained in~\cite{Salvio:2018crh} the eigenstates $|x\rangle$ satisfy
 \be  \langle x'|x\rangle = \delta(x'+x) \label{xx'inner}\ee 
 so, defining the operator $\eta$ through $\eta |x\rangle = |-x\rangle$,
 \be \langle x'|\eta|x\rangle = \delta(x'-x). \label{xx'innerP}\ee
 This equation tells us that the norm operator $P_Q$ is $\eta$. Therefore, we define the wave function corresponding to a generic state $|\psi\rangle$ as
 \be \psi(x) \equiv \langle x|\eta|\psi\rangle = \langle -x|\psi\rangle. \ee   Moreover, the canonical commutator in~(\ref{CanComm}) implies~\cite{Salvio:2018crh} 
 \be P|x\rangle = \frac{d}{dx}|x\rangle  \label{pDPonx}\ee
  It is easy to show $\eta Q \eta = -Q$ and $\eta P \eta = -P$.
  
Should the components of the massive spin-2 field be described by standard or Dirac-Pauli variables? To answer this question 
we apply $a$ in~(\ref{aad}) to the ground state of $H$ in~(\ref{simpleH}). Using~(\ref{DPdef}) and~(\ref{pDPonx}) we find the differential equation
\be\left(x+\frac{1}{\omega}\frac{d}{dx}\right)\psi_0(x)=0,  \ee
where $\psi_0(x)\equiv \langle x|\eta|0\rangle$ is the groundstate wave function. The solution of this equation is 
\be\psi_0(x)\propto \exp(-\omega x^2/2),\label{VacWave}\ee which is normalizable. If we had started from a standard canonical variable, but keeping  $a$ and $a^\dagger$ defined as in~(\ref{aad}), we would have found a non-normalizable wavefunction (as pointed out in~\cite{Woodard:2006nt,Woodard:2015zca}) because $a$ and $a^\dagger$ in~(\ref{aad}) are exchanged with respect to the standard case. This result tells us that the canonical coordinates describing the massive spin-2 field should be Dirac-Pauli variables. On the other hand, the canonical coordinates corresponding to the other fields (with standard kinetic terms) should be described by standard variables. 

Note, finally, that starting  from a Dirac-Pauli canonical variable, and keeping  $a$ and $a^\dagger$ defined as in~(\ref{aad}), leads to normalizable wave functions $\psi_n(x)\equiv\langle x|\eta|n\rangle$ for all $n$, not only for $n=0$. This is because, using the expression of $\hat a$ and $\hat a^\dagger$ in~(\ref{aad}), 
\be \psi_n(x)\propto \langle x|\eta(a^\dagger)^n|0\rangle = (-1)^n\langle  x| (a^\dagger)^n \eta |0\rangle\propto \left(x-\frac{1}{\omega}\frac{d}{dx}\right)^n\psi_0(x), \label{psin0}\ee
where in the second step we used $\eta Q \eta = -Q$ and $\eta P \eta = -P$ and in the third step we used~(\ref{pDPonx}) and~(\ref{DPdef}).

The results in~(\ref{VacWave}) and~(\ref{psin0}) also imply $\psi_n(-x) = (-1)^n \psi_n(x)$ and so $\eta|n\rangle = (-1)^n |n\rangle$, namely (recalling~(\ref{PHdef})) 
\be P_H=\eta. \ee
 In the spin-$(\pm 2)$ sector of agravity we might have different possibilities where $P_H$ differs from $\eta$~\cite{Salvio:2015gsi,Salvio:2018crh}, because that sector contains two types of canonical variables, one corresponding to the graviton and another one corresponding to the massive spin-2 field, which make together a four-derivative canonical variables  analogous to that of the Pais-Uhlenbeck model~\cite{Pais} (see Secs.~2.3.1 and~4.1.2 of~\cite{Salvio:2018crh} for more details). However, the spin-$(\pm 1)$ sector arising from the metric contains only  massive spin-2 components (see Sec.~2.3.2 of~\cite{Salvio:2018crh}) which therefore  must have $P_H=\eta$. Since the spin components of a massive spin-2 particle are all mixed by the Lorentz group, Lorentz invariance implies $P_H=\eta$ for {\it all} spin components.

\subsubsection{Decays and scattering}\label{Decays and scattering}

Suppose that we prepare a state $|g\rangle$ describing the massive spin-2  ghost (eigenstate of a  free-particle Hamiltonian $H_0$).
According to Eq.~(\ref{ProbNorm}), the probability that this state decays after a large time $t$ into a free-particle state $|\sigma\rangle$ (also taken to be an eigenstate of  $H_0$) is 
\be P_{\rm decay}= \frac{|\langle \sigma|P_{H_0} U(t)|g\rangle|^2}{\langle g|U(t)^\dagger P_{H_0} U(t) |g\rangle} >0, \label{gdecay}\ee
where $U(t)=\exp(-i Ht)$ is the corresponding time-evolution operator. At the fully interacting level $U(t)$ is not guaranteed to commute with $P_{H_0}$, but in the free limit, namely when all coupling constants are set to zero, it does (see our discussion of $P_H$ around~(\ref{PHdef})). If we now choose $|\sigma\rangle$ to be a state describing ordinary particles (for example, an electron-positron pair or  two photons) we know that in the free limit the numerator of $P_{\rm decay}$ vanishes (and so does $P_{\rm decay}$ itself). Therefore, at the leading non-trivial order in $f_2$ the quantity   $P_{\rm decay}$ can be simply computed with the standard formula 
\be P_{\rm decay}= |\langle \sigma|U(t)|g\rangle|^2>0, \qquad \label{leadDecay}\ee
where we have normalized $|g\rangle$ such that $\langle g|P_{H_0}|g\rangle =1$ 
and used $P_{H_0}|\sigma\rangle=|\sigma\rangle$ 
as $|\sigma\rangle$ is a state without ghost quanta. Explicit calculations~\cite{Salvio:2016vxi,Anselmi:2018tmf,Salvio:2018kwh} (see also Ref.~\cite{Donoghue:2018izj}) have shown that $P_{\rm decay}\neq 0$ and so the massive spin-2 particle is unstable. Note that the presence of the norm operator $P_{H_0}$ is crucial to have a positive decay probability: without it  we would have a negative denominator as $\langle g|g\rangle <0$.

 The  decay has been computed so far only at the leading non-trivial
 order in $f_2$, where~(\ref{leadDecay}) applies. Going to the next-to-leading order gives (possibly non-standard)  corrections suppressed by higher powers of $f_2$ and other couplings, as dictated by~(\ref{gdecay}).
However, as we will discuss in Sec.~\ref{Predictions}, in order to have any hope to observe the effects of the abnormal graviton, we should have  $f_2 \lesssim 10^{-5}$ therefore  the leading non-trivial order in $f_2$ appears to be more than adequate. 
At this order in $f_2$ and at zero order in the other couplings the decay rate ($P_{\rm decay}$ per unit time) of the massive spin-2 field is\footnote{$\Gamma_2$ differs by an overall minus sign compared to the expression in~\cite{Salvio:2018kwh} because of a different sign convention.}~ \cite{Salvio:2016vxi,Salvio:2018kwh,Anselmi:2018tmf}
\be  \Gamma_2 =  \frac{f_2^2 M_2}{16\pi}  \bigg(\frac{N_s}{120}+\frac{N_f}{40}+\frac{N_V}{10}\bigg), \label{Gamma2}\ee
 in a theory with $N_s$ real scalars, $N_f$ Weyl fermions and $N_V$ vectors
all with masses  well below $M_{2}$.

Since the kinetic term of a ghost appears with the abnormal minus sign, the decay rate appears in the resumed propagator multiplied by a corresponding minus sign: close to the on-shell value $k^2\approx M_2^2$ of the four-momentum $k$, such propagator reads
\be 
 \frac{-1}{k^2-M_2^2 - i M_2 \Gamma_2} \label{resumed} \ee 
modulo Lorentz indices and $k$-independent factors (the full expression involving Lorentz indices can be found in~\cite{Salvio:2018kwh}). 
 The negative sign in front of $ i M_2 \Gamma_2$ signals  a violation of causality at microscopic scales because it is opposed to the usual  sign of the  $i\epsilon$ prescription. The acausality  could  be probed by
an observer   for example through a $2\to 2$ process among matter particles 
mediated by the ghost, 
measuring that the secondary  vertex is displaced
in the unusual direction~\cite{ColemanAcausality,Grinstein:2008bg,Anselmi:2018tmf} (see also Ref.~\cite{Donoghue:2019ecz} for a related work). 
The acausality (if it ever occurs in agravity) does not manifest itself at energies much below $M_2$. So all observational causality bounds are easily satisfied by taking $M_2$ much above the maximal energies we can observationally probe. As we will see in Secs.~\ref{Applications to the hierarchy problem} and~\ref{Classical metastability and chaotic inflation}, the most interesting situation is when $M_2$ acquires a value right in between the Fermi and the Planck scales. Therefore, the causal behavior observed in all experiments we have performed so far is reproduced. In Sec.~\ref{Causality} we will see that the predictions of agravity can be compatible also with the observations that give us information on the early universe, when the typical energies could have reached and exceeded $M_2$.

Long time ago Lee and Wick~\cite{Lee:1969fy}  claimed that the $S$-matrix  is unitary because the ghost is unstable.
The Lee-Wick idea has been studied in the context of quadratic gravity in several papers (see e.g.~\cite{Tomboulis:1980bs,Antoniadis:1986tu,Hasslacher:1980hd,Salvio:2016vxi,Anselmi:2017yux,Anselmi:2017lia,Anselmi:2017ygm,Anselmi:2018kgz,Anselmi:2018ibi,Donoghue:2019fcb}).
In Refs.~\cite{Anselmi:2017ygm} the unitarity of the $S$-matrix has been achieved by treating the propagators and the loop diagrams with unusual prescriptions introduced as first principles (see also~\cite{Donoghue:2019fcb} for a more recent discussion on the unitarity of the $S$-matrix).
 Here we simply note that the full unitarity of the theory, which was established in Sec.~\ref{Calculation of probabilities}, implies in particular the unitarity of the theory applied to scattering experiments and beyond.
 
In the scattering case one considers the transition probability $P(\sigma\to\sigma')$ between two stable free-particle states $|\sigma\rangle$ and $|\sigma' \rangle$, which are eigenstates of $H_0$. According to Eq.~(\ref{ProbNorm}),
 \be P(\sigma\to\sigma')= \frac{\left|\langle \sigma'|P_{H_0} U(t)|\sigma\rangle\right|^2}{\langle \sigma| U(t)^\dagger P_{H_0} U(t) |\sigma\rangle}>0. \label{ScatteringPH0}\ee
 Again $P_{H_0}|\sigma'\rangle = |\sigma'\rangle$, which allows us to simplify the numerator of the expression above. Assuming $|\sigma\rangle$ and $|\sigma' \rangle$ to be distinct, such that $\langle \sigma'|\sigma\rangle = 0$, the probability under study reduces at tree level to the standard formula
  \be P(\sigma\to\sigma')= \left|\langle \sigma'|U(t)|\sigma\rangle\right|^2>0, \ee
 where we have normalized $|\sigma\rangle$ such that $\langle \sigma|\sigma\rangle =1$. Many processes at tree level have been computed in~\cite{Salvio:2018kwh} by using the standard formula above.
Going to the next-to-leading order gives  (possibly non-standard) corrections suppressed by higher powers of $f_2$ and the other couplings, as dictated by~(\ref{ScatteringPH0}).

 
 %

 \section{Softening of gravity}\label{Softening of gravity}
 
 As we have seen in Secs.~\ref{The action}  the quantum consistency of agravity  
 demands the presence of quadratic-in-curvature terms  in the action, which correspond to extra gravitational degrees of freedom beyond the graviton. Once $\bp$ is generated by DT these degrees of freedom form spin-2 and spin-0 massive particles with masses $M_2$ and $M_0$, respectively (see Sec.~\ref{Quadratic gravity}). They are responsible for a softening of gravity at length scales much smaller than a critical gravitational length scale $L_G$ (which we determine below), or, equivalently, for energy scales much larger than $\Lambda_G\equiv 1/L_G$. The reason behind this effect is that the extra gravitational particles tend to erase the effect of the ordinary graviton in the UV, because this is the only way one can avoid the non-renormalizability of Einstein's gravitational force (which becomes stronger and stronger as the energy increases). The scale $\Lambda_G$ depends on how far we are from the   the quasi-conformal case ($f_0\gg 1$ and $\xi_{ab}\approx -\delta_{ab}/6$, see Sec.~\ref{The action}). If we are very far from it,   both the extra graviparticles play a crucial role and 
 \be\Lambda_G=\max\left(M_0, M_2\right),\label{LambdaG1} \qquad \mbox{(non-conformal regime)}\ee while in the quasi-conformal regime the spin-0 graviparticle $\zeta$ decouples and 
 \be \Lambda_G=M_2, \qquad \mbox{(quasi-conformal regime)}.\label{LambdaG2} \ee
   Intermediate regimes are possible, as we will see in Sec.~\ref{Agravity far from the quasi-conformal regime}.
 
 \subsection{The general scenario of softened gravity}
 
 Agravity is a particular realization of a more general scenario called ``softened gravity", which was proposed in~\cite{Giudice:2014tma} as a solution to the hierarchy problem (see the discussion below). The applications of  softened gravity, however, go well beyond the hierarchy problem.
 
\begin{figure}[t]
\begin{center}
 $ \includegraphics[scale=0.7]{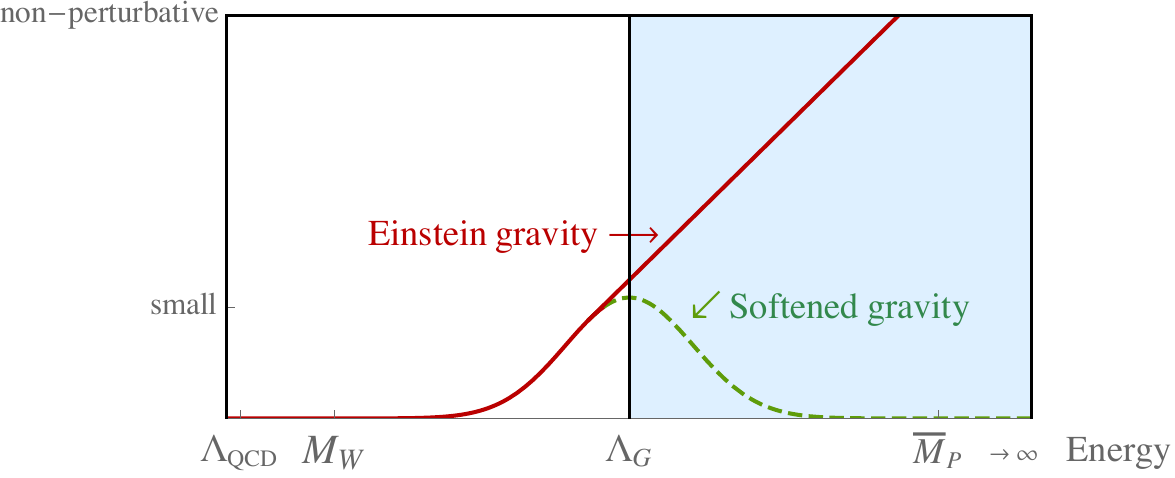} $
  \end{center}
   \caption{\em Pictorial behavior of gravitational interactions in Einstein gravity (red solid line) and in softened gravity (green dashed line).}
\label{softened}
\end{figure}

 The UV issue of Einstein gravity is due to the growth of gravitational interactions as the energy increases, which drives the theory out of control around $\bp$.
A possible solution to obtain a consistent quantum behavior   is to assume that Einstein gravity is replaced by another theory (generically called softened gravity) at energies above some scale $\Lambda_G$ between the EW scale $M_W$ and the Planck scale $\bp$. In this theory the  gravitational interactions get smaller as the energy increases, see Fig.~\ref{softened}. This has important implications (that will be discussed in the next sections). As we have seen above, in agravity $\Lambda_G$ depends crucially on the mass of the extra spin-2 particle, $M_2$. 
 
 \subsection{Applications to the hierarchy problem}\label{Applications to the hierarchy problem}
 
 According to 't Hooft's definition of naturalness~\cite{tHooft:1979rat}, a physical quantity is naturally small when setting it to zero leads to an enhanced symmetry. If a physical quantity is not naturally small a tuning is necessary to keep it small after quantum corrections are taken into account. Indeed, in a QFT all terms compatible with the symmetries must be generically present.  In the SM the Higgs mass parameter $M_h$ is the only dimensionful parameter and setting it to zero makes the SM action scale invariant.  Scale invariance, however, is generically broken by quantum effects regardless of the value of $M_h$. Therefore, the quantity $M_h/\bp$ is not naturally small in the SM; this is known as the hierarchy problem.  
 
It is important to note that   promoting scale invariance to a quantum symmetry is not sufficient {\it per se} to solve the hierarchy problem. Indeed, quantum scale invariance must be spontaneously broken to account for the observed masses and this generically reintroduces a tuning, as we now show following the approach of~\cite{Burgess:2004ib}. Note that if the Higgs potential contains only its scale-invariant part $\lambda_h |h|^4$ then the VEV of $h$ vanishes and EW symmetry breaking does not occur. To cure this problem in a theory with quantum scale  invariance one can try to introduce another scalar field $s$ (which we assume here real for simplicity) and write the scale-invariant potential~\cite{Shaposhnikov:2008xi}
\be W(h,s) =  (\sqrt{\lambda_h}|h|^2-\sqrt{\lambda_s} s^2/2)^2 = \lambda_h |h|^4 + \frac{1}{4}\lambda_s s^4 -\sqrt{\lambda_h\lambda_s}|h|^2 s^2, \label{WHS}\ee
where the quartic couplings $\lambda_h$ and $\lambda_s$ are assumed to be positive to ensure that this potential is bounded from below.
$W(h,s)$ has a flat direction corresponding to $\sqrt{\lambda_h}|h|^2=\sqrt{\lambda_s} s^2/2$ and so one can have a non-vanishing value of the VEV of $h$. The problem is that~(\ref{WHS}) is not the most general potential compatible with scale invariance, which is instead
 \be \lambda_h |h|^4 + \frac{1}{4}\lambda_s s^4 +\lambda_{hs}|h|^2 s^2 \ee
 with a generic $\lambda_{hs}$. In~(\ref{WHS}) $\lambda_{hs}=-\sqrt{\lambda_h\lambda_s}$ and so the tuning $\lambda_{hs}+\sqrt{\lambda_h\lambda_s} = 0$ must be done to preserve the flat direction. In other words the quantity $\lambda_{hs}+\sqrt{\lambda_h\lambda_s} $ is not naturally small.
 
 Given that scale invariance is not sufficient to solve the hierarchy problem another mechanism is required. Several proposals are available in the literature (such as supersymmetry). Softened gravity offers another mechanism. At the scale $\Lambda_G$ Einstein's gravity is still weak, so we can compute the gravitational contribution to $M_h^2$ at the leading order in the Newton constant $G_N$ to find
$$ \delta M_h^2  \lesssim  \frac{G_N \Lambda_G^4}{(4\pi)^2}. $$
The fourth power of $\Lambda_G$ appears for dimensional reasons, while the $(4\pi)^2$ in the denominator is there because this is a  loop effect. 
 Requiring now $\delta M_h\lesssim M_h$  we find that 
 \be \Lambda_G \lesssim  10^{11}~\mbox{GeV}\label{GsoftB}\ee
  in order for the ratio $M_h/\bp$ to be naturally small. If~(\ref{GsoftB}) holds what happens is that the gravitational interactions preserves an approximate shift symmetry acting on the Higgs field, which
  is softly broken by $M_h$ itself (and so all radiative gravitational corrections give a contribution to $M_h$ which is at most of order $M_h$). In order to have a complete solution to the hierarchy problem, it is necessary that the matter sector preserves such shift symmetry. This, together with the requirement of having UV fixed points for the matter couplings (see Sec.~\ref{The action}), leads to the presence of new physics at scales not far from 10~TeV~\cite{Pelaggi:2015kna}, with interesting implications for future colliders.

 \subsubsection{Agravity far from the conformal regime}\label{Agravity far from the quasi-conformal regime}
 
 Let us focus now on agravity. The first case we consider is when we are far from the conformal regime (see~\cite{Salvio:2014soa,Salvio:2017qkx}).
  We assume $f_0\lesssim 1$, so a perturbative expansion in $f_0$ is possible. Looking at the RGE of $M_h^2/\bp^2$ in Eq.~(\ref{eq:RGEm}) we see that only the first term
 \be -\xi_H [5\gt^4+\gs^4(1+6\xi_H)]\ee
 can generate unnaturally large corrections. So a naturally small $M_h/\bp$ requires small values of $\gt^4$ and $\gs^4(1+6\xi_H)$. Inserting the experimental values of $M_h$ and $\bp$  we obtain that both these couplings should satisfy\footnote{One could slightly increase those bounds with specific matter contents, but we quote here the most general ones.}~\cite{Salvio:2014soa,Salvio:2017qkx,Kannike:2015apa}
 \be \gt\lesssim 10^{-8}, ~~ \gs |1+6\xi_H|^{1/4} \lesssim 10^{-8}.\label{softf02}\ee
 In this case, gravity is weak enough to preserve the approximate Higgs shift symmetry softly broken by $M_h$, which we discussed above.
 
 The bound in
~(\ref{softf02}) demands, according to Eqs.~(\ref{M0}) and~(\ref{M2}), that 
 \be M_2\lesssim 10^{10}~\mbox{GeV}, \qquad M_0\lesssim \frac{10^{10}~\mbox{GeV}}{|1+6\xi_H|^{1/4}}, \label{softM2M0} \ee 
and far from the quasi-conformal value, where $(1+6\xi_H)$ is not small, we find agreement with the general bound in~(\ref{GsoftB}) for $\Lambda_G=\max\left(M_0, M_2\right)$. One can increase the value of $M_0$ compatibly with the Higgs mass naturalness by going a bit towards the Weyl-invariant limit (taking $\xi_H+1/6$ moderately small)~\cite{Kannike:2015apa}; this can be considered as an intermediate regime between the one considered in this section and that in the next section~\ref{Agravity in the quasi-conformal regime}. 
However, the bound on $M_2$ in~(\ref{softM2M0}) cannot be changed significantly, even in the quasi-conformal regime, as we now discuss.
 
 Note that, being far from the conformal regime, we can trigger DT not only though the non-perturbative mechanisms mentioned in Sec.~\ref{Non-perturbative mechanisms}, but also through the perturbative mechanism of Sec.~\ref{Perturbative mechanisms}, which (to generate a real Planck mass) requires, as we have seen, a non-minimal coupling $\xi_\varphi$ to be positive (see~(\ref{eq:agravMPl}))  and so far from the conformal value $-1/6$.

 \subsubsection{Agravity in the quasi-conformal regime}\label{Agravity in the quasi-conformal regime}
 
 Let us turn now to the quasi-conformal case ($f_0\gg 1$ and $\xi_{ab}\approx -\delta_{ab}/6$) discussed in Sec.~\ref{The action} (see also~\cite{Salvio:2017qkx} for more details). In this case the effective scalar $\zeta$ corresponding to the $R^2$ term has only negligibly small couplings with the other degrees of freedom. As a result, there are no $f_0$ contributions in the one-loop RGE of $M_h^2/\bp^2$, whose relevant part reads~\cite{Salvio:2017qkx}
\be
(4\pi)^2\mu\frac{d  }{d\mu} \frac{M_h^2}{ \bp^2}= \frac{5}{6} \gt^4 + 
\cdots,\qquad \hbox{for}\qquad M_2 < \mu < M_{\rm Pl}.
\ee
The requirement of a naturally small Higgs mass then leads to $f_2 \lesssim10^{-8}$, like in~(\ref{softf02}), but now there are no constraints on $f_0$ because $\zeta$ is essentially decoupled. Then looking at~(\ref{M2}) we obtain 
\be M_2\lesssim 10^{10}~\mbox{GeV}, \label{softM2M0c}\ee
again in agreement with the general bound in~(\ref{GsoftB}), but this time for $\Lambda_G = M_2$.

 Note that, in the quasi-conformal regime, we cannot trigger DT generating a real Planck mass  through the perturbative mechanism of Sec.~\ref{Perturbative mechanisms}, which requires a non-minimal coupling $\xi_\varphi$ to be positive (see~(\ref{eq:agravMPl})). Therefore, in this case we assume that DT has taken place non-perturbatively as described in Sec.~\ref{Non-perturbative mechanisms}.

 \subsubsection{Implications for the classical metastability}\label{Implications for the classical stability}
 
 In the previous sections we have seen that the Higgs mass naturalness requires a very small coupling $f_2$ of the massive spin-2 particle. Combining this with the result of Sec.~\ref{Classical aspects} we see that the energy thresholds $E_2$ and $E_m$ (above which the classical Ostrogradsky instabilities may take place) are both much larger than $M_2$. On the other hand, we have also seen that the naturalness bound $f_2\lesssim 10^{-8}$ leads to an upper  bound on $M_2$ around $10^{10}$~GeV.  
 
This leads us to a very  interesting situation: there exists an energy range in which the predictions of agravity deviates from those of GR, but without activating runaway solutions. Calling $E$ the typical energy associated with the derivatives of the spin-2 fields and using~(\ref{E2}), this energy range for a small $f_2$  reads
\be M_2\lesssim E\ll \frac{M_2}{\sqrt{f_2}}\quad 
  \iff \quad \frac{f_2}{\sqrt{2}}\bp\lesssim E\ll \sqrt{\frac{f_2}{2}} \bp
  \ee
Setting now for example the maximal value of $f_2$ compatible with Higgs mass naturalness, $f_2\sim 10^{-8}$, we obtain 
\be 10^{-8}\bp\lesssim E\ll 10^{-4}  \bp.
 \label{RangeGhost} \ee
For smaller values of $f_2$ we obtain  an even larger energy range.
In Sec.~\ref{Inflation} we will see some of the predictions which differ in agravity and in GR. In the same section we will also see that setting $f_2$ around $10^{-8}$ allows us to understand why we live in a nearly homogeneous and isotropic universe.
 
 \subsection{The cosmological constant problem}\label{The cosmological constant problem}
  
  In the SM there is another fine tuning problem, the one which affects the cosmological constant $\Lambda$. Observations tell us that $\Lambda$ is about $120$ orders of magnitude smaller than $\bp$, but no known symmetry implies $\Lambda =0$ and can be implemented in a realistic model at the same time. This is  the cosmological constant problem~\cite{Weinberg:1988cp}.  For example, quantum scale invariance would imply $\Lambda=0$ but, as we have seen, a realistic model requires this symmetry to be spontaneously broken and this reintroduces a fine tuning.
  
   In softened gravity a possible solution to the cosmological constant problem would demand $\Lambda_G\lesssim \Lambda$. 
  This condition 
  is, however, only necessary to have a naturally small cosmological constant. Indeed, any mass scale $m_i$ including those in the SM, which are a priori unrelated to gravity, would contribute to the RGE of  $\Lambda$ with a term proportional to $m_i^4$ (see Appendix B of~\cite{Salvio:2017qkx} for the complete RGE of $\Lambda$ in agravity) and all SM particles have $m_i^4\gg\Lambda$. This renders the problem very difficult if not impossible to solve: no symmetry seems to be able to protect $\Lambda$  leaving all $m_i$ at their observed values. As a result, it is not known whether a realistic model with a naturally small $\Lambda$ can be built. 
  
  A possible explanation of the hierarchy $\Lambda\ll \bp$ could be found by noting that a much bigger value would not be compatible with life\cite{Weinberg:1988cp,Weinberg:1987dv}, but in order for this ``anthropic principle" to be a real explanation it is necessary to have a multiverse, where $\Lambda$ and the other parameters vary according to some distributions. Currently, however, it is not known how to derive such a multiverse and the corresponding distributions from first principles without ad hoc assumptions. Therefore, the cosmological constant problem remains an open challenge for future research.

 \subsection{Applications to ultracompact objects}\label{Applications to ultracompact astrophysical objects}
 
 The softening of gravity implies that objects with a sufficiently small mass $M$  (with a small enough Schwarzschild radius
$r_h\equiv 2 G_N M$) can be well described by the linearized theory in $h_{\mu\nu}\equiv g_{\mu\nu}-\eta_{\mu\nu}$ and the corrections in higher powers of $h_{\mu\nu}$ get smaller the more $r_h \Lambda_G$ decreases~\cite{Salvio:2019llz}. This effect, which we refer to as the {\it linearization mechanism}, implies that objects with
\be r_h\ll L_G\equiv \frac1{\Lambda_G} \label{LinCond}\ee
 do not feature an (event) horizon in softened gravity and in particular  in agravity~\cite{Salvio:2019llz} (see also~\cite{Salvio:2014soa} for a previous discussion). This result assumes that the objects in question are generated by a physical energy-momentum distribution. This is because the softening  of gravity prevents all matter to collapse to a point unlike in GR\footnote{For studies of compact and ultracompact objects in the vacuum see Refs.~\cite{Stelle:1977ry,Holdom:2002xy,Nelson:2010ig,Myung:2013doa,Lu:2015cqa,Lu:2015psa,Cai:2015fia,Lin:2016kip,Goldstein:2017rxn,Lu:2017kzi,Kokkotas:2017zwt,Stelle:2017bdu,Podolsky:2019gro}.}. The linearization mechanism implies that the black holes (BHs) of GR are replaced by horizonless ultracompact objects (UCO).
 
 In this section we discuss the main features of the linearization mechanism at the classical level (see~\cite{Salvio:2019llz} for more details). For this purpose we look at the gravitational field equations of quadratic gravity,
\be\label{eq:HD_eom}
	{\cal G}_{\mu\nu}\equiv G_{\mu\nu} + \frac{2}{M_{2}^2} B_{\mu\nu} - \frac{1}{3M_{0}^2} \left[R\left(R_{\mu\nu} - \frac{1}{4}R g_{\mu\nu}\right) + g_{\mu\nu} R^{;\rho}{}_{\rho} - R_{;\mu\nu} \right] = \kappa T_{\mu\nu},
\ee
where $G_{\mu\nu}\equiv R_{\mu\nu}-g_{\mu\nu}R/2$ is the Einstein tensor, $B_{\mu\nu} \equiv \left( \nabla^{\rho} \nabla^{\sigma} + \frac{1}{2} R^{\rho\sigma} \right)W_{\mu\rho\nu\sigma}$ is the Bach tensor, $\kappa\equiv 8\pi G_N$  and $T_{\mu\nu}$ is the (matter) energy-momentum tensor\footnote{As usual the semicolon corresponds to the covariant derivative (e.g. $R_{;\mu\nu} \equiv \nabla_\mu\nabla_\nu R$).}.
 
 A  way to qualitatively understand the origin of the linearization mechanism is to  consider a simple point-like distribution of mass $M$,
 which has an energy-momentum tensor
 \be T^{\mu}_{~\nu} = {\rm diag}(M\delta(\rho),0,0,0),\label{Tpoint}\ee 
 where $\rho$ is a radial coordinate,
and  generates a Newtonian potential~\cite{Stelle:1977ry}
\be\label{VNexpr}
	V_N(\rho) = -\frac{r_h}{2\rho}\left(1-\frac43 e^{-M_2 \rho}+\frac13 e^{-M_0 \rho}\right).
\ee
As shown in~\cite{Salvio:2019llz}, this potential satisfies
\be\label{eq:Vbound_0}
	|V_N(\rho)| 
	\leq \frac{r_h}{6}(4 M_2 + M_0).
\ee
Thus $|V_N(\rho)|$ is much smaller than 1 for any $\rho$ if $r_h\ll \min\left(1/M_0, 1/M_2\right)$ and a horizon cannot form. This occurs precisely because the contribution of the massive spin-2 and spin-0 graviparticles to $V_N$ (the second and third terms in Eq.~(\ref{VNexpr})) cancel the graviton contribution (the first term in Eq.~(\ref{VNexpr})) for small length scales, namely much smaller than $\min\left(1/M_0, 1/M_2\right)$. This agrees with the general condition in~(\ref{LinCond}) for $L_G=\min\left(1/M_0, 1/M_2\right)$. Indeed, as we have seen at the beginning of Sec.~\ref{Softening of gravity}, this is the correct value of $L_G\equiv 1/\Lambda_G$ when we are far from the quasi-conformal regime: Eq.~(\ref{Tpoint}) implies $T^{\mu}_{~\mu} \neq 0$, while Weyl symmetry would require $T=0$ on shell (on the solution of the field equations).

As we have discussed, whenever Condition~(\ref{LinCond}) is satisfied a horizon does not form in agravity, not only for the point-like distribution. To see how this is possible one can look at the static $h_{\mu\nu}$
 generated by a static $T_{\mu\nu}$. In~\cite{Salvio:2019llz} it was shown 
 \be\label{eq:h_bound}
	|h^{\mu}{}_{\nu}| \leq \frac{\kappa }{48 \pi^2} M_{0} |T^{\rho}_{~\rho}|_{\rm int} \delta^{\mu}_{\nu}+ \frac{\kappa }{8 \pi^2} M_{2} \left| T^{\mu}{}_{\nu} -  T^{\rho}_{~\rho}\delta^{\mu}_{\nu}/3 \right|_{\rm int} + \left(\ldots \right),
\ee
where for a generic  function $X$ of the space point $\vec x$
\be
	|X|_{\rm int} \equiv \int d^3 x |X(\vec x)|
\ee
and the ellipsis stands for terms that can be set to zero by a suitable gauge choice.
So up to gauge-dependent terms, which can be set to zero without loss of generality, the metric perturbation is bounded for finite non-singular sources. Moreover, if we take on physical grounds $\rho=T^{0}_{~0} > 0$, and the components of stress energy tensor  bounded by $\rho$ , i.e. there exists a constant $C>0$ so that $|T_{\mu\nu}| \leq C \rho$, e.g. when the dominant energy condition is satisfied, then an upper bound analogous to \eqref{eq:Vbound_0} holds and  the condition in \eqref{LinCond} implies a weak gravitational field and thus no horizon. 

Moreover, looking at~(\ref{eq:h_bound}) we can explicitly see that in the  conformal limit (when $T^{\rho}_{~\rho}=0$ and thus $|T^{\rho}_{~\rho}|_{\rm int}=0$ on shell) the dependence of $L_G$ on $M_0$ disappears and so $L_G = 1/M_2$. So, once again, when we are  in the quasi-conformal regime discussed at the end of Sec.~\ref{The action}, we find that the energy scale above which gravity is softened is simply $M_2$. 

In~\cite{Salvio:2019llz} it was also shown that in agravity one can have a UCO, namely one with linear size $S_l< r_h = 2G_N M$, and avoid the Ostrogradsky instabilities as described in Sec.~\ref{Classical aspects}. When~(\ref{LinCond}) holds this happens for
\be 
	M_2 \ll \frac{S_l}{3r_h}\bp. \label{Crun}
\ee
On the other hand, a study of horizonless ultracompact objects when~(\ref{LinCond}) does not hold was performed in~\cite{Holdom:2016nek,Ren:2019afg}, where the implications for the  information puzzle were discussed.
Note that the condition in~(\ref{Crun}) is easily satisfied for values of $M_2$ that correspond to a natural Higgs mass (see the bound on $M_2$ in~(\ref{softM2M0}) and~(\ref{softM2M0c})). Several horizonless ultracompact objects that are free from any singularity and from Ostrogradsky instabilities have been found in~\cite{Salvio:2019llz}.

Due to the lack of horizons, UCOs do not evaporate. This has several phenomenological implications. Unlike a  BH in GR, such objects can be stable and thus serve as possible candidates for DM~\cite{Salvio:2019llz,Aydemir:2020xfd} as they form, e.g., via the collapse of large primordial fluctuations~\cite{Misner:1974qy,Carr:1974nx}. Heavier UCOs, if they possess a horizon, evaporate, but have to stop doing so after they lose most of their mass and enter the softened-gravity regime, leaving a remnant.  The idea of BH remnants as DM is not recent~\cite{MacGibbon:1987my}, although no concrete realizations of such objects were proposed elsewhere. Softening of gravity implies that the evaporation history of BHs must thus be changed and this can affect the allowed mass window of heavier DM candidates~\cite{Raidal:2018eoo}. 

Another implication of the linearization mechanism is that it can avoid the formation of microscopic BHs with horizons $r_h$ smaller than $1/h_{\rm max}$, where $h_{\rm max}$ is the value of the Higgs radial field for which the effective Higgs potential acquires its maximum. Indeed, these microscopic BHs have been proven to be very dangerous for the SM as they can act as seeds for EW vacuum decay~\cite{Burda:2015isa,Burda:2016mou,Tetradis:2016vqb}, if the EW vacuum is metastable~\cite{Buttazzo:2013uya,Salvio:2017eca}
 (see~\cite{Salvio:2019llz} for a more detailed discussion).

 \section{The early universe}\label{Inflation}
 
 As we have seen in Sec.~\ref{Implications for the classical stability}, having a natural hierarchy between the Higgs and the Planck masses leads to the existence of an energy range where the classical Ostrogradsky instabilities are avoided, but still a deviation from Einstein gravity occurs. This energy range is very high (see~(\ref{RangeGhost})) so the natural arena to test agravity is the early universe. In this section we therefore  focus on this epoch.
 Moreover, as we will see in Sec~\ref{Classical metastability and chaotic inflation}, for values of $f_2$ close to (but still compatible with) the Higgs naturalness bound, $f_2 \lesssim10^{-8}$, the presence of the Weyl-squared term allows us to understand why we live in a homogeneous and isotropic universe.
 

 \subsection{The Einstein frame Lagrangian}\label{The Einstein frame Lagrangian}
 
 In the early universe the scalars $\phi_a$ can acquire very large values, so the non-minimal couplings in~(\ref{Lnonminimal}) can have non-negligible effects. It is therefore  useful to rewrite the theory in a more familiar way. The general scalar-tensor theory in agravity is  (up to total derivatives)
\be S_{st} = \int d^4x  \sqrt{- g} \,\bigg[\frac{R^2}{6f_0^2}- \frac{W^2}{2 f_2^2} -\frac{\mathscr{F}(\phi)}{2}R+ \frac{1}{2} \left(\partial \phi\right)^2 - V(\phi) \bigg],\label{renAction}\ee
where $\left(\partial \phi\right)^2\equiv \left(\partial_\mu \phi^a\right)\left(\partial_\mu \phi^a\right)$ and $\mathscr{F}$ and $V$ are functions of the scalar fields. Before DT occurs these functions are given by $\xi_{ab}\phi_a\phi_b$ and~(\ref{Vns}), respectively. After DT the mass scales, such as the Planck mass $\bp$ and the cosmological constant $\Lambda$, appear and we include them in $\mathscr{F}$ and $V$.

The non-standard $R^2$ term can be removed by introducing an auxiliary field $\cal A$ with a Lagrangian that vanishes on-shell:
\be  - \sqrt{-g} \frac{(R+3f_0^2 {\cal A}/2)^2}{6f_0^2},\ee
which we are therefore free to add to the total Lagrangian. Once we have done so   the action reads 
\be S_{st} = \int d^4x  \sqrt{-g} \,\bigg[ - \frac{W^2}{2 f_2^2} -\frac{f({\cal A},\phi)}{2} R + \frac{1}{2} \left(\partial \phi\right)^2 - V(\phi) -\frac{3f_0^2{\cal A}^2}{8} \bigg], \label{Sststep}\ee
where $f({\cal A},\phi)\equiv\mathscr{F}(\phi)+{\cal A}$. Note that, after DT, the leading term in $f({\cal A},\phi)$ in the field expansion is the Planck mass, $f({\cal A},\phi)= \bp^2 +...$ because $\mathscr{F}$ contains $\bp$. Therefore, for small enough scalar fields 
\be f({\cal A},\phi) > 0. \label{fPos}\ee
For all values of ${\cal A}$ and the $\phi_a$ such that~(\ref{fPos}) is satisfied we can perform a Weyl transformation (sometimes called ``conformal transformation") of the metric 
\be g_{\mu\nu} \rightarrow \frac{\bp^2}{f}g_{\mu\nu}. \label{ConfTransf}\ee
This has the effect of eliminating the  non-standard $fR$ term in~(\ref{Sststep}) 
and the scalar-tensor action becomes 
\be S_{st} = \int d^4x  \sqrt{-g} \,\left\{ - \frac{W^2}{2 f_2^2}  -\frac{\bp^2}{2} R  +\bp^2 \left[\frac{\left(\partial \phi\right)^2}{2f} +\frac{3(\partial_\mu f)^2}{4f^2}\right]- U \right\}, \label{SstTransf}\ee
where 
\be U = \frac{\bp^4}{f^2}\left( V +\frac{3f_0^2}{8} {\cal A}^2\right). \ee
The field $f$ can now be seen as an extra scalar degree of freedom.
 To simplify further the action we define $\zeta\equiv\sqrt{6f}$ (notice that, in order for the metric redefinition in (\ref{ConfTransf}) to be regular,~(\ref{fPos}) must hold and thus we can safely take the square root of $f$) and we obtain 
\be S_{st} = \int d^4x  \sqrt{-g} \,\left\{ - \frac{W^2}{2 f_2^2}    -\frac{\bar M_{\rm Pl}^2}{2} R  +\frac{6\bp^2}{2\zeta^2} \left[\left(\partial \phi\right)^2  + (\partial \zeta)^2 \right]- U(\zeta,\phi) \right\}, \label{EFstaction}\ee
where 
\be U(\zeta,\phi) = \frac{36 \bp^4}{\zeta^4}\left[V(\phi) +\frac{3f_0^2}{8} \left(\frac{\zeta^2}{6} - \mathscr{F}(\phi)\right)^2\right]. \label{UEins}\ee
The form of the scalar-tensor action in (\ref{EFstaction}) is known as the Einstein frame because all the non-minimal couplings have been removed\footnote{The part of the Einstein-frame action for the gauge fields and fermions can be found in~\cite{Kannike:2015apa}  (and in~\cite{Salvio:2018crh}, where the Planck mass and the cosmological constant appear explicitly).}.

Starting from the initial action~(\ref{Lgravity3}) and using the auxiliary field method, it is also possible  to make the massive graviton appear explicitly  in the Lagrangian by performing appropriate field redefinitions~\cite{Hindawi:1995an,Salvio:2019ewf}.

\subsection{Inflation: the classical FRW background}\label{Inflation: the FRW background}

To describe inflation in this theory we introduce as usual 
the homogeneous and isotropic Friedmann-Robertson-Walker (FRW) metric,
 \be ds^2 = dt^2 -a(t)^2 \left[dr^2+r^2(d\theta^2 +\sin^2\theta d\phi^2)\right], \ee
 which approximately describes   the patch of the universe where we live.
Here $a(t)$ is the scale factor
and, for simplicity, we have set to zero the spatial curvature parameter. If this parameter  is zero at some initial time $t$ it remains so for all values of  $t$; moreover, as usual inflation makes this parameter decrease exponentially with $t$ so its effect is eventually negligibly small in an inflationary universe. In this section we only consider the classical aspects of inflation. The FRW metric corresponds to a possible choice of the classical background $g^{\rm cl}_{\mu\nu}$ in~(\ref{metricSplit}). Quantum fluctuations will be studied in Sec.~\ref{Inflation: quantum perturbations}. As we will see in Sec.~\ref{Classical metastability and chaotic inflation}, the classical metastability discussed in Sec.~\ref{Classical aspects} will allow us to understand why the metric of our patch is nearly homogeneous and isotropic, at least for values of $f_2$ close to the Higgs naturalness bound, $f_2 \lesssim10^{-8}$.

For the following description of the quantum fluctuations  it is convenient to introduce here the conformal time $\tau$ defined as usual  by
\be \tau = \int_{t^*}^t \frac{dt'}{a(t')}, \ee
where $t^*$ is some reference time; in the following we will choose $t^* \rightarrow \infty$. The FRW metric in terms of $\tau$  is
  \be ds^2 = a(\tau)^2\left(d\tau^2 -\delta_{ij}dx^idx^j\right).  \label{ConfFRW} \ee
  This is a good point to mention that, because agravity can flow to conformal gravity in the UV (see the discussion at the end of Sec.~\ref{The action}) and the FRW metric is conformally flat, as clear from~(\ref{ConfFRW}), the initial-time cosmological singularity of GR can  be elegantly avoided.

We introduce  as usual  ${\cal H}(\tau) \equiv a'(\tau)/a(\tau)$ (related to $H(t)\equiv \dot a(t)/a(t)$ by ${\cal H}(\tau) = a(t)H(t)$) and a prime denotes a derivative with respect to $\tau$, while a dot is a derivative with respect to $t$.

Since the theory includes several scalar fields ($\zeta$ and the $\phi_a$) it is convenient to include all of them in a set $\phi^i$, where a value of the index $i$  corresponds to $\zeta$ and the remaining ones correspond  to $\phi_a$. With this notation the scalar field metric in~(\ref{EFstaction}) can be  written
 \be K_{ij} =  \frac{6\bp^2}{\zeta^2} \delta_{ij}. \label{KijRen}\ee
Also $K^{ij}$ denotes the inverse of the field metric (which is used to raise and lower the scalar indices $i,j,k, ...$); for example $F^{,i}\equiv K^{ij}F_{,j}$, where,  for a generic function $F$ of the scalar fields, we defined $F_{,i}\equiv \partial F/\partial \phi^i$. Notice that in the case of pure de Sitter space we have $a(\tau)=-1/(H\tau)$, $\phi'^i=0,$ $U_{,i} =0$.

Inflation (a nearly exponential growth of $a(t)$) in general takes place when 
\be \varepsilon \equiv -\frac{\dot H}{H^2}  < 1, \ee 
but great simplifications occur in the so-called slow-roll regime. 
The slow-roll inflation occurs when two conditions are satisfied \cite{Salvio:2016vxi} (see also \cite{Chiba:2008rp} for previous studies):
\be \epsilon \equiv  \frac{  \bar M_{\rm Pl}^2 U_{,i}U^{,i}}{2U^2} \ll 1. \label{1st-slow-roll}\ee 
\be \left|\frac{\eta^{i}_{\,\,\, j} U^{,j}}{U^{,i}}\right|  \ll 1 \quad \mbox{($i$  not summed), }\quad \mbox{where}\quad \eta^{i}_{\,\,\, j}\equiv \frac{\bar M_{\rm Pl}^2 U^{;i}_{\,\,\, ;j}}{U}. \label{2nd-slow-roll} \ee
 It is easy to check that $\epsilon$ and $\eta^i_{\,\,\, j}$ reduce to the well-known single-field slow-roll parameters in the presence of only one field.
The duration of inflation is parameterized by the number of e-folds $N_e$, which  is defined by
\be N_e \equiv \int_{t_e}^{t} dt' H(t'), \label{Ndef2}\ee
where $t_e$ is the time when inflation ends.

\subsubsection{Inflation far from the conformal regime}\label{Inflation far from the conformal regime}

When we are far from the conformal regime, and in particular some of the non-minimal couplings $\xi_{ab}$ are positive, we can trigger  DT through the gravitational perturbative mechanism described in Sec.~\ref{Perturbative mechanisms}. This requires the existence of a scalar field, the Planckion $\varphi$, which is responsible for the Planck mass and the cosmological constant. Then the Planckion can be identified with the inflaton. We do so in this section. This will allow us to illustrate a mechanism to generate naturally flat inflationary potentials (where the conditions in~(\ref{1st-slow-roll}) and~(\ref{2nd-slow-roll}) are  satisfied automatically),  which can be realized in agravity whenever the Planck mass and the cosmological constant are generated perturbatively.

When the role of the inflaton is played only by the Planckion the potential is steeper along the  other scalar-field directions. In particular this means that the second term in the square bracket of~(\ref{UEins}) effectively sets $\zeta^2=6\mathscr{F}(\phi)$ and so the Einstein-frame potential of $\varphi$ is 
\be U = \bp^4 \frac{V}{\mathscr{F}^2}. \ee 
Also, in this case we can substitute $V$ with $\lambda_\varphi \varphi^4/4$ and  $\mathscr{F}$ with $\xi_\varphi \varphi^2$ (see Sec.~\ref{Perturbative mechanisms}) to obtain
\be U = \frac{\bp^4}{4}\frac{\lambda_\varphi}{\xi_\varphi^2}. \label{UPlanckion}\ee 
The potential that $\varphi$ feels is therefore  flat at tree-level. The same radiative corrections that lead to DT also lead to some slope, which, however, is small because we have assumed the theory to be perturbative. The slow-roll parameters in~(\ref{1st-slow-roll}) and~(\ref{2nd-slow-roll}) are therefore  given by the beta functions of $\lambda_\varphi$, $\xi_\varphi$ and their beta functions as~(\ref{2nd-slow-roll}) involves the second derivative of the potential (see~\cite{Salvio:2014soa} for explicit expressions). The slow-roll parameters are therefore  automatically suppressed by loop factors as the theory is assumed to be perturbative in this case; no ad hoc assumptions are needed to make them small.

The exact form of the potential is  model dependent as it depends on the beta functions. However, if we expand $U$ around the field value $\bar \varphi$ introduced in Sec.~\ref{Perturbative mechanisms} and we stop at the quadratic order in $\varphi- \bar\varphi$ we have
\be U = \frac{M_\varphi^2}{2} (\varphi - \bar\varphi)^2, \label{Uplanckionapp}\ee
where we have used 
\be \varphi \frac{\partial U}{\partial\varphi} = \frac{\beta_{\lambda_\varphi}}{\xi_\varphi^2} -\frac{2\lambda_\varphi}{\xi_\varphi^3} \beta_{\xi_\varphi}, \ee
which follows from~(\ref{UPlanckion}),
and the fact that both $\lambda_\varphi$ and $\beta_{\lambda_\varphi}$  vanish at $\bar\varphi$ (see~(\ref{eq:agravMPl})). The quantity $M_\varphi^2$ in~(\ref{Uplanckionapp}) is a non-negative\footnote{$M_\varphi^2$ cannot be negative because, as we have seen in Sec.~\ref{Perturbative mechanisms}, the function $\lambda_\varphi(\varphi)$, which vanishes at $\varphi=\bar\varphi$, must be non-negative at least in a neighborhood of $\varphi$.
} constant defined by
\be M_\varphi^2 \equiv \frac{d^2U}{d\varphi^2}(\bar\varphi), \ee which we interpret as the Planckion squared mass. So in this approximation the form of the Planckion potential is universally quadratic in any model. Information on the  Planckion mass will be obtained in Sec.~\ref{Inflation: quantum perturbations} by comparing the quantum predictions in this model with the observational constraints.

Far from the conformal regime the role of the inflaton can be played by other scalar fields as well, for example, the effective scalar $\zeta$ due to the $R^2$ term (Starobinsky inflation~\cite{Starobinsky:1980te}), the Higgs field (Higgs inflation~\cite{Bezrukov:2007ep}), etc. In the most general case all these scalars can contribute leading to a multi-field inflation (see e.g.~\cite{Kannike:2015apa,Salvio:2015kka,Salvio:2015jgu,Calmet:2016fsr,Ema:2017rqn,Gundhi:2018wyz,Karam:2018mft,Ema:2019fdd}).

\subsubsection{Inflation in the quasi-conformal regime}\label{Inflation in the quasi-conformal regime}

When $f_0\gg1$ and $\xi_{ab}\approx -\delta_{ab}/6$ (for all $a$ and $b$), namely when we are in the quasi-conformal regime, none of the inflationary models mentioned in Sec.~\ref{Inflation far from the conformal regime} can work. Indeed, inflation triggered by the Planckion and/or the Higgs require positive values of $\xi_\varphi$ and/or $\xi_H$~\cite{Bezrukov:2007ep,Isidori:2007vm,Hamada:2014iga,Bezrukov:2014bra,Hamada:2014wna}, respectively, while  Starobinsky inflation demands a small $f_0$~\cite{Salvio:2017xul}.

There are, however, other options that work well for $f_0\gg1$ and $\xi_{ab}\approx -\delta_{ab}/6$~\cite{Salvio:2019wcp}. An example is hilltop inflation~\cite{Boubekeur:2005zm}, where inflation occurs when a scalar field goes down from the top of a potential hill and is also viable in agravity when the non-minimal coupling of the scalar is close to the conformal value~\cite{Salvio:2019wcp}.  

Another interesting option is identifying the inflaton with a pseudo-Goldstone boson, $\phi_n$, a possibility that was proposed in the context of Einstein gravity in~\cite{Freese:1990rb} and then implemented in agravity in~\cite{Salvio:2019wcp}. Indeed, in this case the inflaton does not need to a have a (quasi-)Weyl invariant non-minimal coupling being morally the {\it phase} of some field representation of a spontaneously broken global group. This scenario is called ``natural inflation" because the inflaton potential is protected by the Goldstone theorem  from large quantum corrections.  Natural inflation is particularly relevant also because it admits an elegant UV completion within an asymptotically free QCD-like theory~\cite{Adams:1992bn,Salvio:2019wcp}. This is a very important feature as the quasi-conformal regime is motivated by the UV completion of agravity (see the end of Sec.~\ref{The action}). Furthermore, the asymptotically free theory, which provides the inflaton, might also be responsible at the same time for generating $\bp$ with the non-perturbative mechanisms of Sec.~\ref{Non-perturbative mechanisms}.

 We therefore give some details of this inflationary scenario here. 
The inflaton potential of natural inflation, $U_n$, can be written as~\cite{Freese:1990rb} 
\be U_n(\phi_n) = \Lambda_n^4 \left(1+\cos\left(\frac{\phi_n}{f_n}\right)\right), \label{VInf}\ee
where $\Lambda_n$ and $f_n$ are two energy scales.
This potential is even  and periodic with period $2\pi f_n$ so we restrict ourselves to  the interval $\phi \in [0, \pi f_n]$. 

The slow-roll quantities in~(\ref{1st-slow-roll}) and in the second equation in~(\ref{2nd-slow-roll}) can be written here as 
\be \epsilon_n \equiv\frac{\bp^2}{2} \left(\frac{1}{U_n}\frac{dU_n}{d\phi_n}\right)^2, \quad \eta_n \equiv \bp^2\frac{1}{U_n}\frac{d^2U_n}{d\phi_n^2} .\label{epsilon-eta-def}\ee
and so, using~(\ref{VInf}), 
\be \epsilon_n =\frac{\bp^2  \tan ^2\left(\frac{ \phi_n }{2 f_n}\right)}{2 f_n^2}, \quad \eta_n =-\frac{\bp^2  \cos \left(\frac{\phi_n }{f_n}\right)}{f_n^2 \left(1+\cos \left(\frac{\phi_n }{f_n}\right)\right)} .\label{epsilon-eta}\ee
These expressions show that the scale $f_n$ typically exceeds the Planck scale in order to have slow-roll natural inflation. More precise information on $f_n$, as well as on $\Lambda_n$, will be obtained in Sec.~\ref{Inflation: quantum perturbations} when  the quantum predictions of this model will be  compared with the observations.
 
\subsection{Inflation: quantum perturbations}\label{Inflation: quantum perturbations}

 We now turn to the study of the  perturbations around the FRW background, which can break homogeneity and isotropy and will be treated as quantum fields. With an appropriate gauge choice we can always write the full metric (the FRW background plus the perturbations) as~\cite{Salvio:2017xul}
\be ds^2= a(\tau)^2\left\{ (1+2C) d\tau^2 -2 \left(V_i+\frac{\partial_iB}{a(\tau)}\right) d\tau dx^i  - \left[(1+2\mathcal{R}) \delta_{ij}+h_{ij}\right]dx^idx^j\right\}, \label{dsPert}\ee
where $h_{ij}$, $V_i$, $B$, $C$ and $\mathcal{R}$ are the metric perturbations, which are  functions of both $\tau$ and the $x^i$. Morever,  the tensor perturbations $h_{ij}$  obey
 \be h_{ij}=h_{ji},  \qquad h_{ii} =0, \qquad \partial_ih_{ij}=0\label{Cond2}\ee
 and the vector perturbations $V_i$ satisfy
\be \partial_iV_i=0. \label{Cond1}\ee
Note that $h_{ij}$, $V_i$, $B$, $C$ and $\mathcal{R}$ represent the quantum fluctuations, $\hat h_{\mu\nu}$, in~(\ref{metricSplit}). The energy-momentum tensor should also be perturbed by adding to the homogeneous and isotropic part studied in Sec.~\ref{Inflation: the FRW background} a perturbation term that generically depends on both $\tau$ and the spatial coordinates $x^i$.

Detailed information on the perturbations can be obtained by solving the EOMs. By doing so one finds that at the linear level in the perturbations the tensor sector, the vector sector  and the scalar sector ($B$, $C$ and $\mathcal{R}$) do not mix with each other and can therefore be analyzed separately. Since the rotations are symmetries of the background,  it is convenient to consider the Fourier transforms of the perturbations with respect to the $x^i$ as we will do below.

\subsubsection{Tensor perturbations}\label{Tensor perturbations}

Let us start with the tensor perturbations $h_{ij}$. We can write them in terms of their Fourier transforms as
 \bea h_{ij}(\tau, \vec{x}) =  \int \frac{d^3q}{(2\pi)^{3/2}}  e^{i\vec{q}\cdot \vec{x}} \sum_{\lambda= \pm 2}  h_\lambda(\tau,\vec{q}) e^\lambda_{ij} (\hat q),  \label{ExpMom}  \eea
where $e^\lambda_{ij} (\hat q)$ are the usual polarization tensors for helicity $\lambda= \pm 2$. We recall that for $\hat q$ along the third axis the polarization tensors that satisfy (\ref{Cond2}) are  given by 
\be e^{+2}_{11} = -e^{+2}_{22} = \frac{1}{2}, \quad e^{+2}_{12} = e^{+2}_{21} = \frac{i}{2}, \quad e^{+2}_{3i} =e^{+2}_{i3} = 0, \quad  e^{-2}_{ij} =  (e^{+2}_{ij})^* \label{PolT}\ee 
and for a generic momentum direction $\hat q$ we can obtain $e^\lambda_{ij} (\hat q)$ by applying to (\ref{PolT}) a rotation that connects the third axis with  $\hat q$.
The polarization tensors defined in this way obey 
$ e^\lambda_{ij} (\hat q) (e^{\lambda'}_{ij} (\hat q))^* = \delta^{\lambda\lambda'}.  $

The EOMs of the $h_\lambda$ have been solved in detail on the de Sitter space in Ref.~\cite{Salvio:2017xul} (see also Refs.~\cite{Clunan:2009er,Deruelle:2010kf,Deruelle:2012xv,Myung:2014jha}
 for previous results).  One obtains
\be h_\lambda(\tau, \vec{q}) = a_\lambda(\vec{q}) y_2(\tau, q) +a^\dagger _{-\lambda}(-\vec{q}) y^*_2(\tau, q)+b_\lambda(\vec{q}) g_2(\tau, q) + b^\dagger _{-\lambda}(-\vec{q}) g^*_2(\tau, q), \label{hDecom}\ee
where  the tensor modes $y_2(\tau, q)$ and $g_2(\tau, q)$ are given by
\bea  y_2(\tau, q) &=&   \frac{\sqrt{2} H}{\bp q^{3/2}\sqrt{1+2\frac{H^2}{M_2^2}}}(1+i q\tau) e^{-i q\tau}, \label{GravitonMode} \\ g_2(\tau, q) &=&   \frac{\sqrt{\pi} He^{i\frac{\pi}4\left(1+\sqrt{1-4\frac{M^2_2}{H^2}}\right) }}{\bp q^{3/2}\sqrt{1+2\frac{H^2}{M_2^2}}} (-q\tau)^{3/2}\left(J_{\sqrt{\frac14 -\frac{M_2^2}{H^2}}}(-q\tau) + i Y_{\sqrt{\frac14 -\frac{M_2^2}{H^2}}}(-q\tau) \right), \label{GhostMode}   \eea
$q\equiv |\vec q|$ and  $J_ n(z)$  and $Y_n(z)$ are the   Bessel functions of the first and second kind, respectively.
Since the $h_{ij}$ are quantum fluctuations the coefficients   $a_\lambda(\vec{q})$ and $b_\lambda(\vec{q})$ are quantum operators. The detailed analysis of~\cite{Salvio:2017xul} has shown that, with the definition of the modes in~(\ref{GravitonMode}) and~(\ref{GhostMode}), $a_\lambda(\vec{q})$ and $b_\lambda(\vec{q})$ satisfy the commutation rules
\be [a_\lambda(\vec{q}), a_{\lambda'}^\dagger(\vec{k})] = \delta_{\lambda \lambda'}\delta^{(3)}(\vec{q}-\vec{k}),\qquad  [b_\lambda(\vec{q}), b_{\lambda'}^\dagger(\vec{k})] = -   \delta_{\lambda \lambda'}\delta^{(3)}(\vec{q}-\vec{k}) \label{CCaad2}\ee
and all the other commutators equal to zero. 
Therefore, the $a_\lambda(\vec{q})$ satisfy the standard commutation relations and we interpret the first two terms in~(\ref{hDecom}) as the quantum fluctuations of the massless graviton. On the other hand, $b_\lambda(\vec{q})$ satisfy unusual commutation rules with a minus sign, like in Eq.~(\ref{UnusComm}). As a result we interpret the third and fourth terms in~(\ref{hDecom}) as the quantum fluctuations of the spin-$(\pm 2)$ components of the massive spin-2 field.
 This interpretation is confirmed by the fact that $y_2(\tau, q)$ is proportional to the tensor mode  of the graviton in Einstein gravity, with a proportionality factor that tends to that in Einstein gravity in the limit $H/M_2\to 0$. 
 
So $a_\lambda(\vec{q})$ and $a_\lambda(\vec{q})^\dagger$ are the annihilation and creation operators of the massless graviton, while $b_\lambda(\vec{q})$ and $b_\lambda(\vec{q})^\dagger$ are the annihilation and creation operators of the massive graviton. Moreover, as we have seen in Sec.~\ref{The Dirac-Pauli canonical variables}, the canonical variables corresponding to the massive graviton,
\be h^{\rm DP}_\lambda(\tau, \vec{q}) \equiv b_\lambda(\vec{q}) g_2(\tau, q) + b^\dagger _{-\lambda}(-\vec{q}) g^*_2(\tau, q) \ee
should be treated as Dirac-Pauli variables, 
while those corresponding to the massless graviton,
\be h^0_\lambda(\tau, \vec{q}) \equiv  a_\lambda(\vec{q}) y_2(\tau, q) +a^\dagger _{-\lambda}(-\vec{q}) y^*_2(\tau, q),\ee
should be described by standard variables. This information should be kept in mind in computing the power spectra of the theory, as we will illustrate in Sec.~\ref{Predictions}.

In the limit of large $H/M_2$, we observe a suppression of the tensor perturbations due to the quantity $\sqrt{1+2H^2/M_2^2}$ in the denominator. This will lead to a suppression of the tensor-to-scalar ratio $r$, as we will see in~Sec.~\ref{Predictions}.

Notice that the tensor modes associated with the abnormal graviton, $g_2(\tau, q)$, vanish in the superhorizon limit, $\tau\rightarrow 0$. On the other hand, the modes $y_2(\tau, q)$, associated with the ordinary graviton, remain finite. Since we are phenomenologically interested in the case  $\tau \sim e^{-N_e}\ll  1$ (with a rather large number of e-folds, $N_e\gg 1$) we conclude that the modes $g_2(\tau, q)$ basically do not affect the observations. However, because  of the quantity $\sqrt{1+2H^2/M_2^2}$ in the denominator of $y_2(\tau, q)$, we see that  $M_2$ does affect in general the graviton perturbations. 

In a recent paper~\cite{Anselmi:2020lpp} the tensor perturbations were reanalyzed by using a  quantization in which the third and fourth terms in~(\ref{hDecom}) are projected out. This projection is proposed there as a new first principle. Since $g_2(\tau, q)\to 0$ as $\tau\to 0$ the physical predictions of the theory are unchanged by the projection of~\cite{Anselmi:2020lpp}.

\subsubsection{Vector perturbations}\label{Vector perturbations}

Vector perturbations can be treated similarly. Ref.~\cite{Salvio:2017xul} contains a detailed analysis of this sector too (see also~\cite{Clunan:2009er} for a previous discussion). 

The most important difference with respect to the tensor sector is that the massless graviton does not appear there (as already mentioned in Sec.~\ref{The Dirac-Pauli canonical variables}). The gravitational vector sector then only contains the spin-$(\pm 1)$ components of the massive abnormal graviton, which, as mentioned in Sec.~\ref{The Dirac-Pauli canonical variables}, must be treated as Dirac-Pauli variables. 

Just like the spin-$(\pm 2)$ components of the massive graviton discussed in Sec.~\ref{Tensor perturbations}, the de Sitter modes of the  spin-$(\pm 1)$ components  vanish in the superhorizon limit~\cite{Salvio:2017xul}  and therefore  do not affect the observables. This also implies that the physical predictions in the vector sector are as if the vector perturbations were projected out, as actually done  in~\cite{Anselmi:2020lpp}.

\subsubsection{Scalar perturbations}\label{Scalar perturbations}

The gravi-scalar perturbations  $B$, $C$ and $\mathcal{R}$ are not all independent. In particular $C$ can be expressed as a functional of $B$ and $\mathcal{R}$. Besides $B$ and $\mathcal{R}$, the scalar sector also includes the perturbations of the  4D scalar fields $\phi^i$. A detailed study of the full scalar sector can  be found in~\cite{Salvio:2017xul} as well (see also~\cite{Deruelle:2010kf,Myung:2014jha,Tokareva:2016ied,Ivanov:2016hcm} for previous partial analysis). Here we focus on the gravi-scalar perturbations as these contain the key differences with respect to the Einstein gravity case. 

The quantity $\mathcal{R}$ is the usual curvature perturbation. As shown in~\cite{Salvio:2017xul}, $\mathcal{R}$ is given by its known value in Einstein gravity, at least in the slow-roll approximation. Formul\ae~to compute $\mathcal{R}$ for a generic matter sector can be found in~\cite{Salvio:2017xul}.

The  $W^2$ term, however, can lead to the presence of another degree of freedom, corresponding to $B$, which can be interpreted as the  component with spin 0 (in a given direction) of the massive spin-2 field. As shown in~\cite{Salvio:2017xul}, the survival of $B$ in the superhorizon limit depends on the value of the ratio $M_2/H$ and the number of e-folds, $N_e$. For $M_2^2/H^2 \gtrsim 1/N_e$ the perturbation $B$ vanishes at superhorizon scales and therefore  does not affect the observations. For $M_2^2/H^2 \lesssim 1/N_e$, on the other hand, $B$ survives at those scales and leads to an isocurvature mode, whose compatibility with present observations will be discussed in Sec.~\ref{Predictions}. 

Considering the Fourier expansion of $B$
\be B(\tau, \vec{x})= \int \frac{d^3q}{(2\pi)^{3/2}}  e^{i\vec{q}\cdot \vec{x}} B_0(\tau,\vec{q}) ,  \label{ExpMomB}  \ee
one can decompose the Fourier transform $B_0(\tau,\vec{q})$ in terms of annihilation and creation operators
 ($b_0$ and $b_0^\dagger$, respectively)
\be B_0(\tau, \vec{q}) = b_0(\vec q) g_B(\tau, q) +b_0(-\vec q)^\dagger g_B(\tau, q)^*. \label{B0anncre}\ee 
Just like the annihilation and creation operators of the spin-$(\pm 1)$ and the spin-$(\pm 2)$ components of the massive spin-2 field (see Eq.~(\ref{CCaad2})), these operators satisfy the commutation relation with an unusual $-$ sign:
  \be [ b_0(\vec{k}),  b_0(\vec{q})^\dagger] = -  \delta(\vec{k}-\vec{q})  \label{bComm}\ee
  and all the other commutators equal to zero.
The mode functions $g_B(\tau, q)$ can be computed explicitly at the leading order in the slow-roll expansion and one finds the de Sitter modes~\cite{Salvio:2017xul,Salvio:2019ewf}
\be   g_B(\tau, q) \equiv \frac{H}{\bp\sqrt{12q}} \left(\frac{3}{q^2}+\frac{3i\tau}{q}-\tau^2\right)e^{-iq\tau} + {\cal R}\mbox{-terms}, \label{gBmodes} \ee
where the ``$ {\cal R}\mbox{-terms}$" are contributions due to ${\cal R}$, which, however, vanishes at superhorizon scales and thus their form is not needed to compute the observable predictions. 

Before moving to the predictions, let us note that the recent paper~\cite{Anselmi:2020lpp} projects out $B$ as well as the other components of the massive spin-2 field as an extra first principle. The consistency of this procedure, as discussed in~\cite{Anselmi:2020lpp}, requires $M_2/H >1/2$. Since $N_e\gg 1$, this condition implies $M_2^2/H^2 \gtrsim 1/N_e$, which corresponds to the absence of $B$ at superhorizon scales. As a result, the analysis of~\cite{Anselmi:2020lpp} gives predictions identical to those of~\cite{Salvio:2017xul} (summarized below) in its range of validity, that is when $M_2/H >1/2$ holds. This is the case for all perturbations (tensor, vector and scalar ones) modulo the fact that in~\cite{Anselmi:2020lpp} only Starobinsky inflation was studied (namely no scalars $\phi_a$ were considered), while~\cite{Salvio:2017xul} analyzed a general scalar sector.

\subsubsection{Predictions}\label{Predictions}

The observable predictions can now be extracted by computing the power spectra. These are given by the vacuum quantum averages of products of fields computed at equal times.

As we have seen in Sec.~\ref{The Dirac-Pauli canonical variables}, quantum averages of Dirac-Pauli variables involve an unusual $-i$ (see Eq.~(\ref{QADP})) and quantum averages of the product of two Dirac-Pauli variables involve an unusual $-1$ (see Eq.~(\ref{QA2DP})). We now illustrate how this leads to positive power spectra by compensating the unusual $-1$ appearing in the commutator of the corresponding annihilation and creation operators, Eq.~(\ref{UnusComm}). 

For this purpose we consider the vacuum quantum average of $B(\tau, \vec{x}) B(\tau,\vec{y})$, which we call $ \langle B(\tau, \vec{x}) B(\tau,\vec{y}) \rangle_{\rm vac}$, 
whose Fourier transform with respect to $\vec{x}-\vec{y}$ gives the power spectrum of $B$. Recalling now that $B$ should be treated as a Dirac-Pauli variable and using Eq.~(\ref{QA2DP}), we have
\be \langle B(\tau, \vec{x}) B(\tau,\vec{y}) \rangle_{\rm vac} = -\frac{\langle 0 |\eta B(\tau, \vec{x}) B(\tau,\vec{y})|0 \rangle}{\langle 0|\eta|0\rangle}, \label{QAB} \ee
where $|0 \rangle$ is the vacuum state (annihilated by all annihilation operators) and we used the fact that the norm operator of a Dirac-Pauli canonical variable is $\eta$ (see Eq.~(\ref{xx'innerP})). Now, from the discussion of Secs.~\ref{Calculation of probabilities} and~\ref{The Dirac-Pauli canonical variables} we know that $P_H |0\rangle=|0\rangle$ and $P_H=\eta$ so
\be \langle B(\tau, \vec{x}) B(\tau,\vec{y}) \rangle_{\rm vac} =  -\langle 0 | B(\tau, \vec{x}) B(\tau,\vec{y})|0 \rangle,\ee
where we have normalized the vacuum state in a way that $\langle 0|0\rangle =1$. By using now~(\ref{ExpMomB}),~(\ref{B0anncre}) and~(\ref{gBmodes}) we find 
\be \lim_{\tau \to 0^-} \langle B(\tau, \vec{x}) B(\tau,\vec{y}) \rangle_{\rm vac} = -\int \frac{d^3 q' d^3 q}{(2\pi)^3}  \frac{3H^2}{4\bp^2} \frac{\langle 0|[b_0(-\vec{q'}), b_0(-\vec q)^\dagger ]| 0 \rangle}{\sqrt{q'q}(q'q)^2} e^{i\vec{q}\cdot \vec{y}-i\vec{q'}\cdot \vec{x}}\ee
and, recalling the unusual minus sign in~(\ref{bComm}),
\be \lim_{\tau \to 0^-} \langle B(\tau, \vec{x}) B(\tau,\vec{y}) \rangle_{\rm vac} =  \int \frac{d^3 q}{(2\pi)^3} \frac{3H^2}{4\bp^2}\frac{e^{i\vec{q}\cdot (\vec{y}-\vec{x})}}{q^5} \label{stepPB}\ee
On the other hand, the power spectrum of (the gradient of) $B$, which we call $P_B$, satisfies
\be \lim_{\tau \to 0^-} \langle B(\tau, \vec{x}) B(\tau,\vec{y})\rangle_{\rm vac} = \int  \frac{d^3q}{4\pi q^3}e^{i\vec{q}\cdot (\vec{y}-\vec{x})} \frac{P_B}{q^2}\ee
and, comparing this expression with~(\ref{stepPB}),
\be P_B =\frac{3}{2\bp^2} \left(\frac{H}{2\pi}\right)^2 > 0.  \label{PBspectrum} \ee 
We see that the unusual minus sign appearing on the right-hand side of~(\ref{QAB}) cancels exactly the unusual minus sign in~(\ref{bComm}) to give a positive power spectrum. This is a particular case of the result in~(\ref{QA2DP}), which tells us that that the quantum average of the square of any observable in any state is not negative. 
The spectral index of $B$ defined by  $n_B\equiv 1 +\frac{d\ln P_B}{d\ln q}$, where $P_B$ is computed at horizon exit, $a(\tau) H(\tau) =q$, turns out to be~\cite{Salvio:2019ewf} (at the leading order in the slow-roll expansion) 
\be n_B = 1-2\epsilon, \ee
 where $\epsilon$ is the first slow-roll parameter in~(\ref{1st-slow-roll}). This result holds both in single-field and multi-field inflation. As shown in~\cite{Salvio:2019ewf}, although $B$ is not present in Einstein gravity $P_B$ is consistent with the most recent CMB bounds~\cite{Ade:2015lrj} in some well-motivated   inflationary scenarios as we will review in this section. Future observations of the CMB background, such as CMB-S4~\cite{Abazajian:2016yjj}, will be able to test further these scenarios.

Analogously one can compute all the other power spectra, which turn out to be  positive as they should. As we have mentioned, $\mathcal R$ is the same as in Einstein  gravity (without the $W^2$ term) in the slow-roll approximation and so the same is true for its power spectrum $P_\mathcal R$ (see e.g.~\cite{Salvio:2017xul,Sasaki:1995aw}  for an explicit formula  in the general multi-field case) and the corresponding spectral index $n_s$~\cite{Salvio:2017xul,Chiba:2008rp,Sasaki:1995aw}. The correlation between the power spectrum of this isocurvature mode and $\mathcal{R}$ is  suppressed at superhorizon scales~\cite{Salvio:2017xul}, i.e. $P_{\mathcal{R}B}=0$. It is also convenient to express $P_B$ in terms of $P_\mathcal{R}$ by introducing 
\be r'\equiv \frac{P_B }{P_\mathcal{R}} \label{rppar}\ee
to perform better the comparison with the observations.
   The power spectrum of tensor perturbations, on the other hand, is given by~\cite{Salvio:2017xul} 
\be  P_t = \frac{P_{tE}}{1+\frac{2 H^2}{M_2^2}}, \label{Ptspectrum} \ee 
where 
\be P_{tE}\equiv \frac{8}{\bar M_{\rm Pl}^2} \left(\frac{H}{2\pi}\right)^2 \ee
is the tensor power spectrum in Einstein gravity. The denominator in~(\ref{Ptspectrum}) comes from the quantity $\sqrt{1+2H^2/M_2^2}$ in the denominator of the mode functions $y_2(\tau, q)$  in~(\ref{GravitonMode}). When $M_2\gg H$ that denominator is close to 1 as it should because in that limit the theory approaches Einstein gravity; on the other hand, for $M_2\lesssim H$ it suppresses the tensor power spectrum. 
By taking the ratio between~(\ref{Ptspectrum}) and $P_\mathcal R$ we obtain the tensor-to-scalar ratio~\cite{Salvio:2017xul} 
\be r\equiv \frac{P_t}{P_\mathcal{R}}= \frac{r_E}{1+\frac{2 H^2}{M_2^2}}, \label{rW}\ee
where $r_E$ is the tensor-to-scalar ratio in Einstein gravity. So any model that is in trouble because predicting a  tensor-to-scalar ratio above the observational limits~\cite{Ade:2015lrj} can be saved in quadratic gravity by taking $M_2/H$ small enough (see below for some examples). Also, we can compute the spectral index $n_t$ of tensor perturbations, defined by $n_t =\frac{d\ln P_t}{d\ln q}$, where $P_t$ is computed at horizon exit, $a(\tau) H(\tau) =q$, to find
\be n_t = -\frac{2\epsilon}{1+\frac{2 H^2}{M_2^2}} \ee
independently of the number of inflatons.  This result reduces to
\be n_t = -\frac{r}{8} \ee
in single-field inflation (where $r_E =16\epsilon$). 

In conclusion the predictions of agravity nicely reduce to those of Einstein gravity for $M_2\gg H$.
 Using the observational bound on $H$~\cite{Ade:2015lrj}
\be H< 2.7\times 10^{-5}\bp \quad  (2\sigma~\mbox{level})
 \label{PlanckHboound}\ee
 and Eq.~(\ref{M2}) we see that all experimentally observable predictions of agravity reduce to those of Einstein gravity (in {\it any} inflationary scenario) for 
 $$f_2 \gg \sqrt{2}\,2.7\times 10^{-5}\sim 10^{-5}.$$

We now consider as examples some  single-field inflationary setups and present the corresponding specific predictions. 

\begin{itemize}
\item {\it Planckion inflation}. This is one of the choices of the inflaton that can be realized far from the conformal regime (see Sec.~\ref{Inflation far from the conformal regime}). In the approximation where the inflaton potential is quadratic, Eq.~(\ref{Uplanckionapp}), the inflationary predictions are~\cite{Salvio:2014soa,Salvio:2017xul}
\be n_s\approx 1-\frac{2}{N_e}\stackrel{N_e\approx 60}{\approx}0.967,\qquad
r'\approx \frac{3}{2N_e}\stackrel{N_e\approx 60}{\approx} 0.025 \qquad
r\approx \frac{1}{1+\frac{2 H^2}{M_2^2}}\frac{8}{N_e}\stackrel{N_e\approx 60}{\approx} \frac{0.13}{1+\frac{2 H^2}{M_2^2}} 
\nonumber. \ee
Taking into account the  data in~\cite{Ade:2015lrj}, the observed scalar amplitude $P_R=M_\varphi^2 N_e^2/(6\pi^2 \bp^2)$
is reproduced (taking  $N_e \approx 60$) for
\be M_\varphi\approx 1.4 \times 10^{13}~{\rm GeV}. \label{Ms} \ee 
Quadratic inflation in Einstein gravity is ruled out by the most recent CMB observations because it predicts a value of $r_E$ above the limits. As we have seen, this problem can be cured in agravity for $H/M_2$ large enough, see Eq.~(\ref{rW}). The power spectrum $P_B$, parameterized by $r'$, is within the observational bounds~\cite{Ade:2015lrj} at $2\sigma$ level (see Fig.~7 of~\cite{Salvio:2019ewf}). 

\item {\it Natural inflation}. This is one of the inflationary scenarios that can be realized in the quasi-conformal regime (see Sec.~\ref{Inflation in the quasi-conformal regime}). In Einstein gravity natural inflation is on the verge of being excluded because it predicts a large value of $r_E$: the most recent observations of $n_s$ and $r$ have retricted the parameter $f_n$ to lie in a very narrow band around $f_n\approx 6.6 \bp$~\cite{Salvio:2019wcp}.  Taking $f_n\approx 6.6 \bp$ we have 
\be n_s\stackrel{N_e\approx 60}{\approx}0.962,\qquad
r' \stackrel{N_e\approx 60}{\approx} 0.011 \qquad
r\approx \frac{r_E}{1+\frac{2 H^2}{M_2^2}}\stackrel{N_e\approx 60}{\approx} \frac{0.061}{1+\frac{2 H^2}{M_2^2}} 
,\ee
 \begin{figure}[t]
\begin{center}
 $ \includegraphics[scale=0.7]{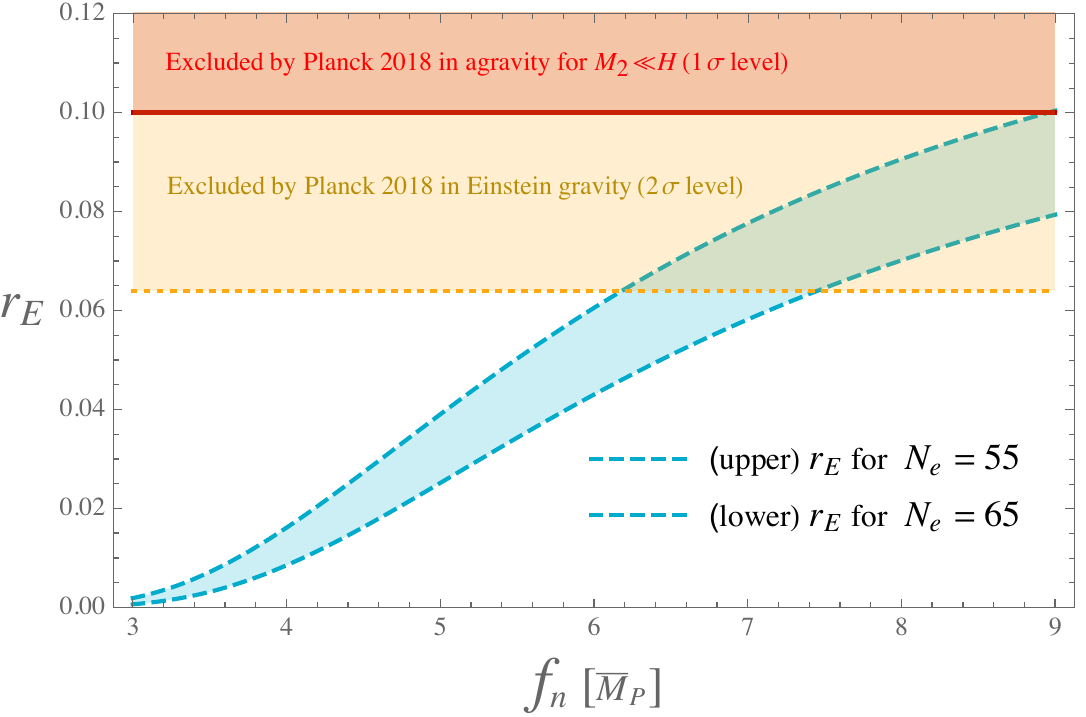} $
  \end{center}
   \caption{\em The Einsteinian tensor-to-scalar ratio $r_E$ in natural inflation as a function of $f_n$. In Einstein gravity $r_E$ is severely constrained by the most recent observations published by the Planck collaboration in 2018~\cite{Ade:2015lrj} because $f_n\approx 6.6 \bp$~\cite{Salvio:2019wcp}. The data also constrain agravity for $M_2\ll H$ as $r_E$ is related to the power spectrum of the extra isocurvature mode, however, in this case the observational bound is much weaker and as a result the model does not suffer from any tension with data.}
\label{rE}
\end{figure}
where we have set $\Lambda_n  \approx 6 \times 10^{-3} \bp$ to fit the observed value of the curvature power spectrum $P_R$.
By taking $H/M_2$ large enough we can have $r$ well below the observational bounds. Moreover, since $r_E\lesssim 0.1$ (see Fig.~\ref{rE}) the isocurvature power spectrum parameterized by $r'$ (see Eq.~(\ref{rppar})) also satisfies the most recent bounds at the 1$\sigma$ level, as illustrated by Fig.~7 of~\cite{Salvio:2019ewf}. This setup is therefore  favored by data~\cite{Ade:2015lrj} compared to Planckion inflation. A more detailed study of natural inflation in agravity is given by Ref.~\cite{Salvio:2019wcp}.
\end{itemize}
Other examples can be found in the literature, such as Starobinsky inflation, hilltop inflation (see~\cite{Salvio:2017xul} and~\cite{Salvio:2019wcp}, respectively, for their study in agravity) and Higgs inflation (whose predictions can be easily extracted by combining the results in Einstein gravity~\cite{Bezrukov:2007ep,Salvio:2013rja,Hamada:2014iga,Bezrukov:2014bra,Hamada:2014wna} with the general formul\ae~we have presented before).  It is  interesting to note that a too large value of $r_E$ in  ``critical" Higgs inflation within the SM~\cite{Hamada:2014iga,Bezrukov:2014bra,Hamada:2014wna,Masina:2018ejw,Salvio:2017oyf}
 can also be cured in agravity by taking $H/M_2$ large enough (but other solutions are possible if critical Higgs inflation is realized in an extended model~\cite{Salvio:2018rv}). Many more inflationary setups are possible.

   \subsubsection{Causality}\label{Causality}

 As mentioned in Sec.~\ref{Decays and scattering}, the fact that the decay rate $\Gamma_2$ appears with a negative sign in the resumed propagator~(\ref{resumed}) signals a violation of  causality at microscopic scales. 
 
However, the possible acausal effects (if they really occur) are extremely diluted by the expansion of the universe~\cite{Salvio:2019wcp}. Indeed,  whenever $M_2\lesssim H$, the width of the massive spin-2 field, $\Gamma_2$, (the only source of acausal behaviors) is   strongly suppressed compared to $H$ (see Eq.~(\ref{Gamma2})). This is because $H$ must satisfy the observational bound in~(\ref{PlanckHboound}) and so $\Gamma_2$ is suppressed compared to $M_2$ by the small factor $f_2^2/(16\pi)\lesssim 10^{-11}$. Therefore, any acausal process is quickly diluted by the expansion of the universe. In the opposite case,  $M_2\gg H$, the Weyl-squared term does not contribute to the experimentally  observable quantities as it essentially decouples (see Secs.~\ref{Tensor perturbations},~\ref{Vector perturbations} and~\ref{Scalar perturbations}). Therefore, the Weyl-squared term respects all observational causality bounds.

  \subsection{Classical metastability and chaotic inflation}\label{Classical metastability and chaotic inflation}

  As we have seen in Sec.~\ref{Classical aspects}, the classical runaways suggested by the Ostrogradsky theorem are avoided if the typical energies are below the bounds in~(\ref{E2}) and~(\ref{Em}). In this case the Ostrogradsky theorem does not imply an instability, but at most a metastability.
  
   In the context of inflation the typical energies we should look at are those associated with inhomogeneities and anisotropies. This is because the homogeneous and isotropic FRW metric is conformally flat, Eq.~(\ref{ConfFRW}), and thus has vanishing Weyl tensor. So only departures from homogeneity and isotropy can activate the $W^2$ term. 
 The homogeneous and isotropic ansatz used in Sec.~\ref{Inflation: the FRW background} (and regularly done in the literature on inflation) to describe the classical part of slow-roll inflation is justified in Einstein gravity when  the energy scales of inhomogeneities ($E_i$) and anisotropies ($E_a$) are small enough~\cite{Weinberg:2008zzc}:
\be  E_i \ll \left|\frac1{\Phi}\frac{dU}{d\Phi}\right|^{1/2}, \qquad E_i, E_a\ll H, \label{ChaosCond}\ee
where $\Phi$ is a generic canonically normalized inflaton field. In agravity we have the same conditions for $M_2\gtrsim H$ and even stronger conditions for $M_2\lesssim H$ as the extra scale $M_2$ appears in the equations.

The second condition in~(\ref{ChaosCond}), together with the observational bound on $H$ in Eq.~(\ref{PlanckHboound}), ensures that both~(\ref{E2}) and~(\ref{Em}) are satisfied for $f_2\gtrsim 10^{-9}$.
The first condition in~(\ref{ChaosCond}) is instead model dependent as the potential $U$ appears explicitly. In Planckion inflation in the quadratic approximation of Eq.~(\ref{Uplanckionapp}), using~(\ref{Ms}), one has 
\be  \left|\frac1{\Phi}\frac{dU}{d\Phi}\right|^{1/2} \sim 10^{-5} \bp \label{Lplanckion}\ee
and we have a similar result in natural inflation for the observationally allowed values of the parameters given  in Sec.~\ref{Predictions} (as well as in Starobinsky, Higgs and hilltop inflation~\cite{Salvio:2019ewf}). So we see that both~(\ref{E2}) and~(\ref{Em}) are satisfied for $f_2\gtrsim 10^{-9}$
 also assuming the first condition in~(\ref{ChaosCond}).  This bound on $f_2$ is very close but compatible with  the upper bound required to have a naturally small ratio $M_h/\bp$ (see Sec.~\ref{Applications to the hierarchy problem}).
 
 The proximity of these two bounds is actually a good thing because it allows us to understand why we live in a nearly homogeneous and isotropic universe~\cite{Salvio:2019ewf}. Indeed, for $f_2$ close to the maximal value required by Higgs mass naturalness, those patches where~(\ref{ChaosCond}) is violated typically have scales of inhomogeneities and/or anisotropies exceeding the bound in~(\ref{E2}) because of~(\ref{Lplanckion}) and the fact that $H$ is typically\footnote{This is the case, for example, for all inflationary scenarios mentioned above: Planckion, Starobinsky, hilltop, Higgs and natural inflation.} around the maximal value allowed in~(\ref{PlanckHboound}).  Life is rendered impossible there by the Ostrogradsky instabilities~\cite{Salvio:2019ewf}. What we have found provides a mechanism to implement the original chaotic inflation idea by Linde~\cite{Linde:1983gd}, which identified the patches compatible with life with those satisfying~(\ref{ChaosCond}). In Einstein gravity it is not  clear if  the patches that were largely inhomogeneous and anisotropic (where Conditions~(\ref{ChaosCond}) were violated) are incompatible with life: the fact that we cannot use an isotropic and homogeneous classical metric in that case is not sufficient to reach this conclusion. On the other hand, the classical runaways that can be triggered when those conditions are violated in the presence of the $W^2$ term certainly render   the universe inhospitable. So we see that taking $f_2$ close to $10^{-8}$ is not only motivated by Higgs mass naturalness, but also allows us to understand the homogeneity and isotropy of the universe.
 
\subsection{Reheating}\label{Reheating}

After inflation has taken place, the universe has to be heated.  
Several articles have shown that it is possible to obtain successful reheating, even keeping the Higgs mass naturally small compared to the Planck mass. Let us give a couple of examples that are relevant for agravity.

In the case of Planckion, Starobinsky and Higgs inflation, which are viable options only when we are far from the conformal regime, reheating can occur because the inflaton(s) couple to the particles of the SM. For Higgs inflation the couplings are simply the Yukawa and gauge couplings of the Higgs field~\cite{Bezrukov:2008ut}. However, Higgs inflation does not occur when $f_0$ is small enough to ensure a small $M_h/\bp$ without fine tuning~\cite{Kannike:2015apa}. In the Planckion and Starobinsky case the inflaton couplings to the SM particles appear because the divergence of the dilatation current $\mathscr{D}^\mu$ and the trace of the energy-momentum tensor are not zero~\cite{Kannike:2015apa}. Indeed, both 
$\partial_\mu \mathscr{D}^\mu$ and $T^{\mu}_{~\mu}$ contain (among others) the fields of the SM. The inflaton(s) can therefore  decay into the SM particles because the typical value of the inflaton mass is the one in Eq.~(\ref{Ms})~\cite{Kannike:2015apa}, which exceeds by several orders of magnitude the masses of all SM particles.

In natural inflation, on the other hand, what allows us to heat the universe is the presence of many weakly coupled scalars, which have sizable couplings to the observed particles and undergo large quantum fluctuations of order $H/(2\pi)$~\cite{Salvio:2019wcp}. Many weakly coupled scalars are, for example, present in asymptotically-free extensions of the SM, which allow us to extend agravity up to infinite energies (as seen in Sec.~\ref{The action}).

\section{Phase transitions and gravitational waves}\label{Phase transitions and gravitational waves}

It is well known that the phase transitions in models featuring a scale invariant or almost scale invariant behavior are generically of first order and can therefore  lead to observable GWs~\cite{Witten:1980ez}. This behavior is typical of theories where all scales are generated through dimensional transmutation, as in this case the breaking of scale invariance only occurs radiatively. Therefore, another important motivation for considering theories of this type is the possibility to test them through GW experiments. These include ground-based interferometers (such as advanced LIGO in Hanford and Livingston~\cite{Harry:2010zz,TheLIGOScientific:2014jea}, Cosmic Explorer~\cite{Evans:2016mbw,Reitze:2019iox} and 
Einstein Telescope~\cite{Punturo:2010zz, Hild:2010id, Sathyaprakash:2012jk})
as well as  space-based interferometers
(BBO~\cite{Crowder:2005nr, Corbin:2005ny, Harry:2006fi}, DECIGO~\cite{Seto:2001qf, Kawamura:2006},
and LISA~\cite{Audley:2017drz}), etc.

To illustrate this very interesting feature we consider the case of CW symmetry breaking described in Sec.~\ref{Perturbative mechanisms}. This case has been amply studied in the literature (see e.g. Refs.~\cite{Jaeckel:2016jlh,DelleRose:2019pgi,vonHarling:2019gme,Ghoshal:2020vud} for some studies of GWs and phase transitions in models with this type of breaking).

 \subsection{Thermal effective potential}\label{Thermal effective potential}
 
 In order to investigate the nature of a given phase transition, one should take into account both quantum and thermal corrections. The thermal effective potential reads
\be V_{\rm eff}(\chi, T) \equiv V_{\rm CW}(\chi) + V_{T}(\chi)+\Lambda_0,  \ee
where $V_{\rm CW}$ is given around the minimum $\chi_0$ in Eq.~(\ref{CWpot}), the term $V_{T}$ is the thermal correction  to the effective potential at finite temperature $T$, which at one-loop level is \cite{Dolan:1973qd} (see also \cite{Quiros:1994dr})
\be V_T(\chi) =  \frac{T^4}{2\pi^2} \left(\sum_b n_b J_B(M^2_b(\chi)/T^2)-\sum_f n_f J_F(M^2_f(\chi)/T^2)\right), \label{Veff} \ee
with  
\be J_{B,F}(x)\equiv \int_0^\infty dy \,  y^2\ln \left[1\mp\exp\left(-\sqrt{y^2+ x}\right)\right], \ee
and we have included  in	 $V_{\rm eff}(\chi, T)$ a constant term $\Lambda_0$ to account for the observed value of the cosmological constant $\Lambda$ when $\chi= \chi_0$: setting $T=0$, from~(\ref{CWpot}) we obtain 
\be \Lambda =\Lambda_0-\frac{\bar\beta_{\lambda_\chi}}{16}\chi_0^4.\ee
In the expression in (\ref{Veff}) the sum over $b$ runs over all bosons (with number of degrees of freedom $n_b$), that over $f$ runs over all fermions (with number of degrees of freedom $n_f$) and $M_{b,f}(\chi)$ are the corresponding background-dependent masses. 

In any QFT all background-dependent squared masses of spinor and vector fields are positive, but in some cases the squared masses  of the scalar fields (which are the eigenvalues of the Hessian matrix of the classical potential)  are negative. This always happens in models where symmetry breaking  is triggered through the Higgs mechanism, via a tachyonic mass term in the classical potential, because in this case the potential is concave for some values of the scalar fields. When this happens  $V_{\rm eff}$ has a non-vanishing imaginary part, which signals the breaking of the perturbative expansion. This cannot happen when symmetry breaking occurs through the CW mechanism: since the classical potential is purely quartic in this case, Eq.~(\ref{Vns}), the Hessian matrix  
\be M_{ab}^2 \equiv \frac{\partial^2V_{\rm ns}}{\partial\phi_a\partial\phi_b} \ee 
evaluated at the CW flat direction must be proportional to $\chi^2$ via some quartic couplings, namely
\be M_{ab}^2(\chi)=\frac12\lambda_{abcd}\nu_c\nu_d\chi^2. \ee
Clearly all eigenvalues of $M_{ab}^2$ must be positive to have a potential bounded from below.  Therefore, the CW  symmetry breaking supports the validity of perturbation theory.

\subsection{Phase transitions}\label{Phase transition}

\begin{figure}[t]
\begin{center}
\includegraphics[scale=0.69]{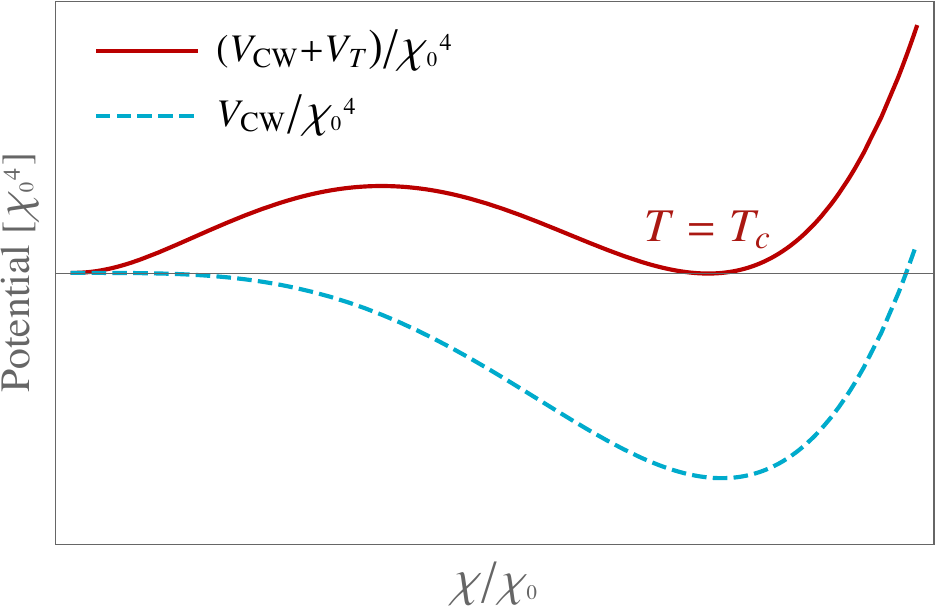}  
\end{center}
	\caption{\em  The effective potential   adding a constant such that it vanishes at $\chi=0$.}
\label{PT}
\end{figure}

The quantum part of the effective potential is very shallow because its departure from flatness is only due to perturbatively small loop effects. Therefore, the thermal correction can dominate the effective potential. By expanding the thermal functions for small values of their argument
one finds 
\bea J_B(x) &=& -\frac{\pi^4}{45}+\frac{\pi^2}{12} x -\frac{\pi}{6} x^{3/2} -\frac{x^2}{32} \ln\left(\frac{x}{a_B}\right) + ...,  \\
J_F(x) &=& \frac{7\pi^4}{360}-\frac{\pi^2}{24} x -\frac{x^2}{32} \ln\left(\frac{x}{a_F}\right) + ... , \eea
where $a_B = 16\pi^2 \exp(3/2-2\gamma_E)$, $a_F = \pi^2 \exp(3/2-2\gamma_E)$ and  $\gamma_E$ is the Euler-Mascheroni constant (the complete series expansion of $J_B$ and $J_F$ can be found in e.g.~\cite{Quiros:1994dr}). So the thermal effects induce a positive quadratic term in $\chi$ near $\chi=0$  and generically lead to two minima, separated by a potential barrier. If one lowers $T$ from very high values, where the symmetry is typically unbroken~\cite{Weinberg:1974hy}, down to some critical value of the temperature, $T_c$, the two minima are degenerate and a first-order phase transition starts. This typical situation is illustrated in Fig.~\ref{PT}, where we show  $V_{\rm eff}$ as a function of $\chi$  for two values of the temperature, the critical temperature $T_c$ and $T=0$. 

 
The absolute minimum of the thermal effective potential is at $\langle \chi\rangle=0$ for $T>T_c$, while, for $T<T_c$, is at a non-vanishing temperature-dependent value. In the latter case the decay rate per unit volume $\Gamma$ of the false vacuum $\chi=0$ into the true vacuum $\chi=\langle\chi\rangle\neq 0$ can be computed with the formalism of~\cite{Coleman:1977py,Callan:1977pt,Linde:1980tt,Linde:1981zj}:
\be  \Gamma\approx \max\left(T^4\left(\frac{S_3}{2\pi T}\right)^{3/2}\exp(-S_3/T)\,, \,\, \frac{1}{R_4^4}\left(\frac{S_4}{2\pi}\right)^{2}\exp(-S_4)\right). \label{GammaOption}\ee
Here $S_{d}$ is the Euclidean $d$-dimensional thermal effective action of $\chi$
evaluated at the regular solution of the corresponding Euclidean  EOM with O($d$) symmetry (which thus only depends on some radial coordinate $\rho$) satisfying the boundary condition $\chi(\rho) \stackrel{\rho\to\infty}{\to} 0$.
 This solution is known as the O($d$) bounce.
Also, $R_4$ is the size of the $O(4)$ bounce. The first entry in the $\max$ appearing in~(\ref{GammaOption}) is typically bigger than the second one when the potential is dominated by thermal effects.

 The phase transition occurs through the nucleation of bubbles of the true vacuum inside a space filled by the false vacuum $\chi=0$.
In the situation we are studying, when $T$ goes below $T_c$, the scalar field $\chi$ is typically trapped in the false vacuum  until  $T$ is much below $T_c$, in other words the universe features a phase of strong supercooling. 

To understand why this happens note that the bubbles created are  diluted by the expansion of the universe and they cannot collide until $T$ reaches the nucleation temperature $T_n$, which is defined as the temperature when $\Gamma/H^4 = 1$, where $H$ the is Hubble rate. When the decay is dominated by the O(3) bounce, $T_n$ can be equivalently defined as the solution of the following equation
\be  \frac{S_3}{T}-\frac32 \ln \left(\frac{S_3/T}{2\pi}\right) = 4 \ln \left(\frac{T}{H}\right)\label{TnEq}\ee
with respect to $T$.
The key quantity that allows us to understand how long cooling lasts is  the inverse of the duration of the phase transition which, following Ref.~\cite{vonHarling:2019gme}, we define through
 \be\beta_n \equiv \left[\frac1{\Gamma}\frac{d\Gamma}{dt}\right]_{T_n}.\ee This quantity, using $dT/T = -H dt$ and the fact that the transition is dominated by thermal effects, can be computed with the formula

\be \frac{\beta_n}{H_n} \approx \left[T\frac{d}{dT}(S_3/T)-4-\frac{3}{2}T\frac{d}{dT}\ln(S_3/T)\right]_{T=T_n},  \label{betaFR}\ee
where $H_n$ is the Hubble rate evaluated  at $T=T_n.$
Now, in a quasi-scale invariant regime $S_3/T$ is nearly temperature independent and so the phase transition lasts a very long time (supercooling). In this situation one can neglect the last term in~(\ref{betaFR}). Supercooling gives often a $T_n$ that is several orders of magnitude below the typical scale of the potential, $\chi_0$. 

During supercooling the energy density  is dominated by the vacuum energy of $\chi$ and the universe grows exponentially  like during primordial inflation, but with Hubble rate 
\be H_f = \frac{\sqrt{\bar\beta_{\lambda_\chi}}\, \chi_0^2}{4\sqrt{3}\bp}\ee
associated with the false vacuum. At the end of this inflationary period the universe has to be reheated so the energy stored in $\chi$ should  be transferred to the SM particles. The details of reheating depend on the model
 and on the specific transition under study so we do not discuss it here.

Finally, the strength of the phase transition is measured by the parameter $\alpha$ defined as the ratio between 
\be \rho(T_n) \equiv \left[\frac{T}{4} \frac{d}{dT} \Delta V_{\rm eff}(\langle\chi\rangle,T)-\Delta V_{\rm eff}(\langle\chi\rangle,T)\right]_{T=T_n}, \ee
where $\Delta V_{\rm eff}(\langle\chi\rangle,T)\equiv V_{\rm eff}(\langle\chi\rangle,T)-V_{\rm eff}(0,T)$, and the energy density of the thermal plasma (more details can be found in Refs~\cite{Caprini:2019egz,Ellis:2019oqb}). So
\be \alpha \equiv \frac{30 \rho(T_n)}{\pi^2 g_*(T_n)T_n^4}, \ee
where $g_*(T)$ is the effective number of relativistic species in thermal equilibrium at temperature $T$.
 The parameter $\alpha$ is typically very large because of supercooling and so the phase transition is very strong.

 \subsection{Gravitational waves}

When the temperature drops below $T_n$ GWs are produced. The production is maximised when the energy scale $\Lambda_G$  above which gravity is softened (see Sec.~\ref{Softening of gravity}) is larger than the relevant scales. These include  $\chi_0$ (which in turn is typically much larger than $T_n$, as we have seen), the values of fields and their derivatives as well as the temperature $T_{\rm RH}$ associated with the reheating after supercooling. Therefore, we  focus from now on  values of $\Lambda_G$ such that this condition is satisfied.

In this case the gravitational corrections to the false vacuum decay studied in Sec.~\ref{Phase transition} are  amply negligible  for $\Lambda_G\ll \bp$, which is necessarily true if one wants to solve the hierarchy problem through softened gravity (see the bound in~(\ref{GsoftB})). These corrections are, indeed, suppressed by factors at least as small as $\Lambda_G^2/\bp^2$~\cite{Salvio:2016mvj,Joti:2017fwe}.

 The dominant source of GWs are typically  bubble collisions that take place in the vacuum: in the era when $T$ reaches $T_n$ the energy density is dominated since a long time by the vacuum energy density associated with $\chi$, which leads to an exponential growth of the cosmological scale factor, as we have seen. This inflationary behavior as usual dilutes preexisting matter and radiation and thus one can typically neglect the GW production due to turbulence and sound waves  in the cosmic fluid~\cite{Maggiore:2018sht}.  

From~\cite{Caprini:2015zlo} (which used, among other things, the results of~\cite{Huber:2008hg}  based on the ``envelope approximation") one finds the following GW spectrum 
due to vacuum bubble collisions (valid in the presence of supercooling and $\alpha\gg 1$)
\be \label{eq:gw_col} h^2 \Omega_{\rm GW}(f_{\rm GW}) \approx 1.29
\times 10^{-6}\left(\frac{H(T_{\rm RH})}{\beta_n}\right)^2\left(\frac{100}{g_*(T_{\rm RH})}\right)^{1/3}\frac{3.8(f_{\rm GW}/f_{\rm peak})^{2.8}}{1+2.8(f_{\rm GW}/f_{\rm peak})^{3.8}},\ee 
where    $f_{\rm peak}$ is the red-shifted frequency peak today and is given by~\cite{Caprini:2015zlo} 
\be f_{\rm peak} \approx 3.79\times 10^2\frac{\beta_n}{H(T_{\rm RH})}\frac{T_{\rm RH}}{10^{10}{\rm GeV}} \left( \frac{g_*(T_{\rm RH})}{100}\right)^{1/6} \, {\rm Hz} . \ee
Some progress has been made to compute the GW spectrum beyond the envelope approximation~\cite{Jinno:2017fby,Konstandin:2017sat,Lewicki:2020jiv}, but Eq.~(\ref{eq:gw_col}) remains to date a reasonable and simple  approximation for the bubble collisions that take place in the vacuum~\cite{Konstandin:2017sat} (which is what we are mainly interested in for quasi scale-invariant theories).


\vspace{0.2cm}

The formalism explained in this and the  previous sections can be used to calculate the key features of the phase transitions and the GW spectrum in a variety of models, where the scales are generated through DT. 
For example, the EW phase transition and the corresponding GW spectrum has been studied in~\cite{Espinosa:2008kw,Sannino:2015wka,Marzola:2017jzl}. The phase transition associated with the breaking of the Pati-Salam gauge group down to the SM gauge group was analyzed in~\cite{Huang:2020bbe}.
The Peccei-Quinn (PQ)~\cite{Peccei:1977hh} case has been investigated in \cite{DelleRose:2019pgi,vonHarling:2019gme}, which focused on effective models, and in~\cite{Ghoshal:2020vud}, which considered a fundamental asymptotically free PQ model previously built in~\cite{Salvio:2020prd}. 
The analysis of Refs.~\cite{Ghoshal:2020vud,Salvio:2020prd} also provides an example of how predictive a theory with CDT can be, especially if combined with requirement of having UV fixed points for all couplings: several couplings of this model are indeed predicted in the IR, where the observations are made. The predictivity of the model in~\cite{Ghoshal:2020vud} leads to 
a rigid dependence of the phase transition (like its duration and the nucleation temperature) and the GW spectrum on the PQ symmetry breaking scale and the QCD gauge coupling.
 
\section{Summary and conclusions}\label{Conclusions}

To conclude, we have provided a review (with some original results) of theories of all interactions where the mass scales are generated quantum mechanically through DT. We have referred to this scenario
as agravity to emphasize that gravitational interactions are included. 

Let us provide here a summary of the main points of the paper, which can be useful after reading the bulk of the review (for a more introductory summary see Sec.~\ref{Introduction}).
\begin{itemize}
\item The general field content and action of agravity has been provided in Sec.~\ref{General no-scale action in field theory}. Quantum consistency requires that, for a given field content, all terms compatible with the symmetries must be present, including, among others, the $W^2$ and the $R^2$ term. The (strong) equivalence principle (as precisely stated in e.g.~\cite{WeinbergGravity}) has been assumed. This fixes $\Gamma_{\mu \, \nu}^{\,\sigma}$ to be the Levi-Civita connection. Any other  connection is either physically equivalent or implies a violation of the equivalence principle.  The matter content includes all types of fields and interactions compatible with renormalizability. 

In Sec.~\ref{General no-scale action in field theory} we have also reviewed how agravity can be UV complete: it can  flow to conformal gravity ($1/f_0\to 0$ and $\xi_{ab}\to -\delta_{ab}/6$) in the infinite energy limit provided that all matter couplings reach UV fixed points (which can be either free or interacting). This allows us to avoid tachyonic instabilities in the gravity sector. 

\item Various mechanisms for DT in theories including gravity have been described in Sec.~\ref{Dimensional transmutation and gravity}.

 In the perturbative approach of Sec.~\ref{Perturbative mechanisms} a gravitational generalization of the CW mechanism has been described, including a novel extension to  the case of multiple scalar fields. The key conditions for the gravitational CW mechanism to successfully occur are given in~(\ref{eq:agravMPl}). Both the standard~\cite{Coleman:1973jx,Gildener:1976ih} and the gravitational~\cite{Salvio:2014soa} CW mechanisms can contribute to $M_h$. 

The non-perturbative generation of mass scales (including $\bp$, $\Lambda$ and $M_h$) has been described in Sec.~\ref{Non-perturbative mechanisms}.

After DT has generated $\bp$, the graviparticles acquire a non-trivial mass spectrum (that of quadratic gravity), which has been described in Sec.~\ref{Quadratic gravity}.

\item One of the graviparticle has spin-2, mass $M_2 = f_2\bp/\sqrt{2}$ and negative classical kinetic energy. Because of the latter abnormal feature this particle is sometimes called ``a ghost". Nevertheless, in Sec.~\ref{The Weyl-squared term} we have provided possible solutions of the related issues both at the classical  and  quantum level. It must be kept in mind, though, that our approach has been intrinsically perturbative because based on the metric split in Eq.~(\ref{metricSplit}). 

At the classical level, we have reviewed how the runaway solutions suggested by the Ostrogradsky theorem can be avoided. This is the case if the energies associated with the derivative of the spin-2 fields respect Condition~(\ref{E2}) and those associated with the derivatives or mass terms of matter fields and/or matter-field values times coupling constants fulfill~(\ref{Em}). Both conditions should be imposed at the space and time boundaries.

At the quantum level the theory has been shown to be unitary even beyond the scattering theory: all probabilities are non-negative and sum up to one. This emerges once a correct definition of probabilities (which takes into account how they are actually introduced  in experiments) is used: the frequency argument of Sec.~\ref{Calculation of probabilities} (which for the first time has been given in a complete  pedagogical form) has shown that the indefinite metric is not the one that should be used to compute probability. Other metrics, which are positively defined and observable dependent, should be used to this purpose. Therefore, the solution proposed here makes use of a modification of quantum mechanics. Another non-standard quantum feature is the fact that the canonical coordinates of the abnormal graviton must be treated as Dirac-Pauli variables, as described in Sec.~\ref{The Dirac-Pauli canonical variables}.
Using this theoretical framework the ghost decay rate is non-vanishing and  positive; it is given at the leading non-trivial order in $f_2$ (and at zero order in the other couplings) in Eq.~(\ref{Gamma2}). Causality may be violated at microscopic scales, but, as discussed in Secs.~\ref{Decays and scattering} and~\ref{Causality}, this is within the observational bounds, at least for the values of $f_2$ of interest.

\item The presence of the extra graviparticles softens gravity  when the energy exceeds a critical value $\Lambda_G$ (see Fig.~\ref{softened}), which generically depends on $M_0$ and $M_2$. Far from the conformal regime $\Lambda_G$ is given in Eq.~(\ref{LambdaG1}). Here, for the first time, we have given $\Lambda_G$ in the quasi-conformal regime of Refs.~\cite{Salvio:2017qkx,Salvio:2019wcp}, that is $f_0\gg 1$ and $\xi_{ab}\approx -\delta_{ab}/6$ (see Eq.~(\ref{LambdaG2})).

The softening of gravity has several phenomenological applications. The small ratio $M_h/\bp$ can be made natural thanks to the Higgs shift symmetry  that can be softly broken at the scale $M_h$ (see Sec.~\ref{Applications to the hierarchy problem}). In order for this to happen not only gravity should be softened at scales $\Lambda_G\lesssim 10^{11}$~GeV, but the matter sector  should preserve this shift symmetry as well. This leads to new physics not far from 10~TeV. On the other hand, the softening of gravity is not sufficient to address the cosmological constant problem, whose solution   should be found somewhere else.

However, the softening of gravity 	does imply that microscopic BHs of Einstein gravity (satisfying Condition~(\ref{LinCond})), which can be considered as the endpoint of gravitational collapse, are replaced by horizonless ultracompact objects that are free from any singularity, as discussed in Sec.~\ref{Applications to ultracompact astrophysical objects}. This has interesting astrophysical implications, e.g. such objects can play the role of DM.  Moreover, in this way one can avoid the formation of microscopic BHs with horizon $r_h$ smaller than $1/h_{\rm max}$, where $h_{\rm max}$ is the value of the Higgs radial field for which the effective Higgs potential acquires its maximum. This is useful because such BHs have been proved to be very dangerous if the EW vacuum is metastable. 

\item An important part of this work is the discussion of the physical implications regarding the early universe in Sec.~\ref{Inflation}. 

We have remarked that, since quadratic gravity can flow to conformal gravity in the UV and the FRW metric is conformally flat, the initial-time cosmological singularity of GR can  be elegantly avoided.

We have mentioned examples of inflationary scenarios that work far from the conformal regime (e.g. Planckion, Starobinsky, etc.) and others that work in the quasi-conformal regime where $f_0\gg 1$ and $\xi_{ab}\approx -\delta_{ab}/6$ (natural inflation and hilltop inflation). 

Sec.~\ref{Inflation: quantum perturbations} was devoted to the study of quantum fluctuations generated during inflation and the corresponding predictions. By using the non-standard quantum theory developed earlier in Sec.~\ref{Quantum aspects}, the power spectra have been proved to be positive (such proof appears here for the first time). When $M_2\gg H$ the abnormal graviton does not affect the predictions, which then turn out to be the same as in Einstein gravity. For $M_2\lesssim H$, on the other hand, we find different predictions that can  be tested with future CMB observations: $r$ is suppressed as shown in Eq.~(\ref{rW}) and (for $M_2/H\lesssim 1/\sqrt{N_e}$) an isocurvature mode of gravitational nature (one of the spin components of the massive spin-2 field) appears and is compatible with the most recent CMB data~\cite{Ade:2015lrj}. We have also briefly reviewed how reheating  can happen in Sec.~\ref{Reheating}.

Moreover, in Sec.~\ref{Classical metastability and chaotic inflation}, we have shown (for a larger class of inflationary models compared to previous papers) that taking $f_2$ close to the maximal value compatible with Higgs naturalness ($f_2 \lesssim 10^{-8}$) allows us to explain the nearly homogeneity and isotropy of our universe: patches where the FRW ansatz in the classical part of inflation is not justified (which do not satisfy~(\ref{ChaosCond})) are rendered inhospitable by the Ostrogradsky instabilities. On the other hand, patches (like ours) where those conditions are satisfied avoid the Ostrogradsky runaway solutions. So the presence of the $W^2$ term gives us a concrete implementation of Linde's chaotic inflation. 

\item Last but not least, theories that are nearly scale invariant (like those where DT is a small perturbative effect)  typically predict very long and strong phase transitions, which can lead to observable GWs at detectors. We have reviewed this topic in Sec.~\ref{Phase transitions and gravitational waves}, which also provided the main theoretical tools to analyse phase transitions and GWs in this type of theories: the effective thermal potential, the theory of vacuum decay at finite temperature and the main source of GWs. It has also been pointed out for the first time that the CW symmetry breaking supports the validity of perturbation theory in studying phase transition because the imaginary part of the thermal effective potential must vanish in that case.
One should keep in mind that in order to maximise the production  of GWs,  $\Lambda_G$ has to be chosen above the typical scale of the potential, the values of the fields and their derivatives as well as $T_{\rm RH}$.
\end{itemize}

Some challenges are left for future research. First, our approach to quantum gravity has been intrinsically perturbative (see Eq.~(\ref{metricSplit})). How to define agravity non perturbatively is not clear. This does not only regards the matter sector\footnote{Currently we do not have a non-perturbative operative definition (such as the lattice) of realistic QFTs that feature chiral fermion representations of the gauge group, such as the SM.}, but also the gravity sector. Perhaps this could be done with some sort of lattice Euclidean path integral with respect to both the metric and the matter fields. 

Another interesting topic for future developments is the cosmological constant problem. Is it possible to have a naturally small $\Lambda/\bp$? If not, is it possible to understand the smallness of  the cosmological constant in some other way?

The list of unanswered question does not stop here. For example, can we have a simultaneous solution to all fine-tuning problems of the SM and GR? Can we understand the flavor and gauge-group structure of the SM? etc. 


\subsection*{Acknowledgments}
I thank the  International Journal of Modern Physics A for inviting me to write this review. I also thank my collaborators on the topics discussed here: A.~Strumia, H.~Veerm\"ae, G.~F.~Giudice, G.~Isidori, K.~Kannike, L.~Pizza, A.~Racioppi, M.~Raidal, G.~M.~Pelaggi, A.~Ghoshal, A.~D.~Plascencia, F.~Sannino, J.~Smirnov.

 \footnotesize
\begin{multicols}{2}

\end{multicols}

\end{document}